\documentclass[onecolumn,authoryear]{els-mrw} 

\usepackage{amsmath,amssymb,amsfonts,amsthm,makeidx,graphicx}
\usepackage{txfonts}
\usepackage{helvet}
\usepackage{siunitx}
\usepackage{sidecap}
\usepackage{hyperref}
\usepackage{grffile}
\usepackage{tcolorbox}
\sidecaptionvpos{figure}{t}
\newcommand{\HeI}[1]{\mbox{He\,{\sc i}~$\lambda${#1}}}
\newcommand{\HeII}[1]{\mbox{He\,{\sc ii}~$\lambda${#1}}}
\newcommand{\LiI}[1]{\mbox{Li\,{\sc i}~$\lambda${#1}}}
\newcommand{\CII}[1]{\mbox{C\,{\sc ii}~$\lambda${#1}}}
\newcommand{\CIIIt}[1]{\mbox{C\,{\sc iii}~$\lambda\lambda\lambda${#1}}}
\newcommand{\CIV}[1]{\mbox{C\,{\sc iv}~$\lambda${#1}}}
\newcommand{\NII}[1]{\mbox{N\,{\sc ii}~$\lambda${#1}}}
\newcommand{\NIIIt}[1]{\mbox{N\,{\sc iii}~$\lambda\lambda\lambda${#1}}}
\newcommand{\NIV}[1]{\mbox{N\,{\sc iv}~$\lambda${#1}}}
\newcommand{\OII}[1]{\mbox{O\,{\sc ii}~$\lambda${#1}}}
\newcommand{\SiIId}[1]{\mbox{Si\,{\sc ii}~$\lambda\lambda${#1}}}
\newcommand{\SiIII}[1]{\mbox{Si\,{\sc iii}~$\lambda${#1}}}
\newcommand{\SiIV}[1]{\mbox{Si\,{\sc iv}~$\lambda${#1}}}
\newcommand{\NaId}[1]{\mbox{Na\,{\sc i}~$\lambda\lambda${#1}}}

\newcommand{\MgIt}[1]{\mbox{Mg\,{\sc i}~$\lambda\lambda\lambda${#1}}}
\newcommand{\MgII}[1]{\mbox{Mg\,{\sc i}~$\lambda${#1}}}
\newcommand{\KId}[1]{\mbox{K\,{\sc i}~$\lambda\lambda${#1}}}
\newcommand{\FeI}[1]{\mbox{Fe\,{\sc i}~$\lambda${#1}}}
\newcommand{\CaII}[1]{\mbox{Ca\,{\sc ii}~$\lambda${#1}}}
\newcommand{\CaIIt}[1]{\mbox{Ca\,{\sc ii}~$\lambda\lambda\lambda${#1}}}
\newcommand{\CaI}[1]{\mbox{Ca\,{\sc i}~$\lambda${#1}}}
\newcommand{\MnI}[1]{\mbox{Mn\,{\sc i}~$\lambda${#1}}}
\newcommand{\SrII}[1]{\mbox{Sr\,{\sc ii}~$\lambda${#1}}}
\newcommand{\YII}[1]{\mbox{Y\,{\sc ii}~$\lambda${#1}}}
\newcommand{\BaII}[1]{\mbox{Ba\,{\sc ii}~$\lambda${#1}}}

\newcommand{\CrI}[1]{\mbox{Cr\,{\sc i}~$\lambda${#1}}}
\def\Teff{\mbox{$T_{\rm eff}$}}
\def\logg{\mbox{$\log g$}}
\def\logT{\mbox{$\log T_{\rm eff}$}}
\def\vsini{\mbox{$v\,\sin i$}}
\def\Msol{\mbox{M$_\odot$}}
\def\BPRP{\mbox{$G_{\rm BP}-G_{\rm RP}$}}

\begin{document}

\chapter{Spectral classification}\label{chap1}

\author[1]{Jes\'us Ma{\'\i}z Apell\'aniz}%
\author[2]{Ignacio Negueruela}%
\author[1]{Jos\'e A. Caballero}%

\address[1]{\orgname{Centro de Astrobiolog{\'\i}a}, \orgdiv{Campus ESAC}, \orgaddress{Camino bajo del castillo s/n, \num[detect-all]{28692} Villanueva de la Ca\~nada, Madrid, Spain}}
\address[2]{\orgname{Universidad de Alicante}, \orgdiv{Departamento de F{\'\i}sica Aplicada}, \orgaddress{Carretera de San Vicente s/n, 03\,690 San Vicente del Raspeig, Alicante, Spain}}


\maketitle

\vspace{-5mm}

\begin{glossary}[Glossary]
\term{Dwarf} or class V is a spectroscopic luminosity class in the MK system for stars with low luminosity for their spectral type.

\term{Giant} or class III is a spectroscopic luminosity class in the MK system for stars with intermediate luminosity for their spectral type.

\term{Harvard system} is the spectral classification system designed by the Harvard group between the 1890s and the 1910s.

\term{MK system} is the spectral classification system designed by William W. Morgan and Philip C. Keenan in the 1940s.

\term{Secchi system} is the first spectral classification system, designed by Angelo Secchi in the 1860s.

\term{Spectral classification} is the division of stars into classes based on their spectra, with the most common system being the MK one.

\term{Supergiant} or class I is a spectroscopic luminosity class in the MK system for stars with high luminosity for their spectral type.
\end{glossary}

\vspace{-5mm}

\begin{glossary}[Nomenclature]
\begin{tabular}{@{}lp{34pc}@{}}
AGB, pAGB  &(Post-) Asymptotic Giant Branch, evolutionary phase for a low- or intermediate-mass star burning He in a shell\\
CMD, CAMD  &Color-(Absolute)-Magnitude Diagram\\
CTTS/WTTS  &Classical/Weak-lined T-Tauri Star, low-mass PMS star with strong/weak emission lines\\
CV         &Cataclysmic Variable, binary system composed of a white dwarf and a mass-donor star\\
DIB        &Diffuse Interstellar Band\\
EW         &Equivalent Width, area of a spectral line with respect to the continuum in a rectified spectrum expressed in \AA\\
FWHM       &Full Width at Half Maximum\\
GOSSS      &Galactic O-Star Spectroscopic Survey, \citet{Maizetal11}\\
HB         &Horizontal Branch, evolutionary phase for a low-metallicity low-mass star burning He in its core\\
HRD        &Hertzsprung-Russell Diagram\\
LBV        &Luminous Blue Variable, evolutionary phase for massive stars with high spectroscopic and photometric variability \\
LPV        &Long Period Variable, red luminous star characterized by its large variability in long time scales\\
LiLiMaRlin &\textbf{Li}brary of \textbf{Li}braries of \textbf{Ma}ssive-Star High-\textbf{R}eso\textbf{l}ut\textbf{i}o\textbf{n} Spectra, \citet{Maizetal19a}\\
LMC        &Large Magellanic Cloud\\
\logg      &Logarithmic surface gravity\\            
MS         &Main Sequence, evolutionary phase during which a star burns $^1$H in its core\\
PMS        &Pre-Main Sequence, evolutionary phase prior to $^1$H ignition in its core\\
PNN        &Planetary Nebula Nucleus, evolutionary phase for a low- or intermediate-mass star after ejecting its outer layers\\
RC         &Red Clump, evolutionary phase for a high-metallicity low-mass star burning He in its core\\
SB2, SB3   &Double- or triple-lined spectroscopic binary (multiple) star\\
sd         &Subdwarf star, spectroscopic luminosity class located below the dwarf sequence\\
SMC        &Small Magellanic Cloud\\
\Teff      &Effective temperature\\
WD         &White Dwarf, evolutionary end phase of most low- and intermediate-mass stars\\
WR         &Wolf-Rayet star. spectroscopic class with broad emission lines of H, He, C, N, and/or O \\
ZAMS       &Zero-Age Main Sequence, evolutionary phase at which a star ignites $^1$H in its core
\end{tabular}
\end{glossary}

\begin{abstract}[Abstract]
Spectral classification is the division of stars into classes based on their spectral characteristics. Different classification systems have existed since the 19th century but the term is used nowadays mostly to refer to the Morgan-Keenan (MK) system, which was established in the 1940s and has been developed since then. An MK classification has three components: a spectral type, a luminosity class, and (possibly) suffixes or qualifiers. The first two components represent temperature and luminosity sequences (a 2-D grid), respectively, and the third one includes additional information. The MK system is an Aristotelian morphological classification system that uses the MK process, an inductive approach based on specimens (standard stars) that define the system. In that respect, it is different from and a required preliminary step for quantitative spectroscopy, whose goal is to extract the physical properties and chemical composition of the stars. In this entry we provide a brief history of spectral classification, describe the general properties of the classification criteria of the MK system, analyze the specific spectral type and luminosity class criteria used for each spectral class, and introduce the different peculiar types that can be found within the 2-D classification grid and at its edges.
\end{abstract}

\begin{keywords}
Spectral classification, Morgan-Keenan classification, Luminosity classification, Harvard classification, O stars, B stars, A stars, F stars, G stars, K stars, M stars, ultracool dwarfs
\end{keywords}

\vfill

\eject

\begin{tcolorbox}[colback=black!20!white]
\textbf{Key points/Objectives box}

\bigskip

\begin{itemize}
 \item The MK system, an Aristotelian morphological classification system based on specimens, is the most common method of spectral classification used in astronomy.
 \item Spectral classification is not obsolete. It is as needed as it was a century ago as a preliminary and complementary procedure to quantitative spectroscopy to obtain the correct image of nature.
 \item An MK classification has three components: a spectral type that corresponds to an effective temperature scale, a luminosity class that corresponds to a luminosity or gravity scale, and suffixes/qualifiers that correspond to other parameters. 
 \item It is important to follow the MK process correctly, especially regarding notation and the (un)certainty of the classification. Otherwise, the author is likely to contribute more noise than signal to astronomical knowledge. 
\end{itemize}
\end{tcolorbox}

\section{Introduction}\label{sec1}

Spectral classification (sometimes referred to as stellar classification) is a generic term for the division of stars into classes based on their spectral characteristics. In principle, one can do spectral classification 
based on any of the stellar properties reflected on the spectrum but, in most circumstances, the term is currently used to refer to the Morgan-Keenan (MK) system. The MK classification system is described in more detail 
below, but it basically consists of three components: 

\begin{enumerate}
 \item The spectral type using the seven letters OBAFGKM as a temperature sequence from hot to cool, with each letter possibly followed by a number, called the spectral subtype. Those seven letters form the original standard sequence, but others are also used nowadays.
 \item The luminosity class, added after the spectral type and separated by a space, as a luminosity sequence designated by a Roman numeral. Originally, the sequence ran from I (supergiants, more luminous) to V (dwarfs, less luminous), but it was eventually expanded at both ends and class subdivisions also appeared.
 \item Possibly suffixes (or qualifiers) to represent other characteristics, such as line width, composition anomalies, the presence of emission lines, uncertainty in the classification, and other peculiarities.
\end{enumerate}

In addition, two or more stars can be identified in a single spectrum if they belong to very different classes or if they can be separated in velocity, as in the case of double-lined spectroscopic binaries (SB2), for which 
the spectral classification of each component is separated by a plus sign i.e. +. As an example, the SB2 system BD~$+$60~497 was classified as O6.5~V(n)((f))z~+~B0:~V(n) by \cite{Maizetal19b}. The primary is a star with a spectral 
type of O6.5 and a luminosity class of V, with the qualifiers (n), ((f)), and z respectively indicating [a] slightly broadened lines, [b] strong \HeII{4686} absorption\footnote{Throughout this entry, spectral lines are identified by their ion followed by the wavelength in \AA\ rounded to the nearest integer value.} accompanied by 
weak emission on the \NIIIt{4634-40-42} triplet, and [c] the \HeII{4686} absorption being more intense than both \HeI{4471} and \HeII{4542}. The secondary has an uncertain (indicated by the : qualifier) B0 spectral type and a luminosity class of V, with (n) indicating again slightly broadened lines. This example shows the possible complexity of spectral classification.

Spectral classification should be differentiated from quantitative spectroscopy, whose aims are to measure physical properties such as effective temperature (\Teff) and logarithmic surface gravity (\logg) and to determine the 
chemical composition of stars. Spectral classification is a morphological process whose aim is to provide an image of nature as an intermediate step prior to its understanding by a subsequent analysis via quantitative 
spectroscopy \citep{Walb11b}. Such a process is necessary when the objects of scientific analysis constitute a complex and diverse sample. If it is not properly followed, our image of nature will be inadequate and physics will not be able to provide a correct understanding. In that way, spectral classification yields a category for a star that does not require a computation and is (unless the classification system changes) permanent, while the physical 
properties obtained from quantitative spectroscopy change according to the input physics of the model used to derive them (see page 31 in \citealt{MorgKeen73}, written by Dimitri Mihalas). The MK system is defined by a series of standard stars that define the spectral classes, of which 
the most robust ones are called anchor points \citep{Garr94}.
For example, $\alpha$~Psc has had an A3~V spectral classification and $\rho$~Boo a K3~III one since the MK system was established in the 1940s, but their values of \Teff\ and \logg\ determined from quantitative spectroscopy have changed since then (see \citealt{Holgetal18} for some examples with O stars). 
Spectral classification has given us the same image of nature during that period and quantitative spectroscopy has shown how our understanding has improved.

A common confusion regarding spectral classification arises from most stellar classes having two definitions: a spectroscopic (or morphological) one and an evolutionary (or astrophysical) one. For example, the answer to ``what is a red supergiant?'' can be ``a star with spectral type K or M and luminosity class Ib to Ia" or "a massive star (of a certain mass range) in its final stages before core collapse during which it greatly expands until reaching low surface temperatures''.  There are several problems with such a double definition:

\begin{itemize}
 \item A one-to-one correspondence between observational and evolutionary definitions is prone to have exceptions. Using the example above, some intermediate-mass stars may have phases during which their spectra mimic that of a red supergiant and a massive star during their final stages may at some point have a different spectral classification. \item Taking the previous point to an extreme, some objects with spectroscopic characteristics similar enough to be grouped in the same morphological category (e.g. Wolf-Rayet stars) correspond to objects of quite different nature.
 \item Most stars have a fixed spectral type during a human lifetime. However, the spectral types of some stars vary, either (quasi)-periodically, such as spectroscopic binaries, cepheids, and Of?p stars, or otherwise, such as Luminous Blue Variables and OBe stars. In those cases, one should indicate that the spectral classification corresponds to a given phase or epoch of the star, e.g. ``the cepheid was observed in an F8~Ib phase'' or ``the cepheid had an F8~Ib spectrum on Julian date X'', but not ``the cepheid is an F8~Ib star''. Excluding such variable cases, one can say e.g. ``the Sun is a G2~V star''.
 \item Some commonly used broad generalizations are wrong or misleading. For example, the vast majority of dwarf stars (MK luminosity class V) are in the main sequence phase (burning hydrogen in their cores) and viceversa, but some exceptions exist in both directions. The term ``red giant'' is especially misleading, as in some contexts it refers to ``cool stars of luminosity class III'' (spectroscopic definition) and in others to ``stars in the red giant branch'' (evolutionary definition that is a subset of the spectroscopic one).
\end{itemize}

For those reasons, in this entry we will attempt to describe each stellar class in both spectroscopic and evolutionary terms and make explicit those cases where the definitions may be in conflict.

\section{Early history of spectral classification}\label{sec2}

\begin{figure}[t]
\vspace{-4.0mm}
\centerline{\includegraphics[width=1.00\linewidth]{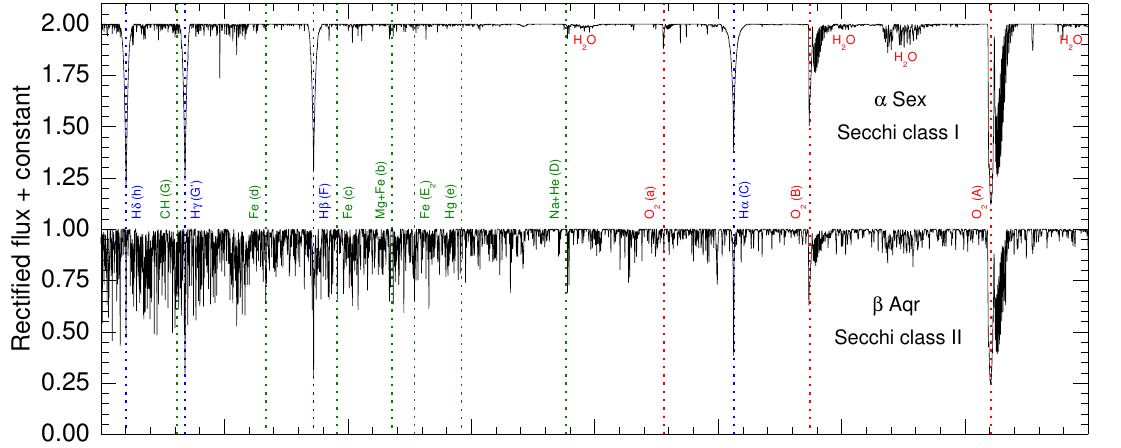}}
\vspace{-2.0mm}
\centerline{\includegraphics[width=1.00\linewidth]{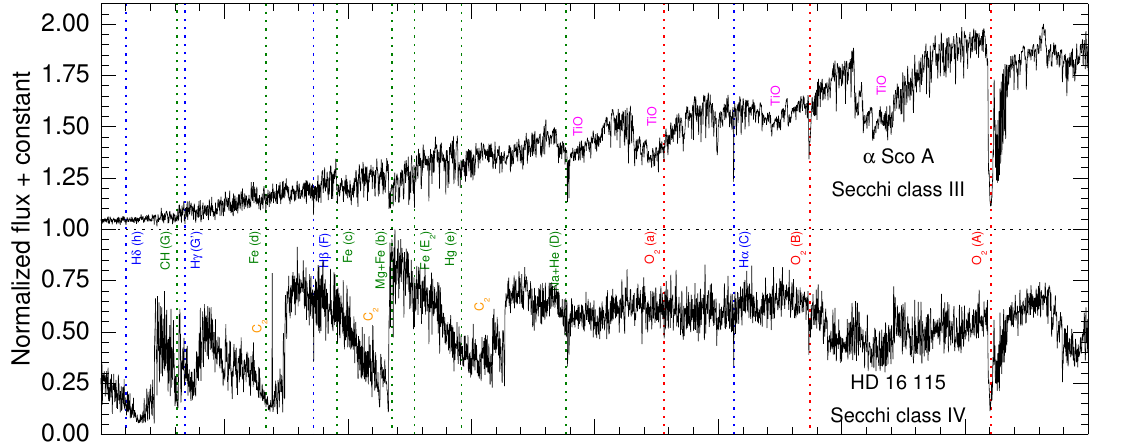}}
\vspace{-2.7mm}
\centerline{\includegraphics[width=1.00\linewidth]{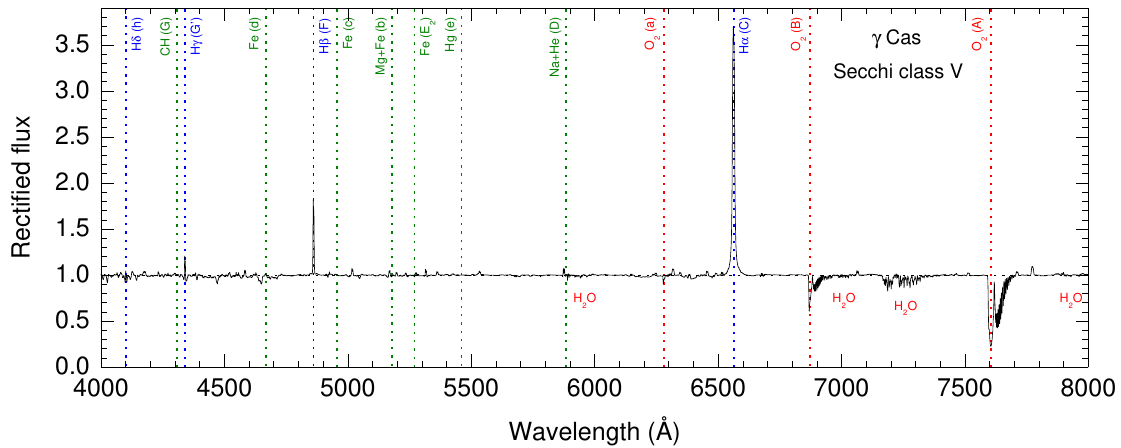}}
\caption{Spectrograms at a spectral resolution of 4000 of examples of the five Secchi classes. The spectra in the top and bottom panels are rectified and those in the middle panel are normalized to their maxima, in all cases 
         separated by a continuum unit. The bottom panel has a different vertical scale to accommodate the H$\alpha$ emission. Blue lines show the Fraunhofer lines from hydrogen and green lines the rest of the Fraunhofer lines. Red, magenta, and orange text is used for telluric, TiO, and C$_2$ bands, respectively.}
\label{Secchi}
\end{figure}

In 1811-1814, the Bavarian physicist Joseph von Fraunhofer used a prism to decompose the solar spectrum and discovered the existence of hundreds of absorption lines, labelling the strongest using letters of 
the alphabet (Fig.~\ref{Secchi}, see \citealt{Curt32} and \citealt{Hear90} for a more detailed account of this and other aspects of this section). He did not try to explain the origin of the lines in the spectrum, most of which were discovered in 
the following decades to originate in the solar photosphere and the rest in our own atmosphere (the so-called telluric lines). The following decades saw further progress in the extension of the spectrum to the
infrared, the introduction of photography to record the solar spectrum, and, most importantly, the work by Robert Bunsen and Gustav Kirchoff that allowed the identification of lines in the solar spectrum by comparing it with the
emission lines seen in the spectra of flames, arcs, and sparks. In that way, it was shown that it was possible to identify the elements the Sun\footnote{And later on, the stars.} was made of,  a question that had been deemed impossible
by the French philosopher Auguste Compte in 1835. Astronomy started its evolution into astrophysics.

Several astronomers observed stellar spectra in the decades after Fraunhofer, but the establishment of spectral classification did not take place until the 1860s, when the Italian Jesuit astronomer Angelo Secchi observed 
and classified at least 4000 stars at the Collegio Romano at the Vatican. His key idea, later exploited at great lengths by subsequent observers, was to study a large sample of spectra in order to be able to group them 
into classes with a significant number of objects in each group. In this way, Secchi first introduced three classes, I, II, and III, with blue/white, yellow, and red respective (intrinsic) colors that we know today  
correspond to a decreasing temperature scale (Fig.~\ref{Secchi}). In addition, he noted the existence of an Orion subtype for class I, where the lines were narrower and that today are known to be the hotter ones of the 
group, and later included a class IV that corresponds to carbon stars. Secchi also produced insights into the nature of the Balmer lines, correctly identifying hydrogen as the element responsible for them, suggesting that 
exceptional pressure caused their broadening in Sirius, and detecting them in emission in some active stars. For the latter type of stars with Balmer lines in emission he defined a class V. His class I stars have a spectrum dominated by those Balmer lines in a relatively clean continuum; 
his class II stars by a solar-like collection of many narrow metallic lines; his class III and IV stars by wide molecular bands of metallic oxides and carbon compounds, respectively; and his class V stars are similar to his
class I stars but with the Balmer (and possibly others) lines in emission. The distinct morphology of the five Secchi classes is on display in Fig.~\ref{Secchi} using modern data from the LiLiMaRlin project \citep{Maizetal19a}.

Between the 1860s and the end of the century, other types of astronomical objects had their spectra identified: Wolf-Rayet (WR) stars by Charles Wolf and Georges Rayet, emission nebulae and novae by William Huggins, and the
first supernova by Nicholas von Konkoly. William Huggins also established two pillars for future developments in stellar spectroscopy by ascribing lines to several elements and starting the field of spectrophotography (spectral classification was done visually until about 1890).
Another major development was the discovery of helium in the Sun by Norman Lockyer in 1868 and on Earth in 1895 by William Ramsay. This led to the identification of Fraunhofer's D$_3$ line as \HeI{5876} and to an evolution of Secchi's original
classification by Hermann Carl Vogel to include the presence or not of helium lines in class I stars, which we now know is a temperature effect.

The period between 1885 and 1924 saw the development of the Harvard system of spectral classification, led by Edward C. Pickering and carried out for the most part by Williamina Fleming, Antonia Maury, and Annie J. Cannon.
Using objective-prism spectroscopy, they eventually published spectral classifications for \num{225300} stars using a new classification system that led to the now famous OBAFGKM sequence, which is remembered by many using the English mnemonic ``Oh, Be A Fine Girl/Guy, Kiss Me''\footnote{There are many alternatives, such as the English ``Our Bright Astronomers Frequently Generate Killer Mnemonics'', the Spanish  ``Obesos, Bebed Aceite Filtrado, Ganar\'eis Kilos Masivamente'', the French ``Observez Bien Au Firmament: Grandiose Kal\'eidoscope Multicolore'', and the German ``Offenbar Benutzen Astronomen Furchtbar Gerne Komische Merks\"atze''.}. To get there:

\begin{figure}[t]
\vspace{-4.0mm}
\centerline{\includegraphics[width=1.00\linewidth]{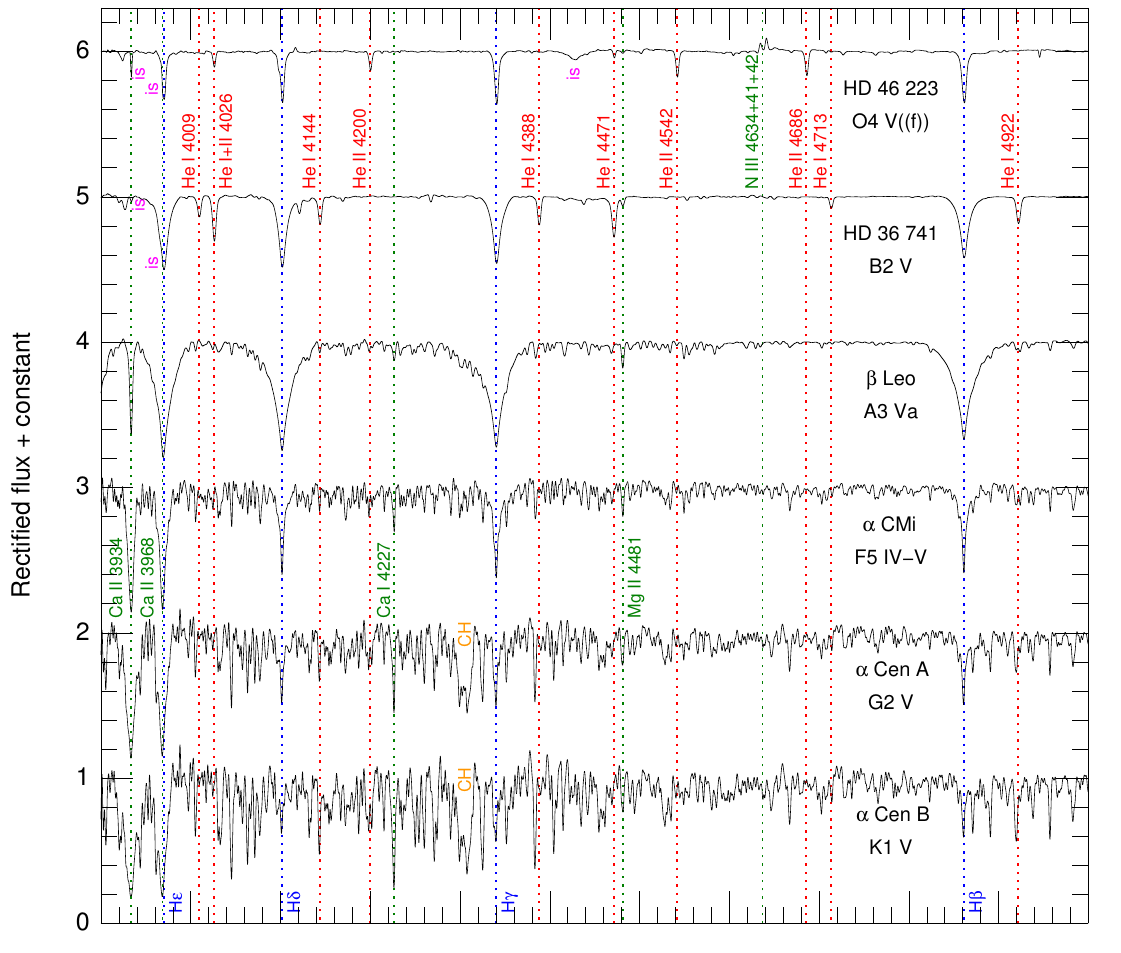}}
\vspace{-4.0mm}
\centerline{\includegraphics[width=1.00\linewidth]{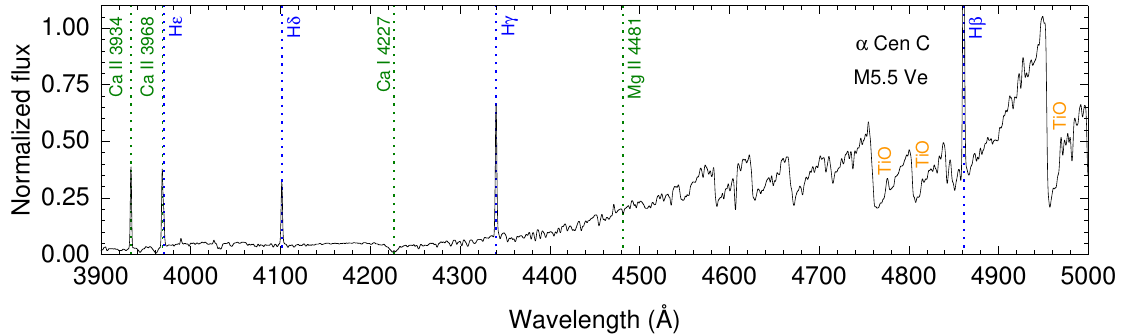}}
\caption{Blue-violet spectrograms at a spectral resolution of 2500 for seven dwarfs representing the OBAFGKM sequence from the LiLiMaRlin project \citep{Maizetal19a}. The spectra in the top panel are rectified and separated by 
         a continuum unit, while the one in the bottom panel is normalized. Blue, red, and green lines show the most relevant stellar hydrogen, helium, and metallic lines, respectively. Orange text is used for molecular bands and 
         magenta text for interstellar lines.}
\label{OBAFGKM}
\end{figure}

\begin{itemize}
 \item A first catalog with \num{10351} stars, the Draper Memorial Catalogue \citep{Pick90}, was first published with most of the classifications done by Williamina Fleming. This first classification system divided the Secchi 
       types as follows: I into A+B+C+D, II into E+F+G+H+I+K+L, III as M, and IV as N. In addition, O was used for Wolf-Rayet stars (not the current notation), P for planetary nebulae, and Q for objects that did not fall into any of the previous 
       categories. Given the limitations in apparent magnitude and declination (all data were obtained from the Harvard College Observatory in Cambridge, Massachusetts, USA), no N-type stars and only one O-type star were included. 
 \item In 1889 Harvard started operating a telescope in Arequipa, Per\'u, that allowed access to the southern hemisphere, where a number of star-forming regions, the Large Magellanic Cloud (LMC), and the Small Magellanic
       Cloud (SMC) led to a broader sample of the hotter stars. That availability allowed Fleming to drop some letters and reorder the sequence into ABFEHGKMNO in her study of stellar clusters \citep{PickFlem97}. 
 \item In 1897 Antonia Maury was the first to introduce a second dimension in spectral classification when she found out stars whose lines were narrow and others that were wide and hazy \citep{MaurPick97}. The latter category 
       was a mixed bag of objects with rapid rotators and unresolved SB2 systems, but the former category (her class c) was more interesting, as it included objects of high luminosity, as proposed later in 1905 by Ejnar 
       Hertzsprung.
 \item Edward C. Pickering observed that some of the unknown lines in the spectrum of $\zeta$~Pup (the earliest O-type star he had available, even though that was unknown to him) could be reproduced by the same pattern 
       derived by Balmer for hydrogen but allowing for half-integer values \citep{Pick97}: such lines are now known as the Pickering series. In 1913 Niels Bohr showed that those lines were produced by ionized helium, 
       opening the path to the understanding of O stars as the hotter end of the spectral sequence.
 \item Annie J. Cannon \citep{CannPick01} exploited the southern spectra from Per\'u to make two contributions that have survived to the present: the final reordering of the sequence into OBAFGKM (Fig.~\ref{OBAFGKM}, plus P and Q 
       as above and N and R that she added later on), thanks to the availability of a large sample of O stars in the southern hemisphere; and the introduction of spectral subtypes, which were originally written as e.g. B3A 
       instead of B3, something that she changed afterwards. Note that in spectral classification it is customary to define hotter spectral types or subtypes as ``earlier'' and cooler ones as ``later''.
 \item During the first two decades of the 20th century, W. Fleming first and A. J. Cannon later extended the sample and studied objects that were rare or anomalous: O, N, and R stars; variables; spectroscopic binaries;
       emission-line stars; novae; gaseous nebulae; and other peculiar objects.
\end{itemize}

Finally, the Henry Draper (HD) Catalogue was published in the period 1918--1924, covering with relative uniformity the whole sky. A.~J.~Cannon continued classifying stars until her death, and the Henry Draper Extension (HDE)
was published a quarter of a century later \citep{CannMaya49}. 

The spectral classifications of the Harvard group allowed for the modern science of stellar astrophysics to appear in the first three decades of the 20th century. Ejnar Hertzsprung and Henry Norris Russell independently
derived the existence of luminosity effects for a given spectral type, producing the first diagrams that carry their names (HRDs). The first stellar temperatures were determined through spectrophotometry, which is
affected by interstellar and atmospheric extinction. In 1921 the Indian astronomer Megh Nad Saha proposed the idea that the stellar spectral sequence was the consequence of the progressive ionization of the different
species as \Teff\ increases and was able to derive a reasonably accurate temperature scale based on it \citep{Saha21}. The application of Saha's work led to the discovery that the ionization
state also depends on pressure, making giants of a given spectral type cooler than the equivalent dwarfs due to their lower atmospheric pressures (and \logg), and supergiants even cooler.
Even more interesting, Saha's work led to Cecilia Payne's PhD thesis conclusion that most stars are primarily made of hydrogen and helium and have relatively similar compositions despite their widely different spectra \citep{Payn25}.

\section{The MK system}\label{sec3}

\subsection{Historical development and summary}

As early as 1922, \citet{AdamJoy22a} had established that Balmer lines become narrower as luminosity increases, which \citet{Stru29} correctly identified as a consequence of the Stark effect, where pressure broadening makes Balmer lines in high-gravity stars ``nebulous'' (using the terminology applied back then, meaning broad; note that giants and supergiants have lower gravities due to their large sizes). William.~W. \citet{Morg37} was the first to determine surface gravities for a large sample of stars and to plot a \logg\ vs. \logT\ diagram. Such a diagram is an equivalent form of an HRD and its use opened the possibility towards 2-D classification by defining Morgan's luminosity classes as a second dimension to be added to the spectral type sequence from the Harvard system. The next year he used those results to define the classes as supergiants (I), bright giants (II, originally ``intermediate between giants and supergiants''), giants (III), subgiants (IV), and dwarfs (V, originally ``main-sequence stars''\footnote{For the sake of consistency, throughout this entry we will consider luminosity classes as spectroscopic or morphological categories. Therefore, we will avoid equating dwarfs or luminosity class V with the main sequence (as originally done), for which we will reserve its evolutionary or astrophysical meaning of being composed of stars burning $^1$H in their cores.}) for the first time \citep{Morg38a}. The original division extended only from F4 to K5, with class IV only for G5--K2, but also included the subdivision of F4--F8 supergiants into Ia (higher luminosity) and Ib (lower luminosity). The work of Morgan and others culminated in the seminal atlas of Morgan, Keenan, and Kellman, widely known as the MKK atlas \citep{Morgetal43}, which included 55 prints with photographic spectrograms for the ``practical astronomer'' that became the basis for spectral classification in the following decades using the MK system defined from them. The MKK atlas could be easily used today with only minor modifications (e.g. it included no M dwarfs, carbon or S stars, and luminosity classes for types earlier than O9 were not defined). 

\begin{figure}[t]
\vspace{-15mm}
\centerline{\includegraphics[width=1.00\linewidth]{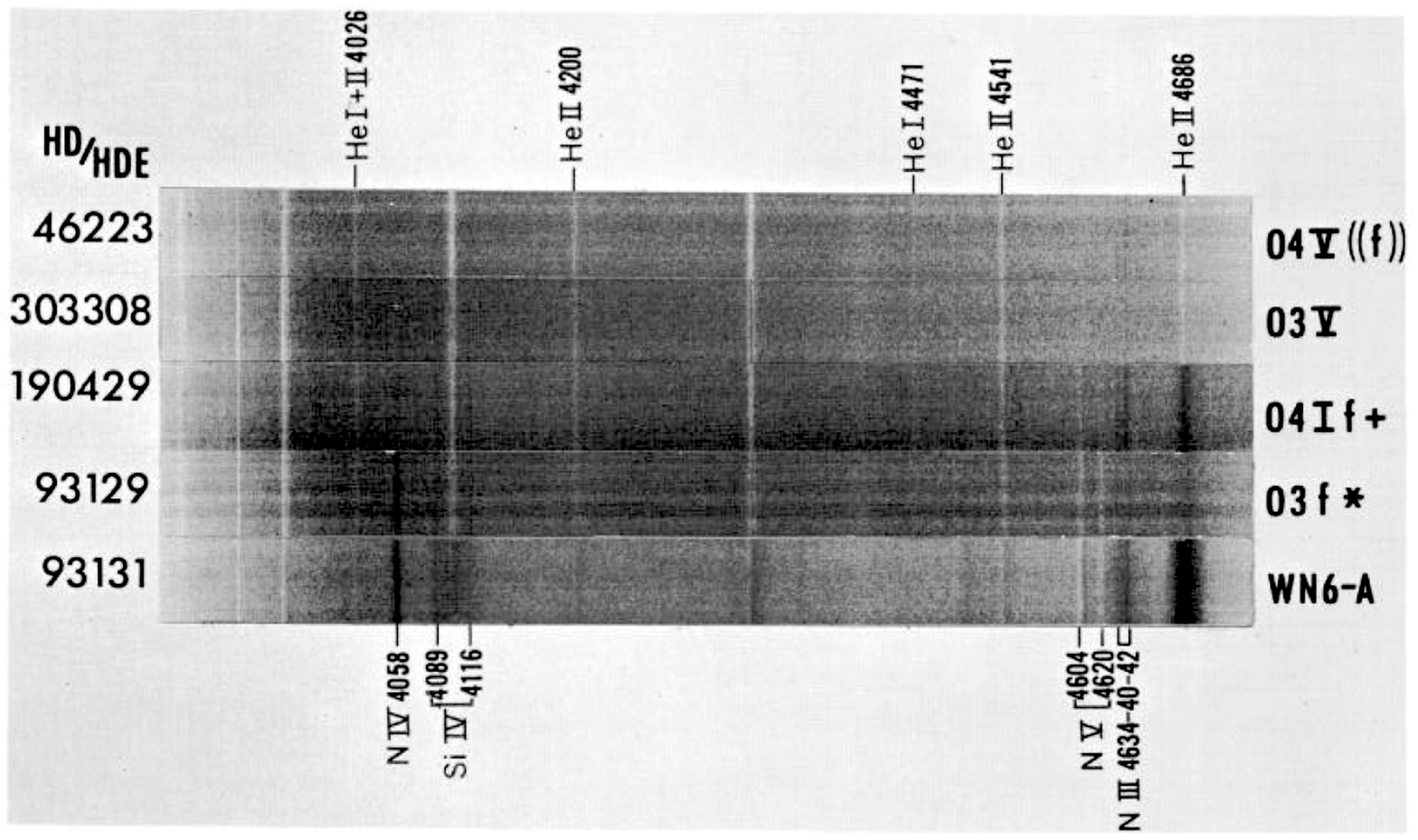}}
\vspace{-10mm}
\caption{Photographic spectra of very early type stars from \citet{Walb71b}. The first star from the top (HD~\num[detect-all]{46223}) is the same as the first one in Fig~\ref{OBAFGKM}. The photographic and digital spectra yield the same spectral type but the digital one is more sensitive to faint and/or diffuse structures. As this is a negative, absorption bands are light and emission lines are dark.}
\label{photographic}
\end{figure}

The MK system was described by \citet{MorgKeen73} and \citet{Garr94}. In this section we provide a description of its characteristics and in subsequent ones we briefly detail it for different types of stars. For a more thorough analysis, the reader is referred to \citet{GrayCorb09}. The MK system is an Aristotelian classification system that uses the MK process, an inductive approach based on specimens that define the system. The specimens are the spectra of standard stars arranged in a 2-D grid, with one axis being the spectral type and the other one the luminosity class. The grid is finite, i.e. one or several stars define a given box and the classification process puts a given spectrum in one box or another. The 2-D grid is complemented by the third component of MK spectral classification: suffixes or qualifiers that provide additional information detected in the spectrum.

The MK system evolves in time, as better data become available, in at least two ways: by expanding the 2-D grid outside its boundaries in spectral type and luminosity class and by making a finer grid dividing former large boxes into smaller ones. The latter is needed because the original grid had many boxes without a standard, so stars that would be assigned to them had to be compared with ``interpolated'' spectra, making the need to enlarge the sample of standards clear from the beginning. However, when enough stars are obtained that could fill those boxes, the classifier may experience the temptation to define a new intermediate box and, indeed, this has happened in the past (e.g. by introducing a full O9.7 spectral subtype between O9.5 and B0, see \citealt{Sotaetal11a}). Four warnings about updating the MK grid are in order: (a) new standards should be chosen with care and old ones may be dropped but only with good reason (e.g. when shown to be composite spectra by being double- or tripled-lined  spectroscopic binaries), (b) new boxes should be sufficiently differentiated from the old ones, (c) one should try to avoid the ``one known star per box'' syndrome when covering parts of the grid with few stars, and (d) more importantly, anchor points should not be moved, as there is a hierarchy of standards that fixes the MK system in place \citep{Garr94}.

The system was developed using photographic spectra in the blue-violet region (approximately, 3900--5000~\AA, Fig.~\ref{photographic}). In the last part of the 20th century, the MK system was transformed into one based on digital spectra (see e.g. \citealt{WalbFitz90}) to ensure its continuity and extended to other wavelength ranges (see below). Morgan and Keenan specified that ``These standard reference points do not depend on values of any specific line intensities or ratios of intensities; they have come to be defined by the appearance of the totality of lines, blends, and bands in the ordinary photographic region" \citep{MorgKeen73}. This text refers to two important points. First, the whole spectrum has to be used to define spectral classifications and specific line strengths or ratios can be a primary or secondary criterion, but should be always analyzed in the context of a full comparison with the standard star. Morgan and Keenan wrote that text in a moment when intermediate and narrow-band photometry was producing alternate quantitative classifications that some thought would render the MK system obsolete, something they specifically said was a false conclusion. The current danger Morgan and Keenan would see would be different: digital spectra are sometimes used to automatically determine line strengths or ratios and assign a numerical MK classification (sometimes ignoring the division in boxes and inventing spectral subclasses with two digits after the decimal point, which should be avoided in the MK system) without considering whether the spectrum fits that of the standard: that is not a proper MK process. Second, the criteria used to determine membership to a class can be ``lines, blends, or bands'', meaning that some will correspond to individual transitions and some to a combination of them. The more relevant issue is that the classification criteria are based on line strengths (``intensities'' in the Morgan and Keenan text, or equivalent widths or EWs in numerical terms) and strength ratios, not in depths, and should be kept that way (see below for more on this).


\subsection{Luminosity classes}

The spectral type sequence of the MK system is essentially the same as for the Harvard system, with the Saha ionization equation determining which ionization state is dominant for a given element at a given \Teff\ (and, to a second order, pressure). Luminosity classification, on the other hand, depends on specific criteria for the spectral type of the star. For example, the width of the Balmer lines is a primary criterion for B and A stars but is irrelevant for M stars. For that reason, the criteria are presented in the specific sections for each spectral type. If a spectral feature strengthens with increasing luminosity (or decreasing luminosity class), it has a positive luminosity effect. If it weakens with increasing luminosity, it has a negative luminosity effect (as it is the case with the width of Balmer lines in B and A stars, see below).

The original five or six classes (depending on the division of I into Ia and Ib or not) retain their basic structure with some minor modifications. Supergiants are nowadays divided into three subclasses (Ia, Iab, and Ib in order of decreasing luminosity) but the subdivision is patchy in coverage (for example, it does not include the earliest O
supergiants). There is also a luminosity class above Ia, the hypergiants, designated as Ia+ or 0 (numeral zero, not capital letter O). For O stars there are two classes of very luminous stars \citep{Sotaetal14} that have characteristics that are intermediate between O supergiants and Wolf-Rayet types called early Of/WN stars (or ``hot slash'') and O Iafpe stars (formerly known as ``cool slash''), but they do not form a proper luminosity class, at least at this stage. For late-B and early-A dwarfs, where the Balmer lines are stronger, it is possible to subdivide the V class into Vb (Zero Age Main Sequence, ZAMS) and Va (normal dwarfs). In a similar manner, the ZAMS can be distinguished for O2-O8 stars, but in that case a z suffix is used instead of a luminosity subclass. Finally, two luminosity classes can also be defined below the dwarf one: subdwarfs and white dwarfs, which in the past were assigned classes VI and VII, respectively. However, that is mostly deprecated and subdwarfs are currently designed with the prefix sd attached to the spectral type (e.g. sdB0, see also below for M and ultracool dwarfs) and white dwarfs by the prefix D attached to a composition symbol (A, B, C, O, Z, or Q). 

\begin{table}[ht!]
\caption{Suffixes/qualifiers frequently used for spectral classification.}
\label{qualifiers}
\centering
\begin{tabular}{lcl}
\\
\hline
Qual.   & Range     & Description \\
\hline
:       & All       & Uncertain classification, may be applied to either the spectral type or the luminosity class, \\
        &           & $\;\;$ sometimes ? also used (see below for f?p)  \\
comp    & All       & Composite spectrum \\
k       & All       & Interstellar absorption present, commonly ignored for O+B+WR stars due to ubiquity \\
p	      & All       & Peculiar spectrum (general, see below for B+A stars), sometimes \ldots\ and ! also used \\
var	    & All       & Variation in line spectrum intensities or content, v sometimes also used \\
w       & All       & Weak lines (metal poor), wl and wk sometimes also used \\ 
\hline
(n)	    & O-F       & Slightly broadened (``nebulous'') absorption lines (O stars: $\vsini\sim 200$~km/s), Fig.~\ref{nindex} \\
n	      & O-F       & Broadened absorption lines (O stars: $\vsini\sim 300$~km/s), Fig.~.\ref{nindex} \\
nn	     & O-F       & More broadened absorption lines (O stars: $\vsini\sim 400$~km/s), Fig.~\ref{nindex} \\
nnn     & O-F       & Even more broadened absorption lines (O stars: $\vsini\sim 500$~km/s) \\
$[$n$]$ & O+B       & H absorption lines more broadened than He lines \\
s       & All       & Very narrow (``sharp'') lines \\
sh      & O-F       & Shell star, very broad profiles in some lines, narrow in others, \\
        &           &  $\;\;$ possible Balmer (and others) lines in emission \\
\hline
e	      & All       & Emission components in H lines and possibly in He\,{\sc i} and even some metallic lines, Fig.~\ref{esuffix} \\
(e)	    & All       & Weak version of above e.g. probable H$\alpha$ emission but no red spectrogram available, Fig~.\ref{esuffix} \\
$[$e$]$ & All       & Emission spectrum including Fe forbidden lines \\
eq      & All       & One or several emission line(s) with P Cyg profile in the blue-violet region, Fig~.\ref{esuffix} and \citet{Feasetal60} \\
neb     & All       & Nebular emission spectrum present \\
\hline
b       & WN        & Broad lines, FWHM (\HeII{4686}) $>$ 30~\AA\ or FWHM (\HeII{5412}) $>$ 40~\AA \\
d       & WC        & Star with dust-production episodes \\
o       & WN        & No hydrogen component in non-pure-Pickering He\,{\sc ii} wind emission lines (usually left off) \\
(h)     & WN        & Weak hydrogen component in non-pure-Pickering He\,{\sc ii} wind emission lines \\
h       & WN        & Strong hydrogen component in non-pure-Pickering He\,{\sc ii} wind emission lines \\
ha      & WN        & Hydrogen lines from wind with a combination of emission (blue-shifted) and absorption \\ 
\hline
((f))   & O3.5-O8.5 & Weak \NIIIt{4634-40-42} emission, strong \HeII{4686} absorption \\
(f)     & O4-O8.5   & Medium \NIIIt{4634-40-42} emission, neutral or weak \HeII{4686} absorption \\
f       & O4-O8.5   & Strong \NIIIt{4634-40-42} emission, \HeII{4686} emission above continuum \\
((f*))  & O2-O3     & \NIV{4058} emission $\ge$ \NIIIt{4634-40-42} emission, strong \HeII{4686} absorption \\
(f*)    & O2-O3.5   & \NIV{4058} emission $\ge$ \NIIIt{4634-40-42} emission, weaker \HeII{4686} absorption \\
f*      & O2-O3.5   & \NIV{4058} emission $\ge$ \NIIIt{4634-40-42} emission, \HeII{4686} emission \\
c       & O2-O8.5   & Added to one of the six previous cases as e.g. ((fc)) or fc* when \\
        &           & $\;\;$ \CIIIt{4647-50-51} emission $\ge$ \NIIIt{4634-40-42} emission \\
f?p	    & O6-O8.5   & Variable \CIIIt{4647-50-51} emission $\ge$ \NIIIt{4634-40-42} emission at maximum; \\
        &           & $\;\;$ variable sharp absorption, emission, and/or P Cygni features at H and He\,{\sc i} lines \\
(n)fp   & O         & \HeII{4686} centrally reversed emission, slightly broadened absorption lines \\
nfp     & O         & \HeII{4686} centrally reversed emission, broadened absorption lines \\
z	      & O2-O8     & \HeII{4686} in absorption and $>$ than both \HeI{4471} and \HeII{4542} \\
\hline
N	      & O+B       & N absorption enhanced, C and O deficient \\
N str   & O+B       & Moderate case of above. Nstr sometimes also used \\
C       & O+B       & C absorption enhanced, N deficient \\
N wk    & O+B       & Moderate case of above, Nwk sometimes also used \\
\hline
He str  & B0-B3     & Exceptionally strong He lines \\
He wk   & B3-B9     & Exceptionally weak He lines \\
p	      & B+A       & Abnormally strong selected metal lines, additional notation for different elements \\
m	      & A         & Abnormally strong metal lines, additional notation for Ca, H, and metallic spectral types \\
$\lambda$ Boo & A+F & $\lambda$ Boo population I stars, additional notation for Ca, H, and metallic spectral types \\
shell   & B-F       & Shell star \\
\hline
CN      & G+K       & Abnormal CN bands, followed by a numerical qualifier\\
CH      & G+K       & Abnormal CH bands, followed by a numerical qualifier\\
Ba      & G-M       & Abnormally strong Ba lines (and other s-process element lines), followed by a numerical qualifier\\
Fe      & G-M       & Abnormal Fe lines, followed by a numerical qualifier\\
\hline
\\
\end{tabular}
\end{table}


\begin{SCfigure}
\centering
\includegraphics[width=0.70\textwidth]{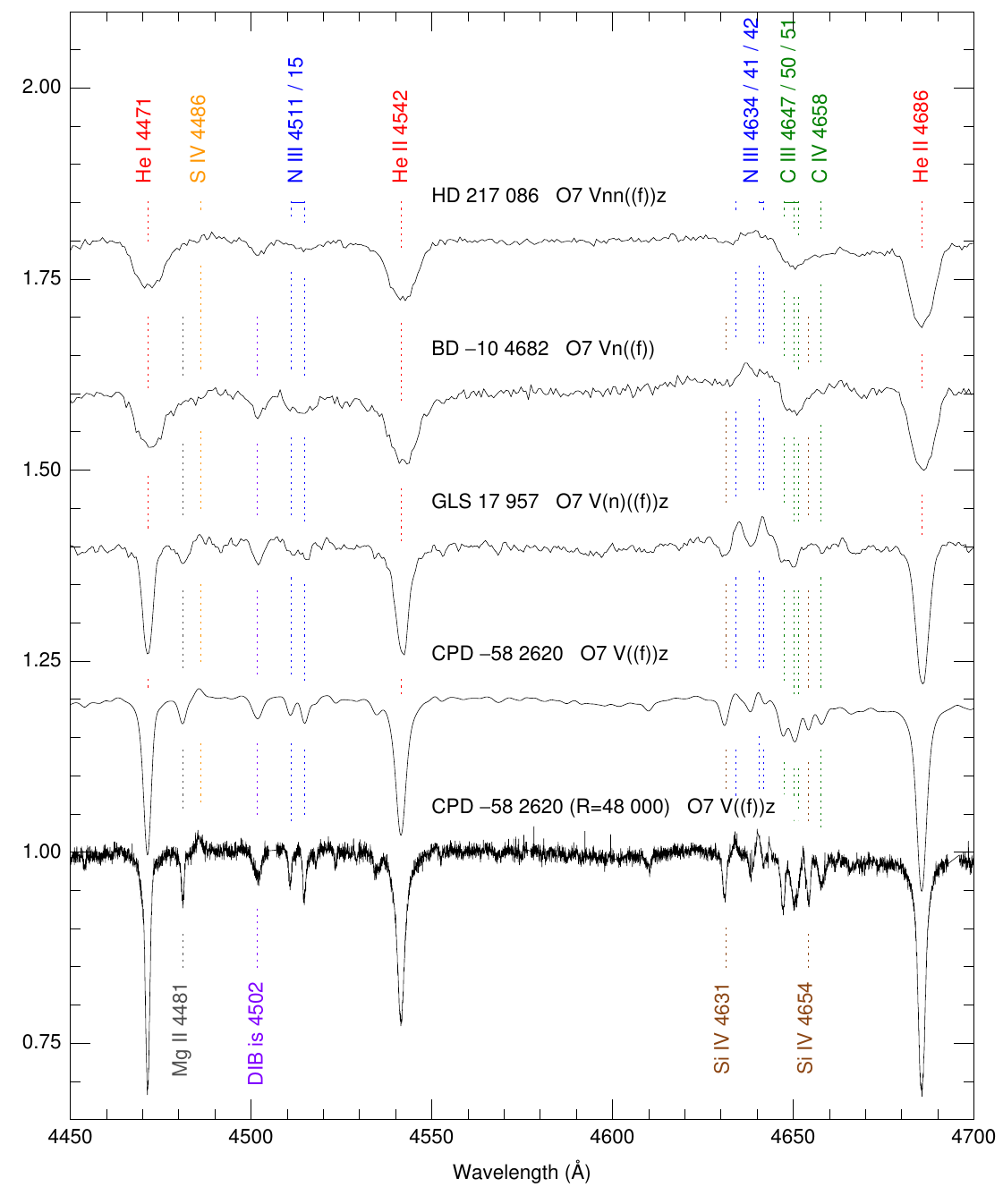} 
\caption{Examples of line broadening or n index. We show four stars with very similar spectral classifications except for their n indices. The bottom star is shown at high spectral resolution and at the standard $\mathcal{R}\approx$~2500 resolution for spectral classification, which is the one used for the other three stars. All spectra are from GOSSS
 or LiLiMaRlin and are rectified and separated by 0.2 continuum units. Rotation broadening dilutes lines while approximately maintaining their EWs, something that can be appreciated in the constancy of the ratio of \HeII{4542}/\HeI{4471} ($\sim$1, as required for the O7 subtype) despite the different ratio of the depths  (a consequence of how Stark
 broadening acts for neutral and ionized He). Weak lines are hard to detect for high \vsini\ values (increase in n index). The diffuse interstellar band (DIB) maintains its shape but has a different EW for each star.}
\label{nindex}
\end{SCfigure}

\begin{figure}
\centering
\includegraphics[width=1.00\linewidth]{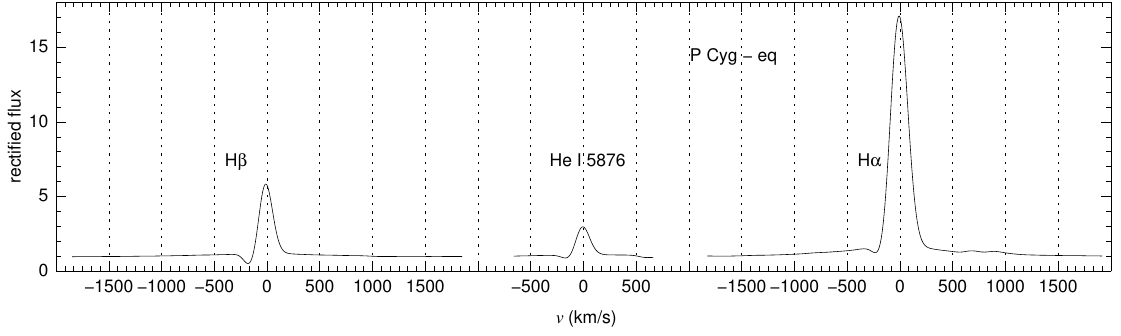}
\includegraphics[width=1.00\linewidth]{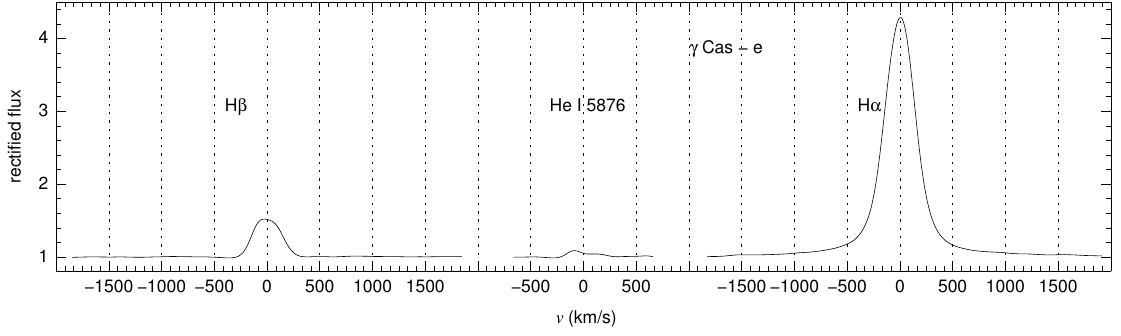}
\includegraphics[width=1.00\linewidth]{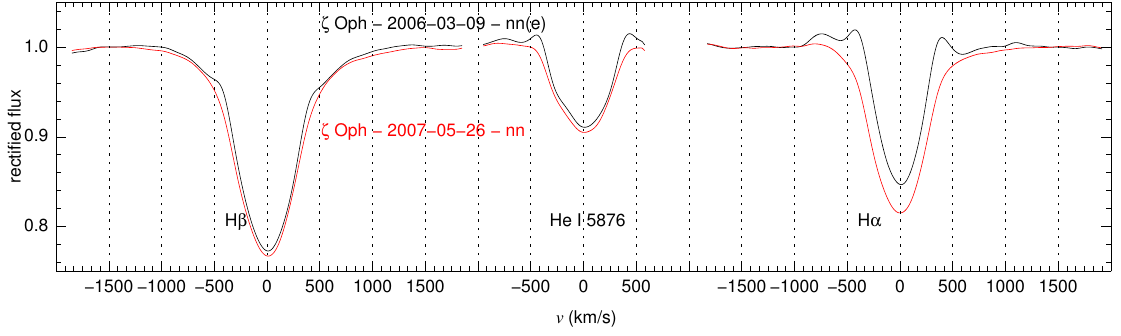}
\caption{Examples of e emission suffix and its variants (P Cygni, strong, and weak) for three stars for which we show relevant lines at $\mathcal{R}\approx$~2500 resolution using the same velocity scale. All spectra are from LiLiMaRlin and are rectified. Note the very different vertical scales for each star. $\zeta$~Oph goes through nn(e) and nn states, one example is given of each (dates indicated). The use of the suffix and its variants is not uniform among authors, with some reserving e for when the line rises above the continuum.}
\label{esuffix}
\end{figure}

We make one final note regarding nomenclature for luminosity classes. Intermediate classes are sometimes used, with e.g. III-IV meaning intermediate between III and IV. However, the nomenclature IV-III means something different: the luminosity range between classes IV and III, for example, due to an uncertainty in the classification. However, this rule is not always used in the literature.

\subsection{Suffixes/qualifiers: line broadening, emission lines, and others}

The possible third component of the MK system is a suffix or qualifier. A list of the most common ones is given in Table~\ref{qualifiers}, note that the Aristotelian morphological nature of the MK process leads to an organic growth in their number with time, as more stars are observed and in more detail. Most suffixes qualify the whole classification
except for the colon qualifier (:) used to express uncertainty in either the spectral type or the luminosity class. For that reason, suffixes are written at the end of the spectral classification but ``:'' may be written after the spectral type: As a result, G2: V means that the spectral type is uncertain and G2 V: that the luminosity class is uncertain. The rest of the qualifiers indicate additional characteristics that cannot be described by the spectral type and luminosity class. In Table~\ref{qualifiers} the suffixes are grouped by theme, first the general ones that apply to all or most spectral types divided in miscellaneous, line width, and emission line presence; and then the ones that apply to certain spectral ranges. Here we discuss some aspects of the general suffixes and leave the specific ones for later.

In general, cool-star spectra (Secchi class II, see Fig.~\ref{Secchi}) are dominated by numerous narrow absorption metallic lines while hot stars (Secchi class I, see Fig.~\ref{Secchi}) have fewer lines of different intrinsic widths: narrow metallic lines followed by lines of He\,{\sc i}, He\,{\sc ii}, and H of increasing intrinsic width due to the Stark (pressure) broadening effect. In addition to intrinsic effects, lines may be broadened by extrinsic effects, of which the three most common ones are double (or triple) spectroscopic binarity, stellar rotation, and the Zeeman effect. For SB2/SB3 systems whose observed radial orbital velocity difference is large enough, one can provide two or three spectral classifications, one for each stellar component. However, when the observed velocity values are small (because the system was observed in an unfavorable orbital phase or the radial velocity amplitude is small), only a broadened profile is observed for all or some of the absorption lines. In those cases, a broadening index is used for O-F stars (see Table~\ref{qualifiers}) in the temporary spectral classification until one with multiple components can be provided (sometimes through techniques such as spectral disentangling,  e.g. \citealt{SimoStur94}). 

The second major effect that can cause line broadening is stellar rotation, which is usually a small effect for late-type unevolved stars but can be very significant in other cases such as early-type stars of different evolutionary stages. For those, the rotation speeds can be measured in hundreds of km/s. When the broadening becomes significant, we start with an (n) index (some authors even include an ((n)) index for a broadening detectable only at high spectral resolution) and we progress to n, nn, and nnn as the projected rotational velocity (\vsini) becomes larger (Fig.~\ref{nindex}). An approximate correspondence between n-type indices and \vsini\ for O stars is given in Table~\ref{qualifiers} \citep{Sotaetal11a,Holgetal22}. The effect of rotation broadening has to be considered because it affects lines of different intrinsic width differently: it is more easily noticeable in intrinsically narrow lines (e.g. metallic) than in intrinsically broad ones (e.g. hydrogen). Rotation broadening should not significantly affect line ratios measured from line strengths (determined as area eye estimates or equivalent widths) but can affect those measured from line depths (a notorious example is \HeI{4471}/\HeII{4542}, the primary spectral subtype criterion for O stars, see Fig.~\ref{nindex}), which is the reason why the latter should be avoided. Modern software used to classify spectra by comparing with standards (e.g. MGB, \citealt{Maizetal12}) includes this effect by selecting only standards with low \vsini, which can be artificially broadened for comparison purposes.  

A caveat that affects the determination of line broadening is the dependence on the spectral resolution of the spectrograph (Fig.~\ref{nindex}). Unless mentioned otherwise, we assume a spectral resolution $\mathcal{R} \equiv \Delta\lambda/\lambda \approx 2500$, which is the closest one to the original one in the MK atlas\footnote{The standard $\mathcal{R}$ is defined as approximately 2500 (meaning in the 2200-2800 range) because, in practical terms, it is not feasible to have spectra with exactly $\mathcal{R}=$~2500 for a large ground-based sample, either because of calibration issues or because one or several of the spectrographs provide a constant $\Delta\lambda$ instead of a constant $\mathcal{R}$.}, though other works have experimented with a higher $\mathcal{R}$ of 4000 \citep{Walb71a,Neguetal24}. It is possible to compare standards at that resolution with stellar spectra at higher resolution if the latter are degraded to $\mathcal{R} =$~2500 (easy to do with digital data, Fig.~\ref{nindex}) and one gets the added benefit of an improved S/N (Signal-to-Noise ratio) in the process. If the stellar spectra are of lower resolution, limitations arise due to the difficulty in detecting weak lines, separating them, and correctly assigning an n-type index, so the spectral classifications will be of lower quality.

The most common emission lines present in stellar spectra are the Balmer lines, either of wind, chromospheric, circumstellar (e.g. disk accretion), or nebular origin, but others may be present. The first three origins are considered intrinsic to the stellar spectrum and the fourth one is regarded as a contaminant and neb is used in that case (Table~\ref{qualifiers}). The e emission suffix and its variants are used to indicate the presence of emission lines (mostly Balmer lines but possibly others, see Table~\ref{qualifiers} and Fig.~\ref{esuffix}) and the f and related indices are reserved for O stars, where weak emission features from strong winds are common. In later-type stars, emission is sometimes related to youth (pre-main-sequence stars, most commonly Herbig Ae/Be and T~Tauri stars), chromospheric activity, or both. Chromospheric activity manifests itself preferentially in the Ca H \& K lines in the blue-violet range but in the reddest cooler stars and with the newest instrumentation, this chromospheric emission is better investigated in H$\alpha$ and \CaIIt{8498,542,662}. In the system devised by \citet{Grayetal03}, a weak emission reversal in these lines is indicated by the qualifier (k), while a stronger emission core is marked by k. When the emission goes above the continuum, the standard qualifier e is added, but in cases with strong emission and accompanying Balmer emission, kee is used.  A different issue is that of stars dominated by emission lines, such as Wolf-Rayet stars, for which the classification criteria are determined by the emission lines themselves and the few absorption lines (usually from hydrogen) that may be present are included in the qualifier.
 \subsection{Beyond OBAFGKM and 2-D classification}

The textbook definition of the MK system as a 2-D classification system with seven spectral types is challenged from three points of view:

\begin{itemize}
 \item \textbf{More than seven spectral types.} A century after the beginning of the development of the Harvard system, three new spectral types 
       (L, T, and Y) were added to the sequence at the low-temperature end thanks to the discovery of brown dwarfs. Therefore, the current sequence is OBAFGKMLTY, which has made the famous English mnemonic evolve into ``Oh, Be A Fine Girl/Guy, Kiss My Lips Tonight, Yo'' or ``... Kiss Me, Less Talk, Yo'' (and the Spanish one to ``Obesos, Bebed Aceite
       Filtrado, Ganar\'eis Kilos Mirando La Televisi\'on Ya(ciendo)''). The new types are described in the ultracool spectral types section below.
 \item \textbf{Evolved stars with anomalous compositions.} From the pioneering times of the 1860s, two types of stars were discovered that could 
       never be fit into the normal spectral-type sequence: Carbon (C) stars and Wolf-Rayet stars. Both cases turned out to be objects with large anomalous surface compositions (and conditions, especially so for WRs) that resulted from being evolved objects (especially so for carbon stars) at the cool and hot edges of the \Teff\ scale. Later on, other types of stars with different large anomalous compositions caused by stellar evolution such as S stars and some white dwarfs (WDs) were added to the list. In contrast, some other objects have relatively minor composition anomalies that allow them to be included in the standard 2-D grid with the help of qualifiers (see examples in Table~\ref{qualifiers} and below), but the first four classes (C and S stars, WRs, and WDs) deserve their own independent categories.
 \item \textbf{A third dimension?} The obvious candidate for a third dimension in spectral classification is metallicity. Starting with  
       \citet{Roma50}, spectral differences that depend on the initial composition of the star were determined and a rich field that eventually measured metallicities in Galactic metal-poor stars was established. For spectral classification purposes, anomalous compositions or metallicity differences are determined by comparing ratios of lines from the same species (composition or metallicity independent) with ratios of lines from different species (dependent) and establishing classification criteria from them. Such systems that add a third dimension through suffixes exist but with limitations \citep{GrayCorb09}: they are mostly confined to the regions of the HRD that can be sampled on nearby Galactic populations and the criteria may rely on weak lines only easily observable at  resolutions higher than 2500 or be confined to a spectral-type range. The efforts to study low-metallicity stars have recently gained strength with the advent of large telescopes equipped with multi-fiber spectrographs, which allow for the execution of massive spectroscopic surveys of Galactic metal poor populations (e.g. Gaia-ESO, \citealt{Gilmetal12,Randetal22}) and reach nearby galaxies containing large numbers of metal-poor stellar types that cannot be easily found in the Milky Way. Most of the LMC stars are likely not metal poor enough to detect a large effect \citep{Walbetal14} but we may have better luck with the SMC \citep{Lenn97,Lenn99,EvanHowa03,Vinketal23}. Such efforts are mostly directed towards quantitative spectroscopy, but they can in principle be used for spectral classification as well.
\end{itemize}

\subsection{Other wavelength ranges}

It is possible to transport the MK system to other wavelengths outside the 3900--5000~\AA\ reference range by observing the standards in both ranges and establishing the corresponding classification criteria. However, it is beyond the scope of this entry to review the large field of spectral classification outside the blue-violet range for lack of space, so the reader is referred to \cite{GrayCorb09} for details on how this is done. Here we just briefly mention some of the advantages and disadvantages of this field.

The advantages of using other wavelength ranges can be divided in intrinsic and extrinsic ones. Intrinsic advantages are those associated with the stars themselves. For example, OB stars are bright and rich in lines (some without correspondence in the optical) in the ultraviolet, so their UV spectral classification allows us access to new characteristics such as wind effects \citep{Walbetal85,Walbetal95a,RounSonn91,SmiNBruh97,SmiNBruh99}. For M and cooler objects, there is very little flux shortward of 5000~\AA, making the red and infrared ranges preferred. Extrinsic advantages are those not associated with the targets, of which the most obvious example is the use of the infrared to observe targets with high extinction. The disadvantages of using other wavelength ranges are the possible difficulty or impossibility in accessing them (e.g. the UV or regions of the mid-infrared from the ground) and the lack of lines required to accurately classify some spectral types. One should keep in mind that the blue-violet region is rich in spectral lines (especially for some spectral types) and without telluric lines (Fig.~\ref{Secchi}), which are two of the reasons for the success of the original MK system.

\begin{figure}
\centering
\includegraphics[width=0.450\linewidth]{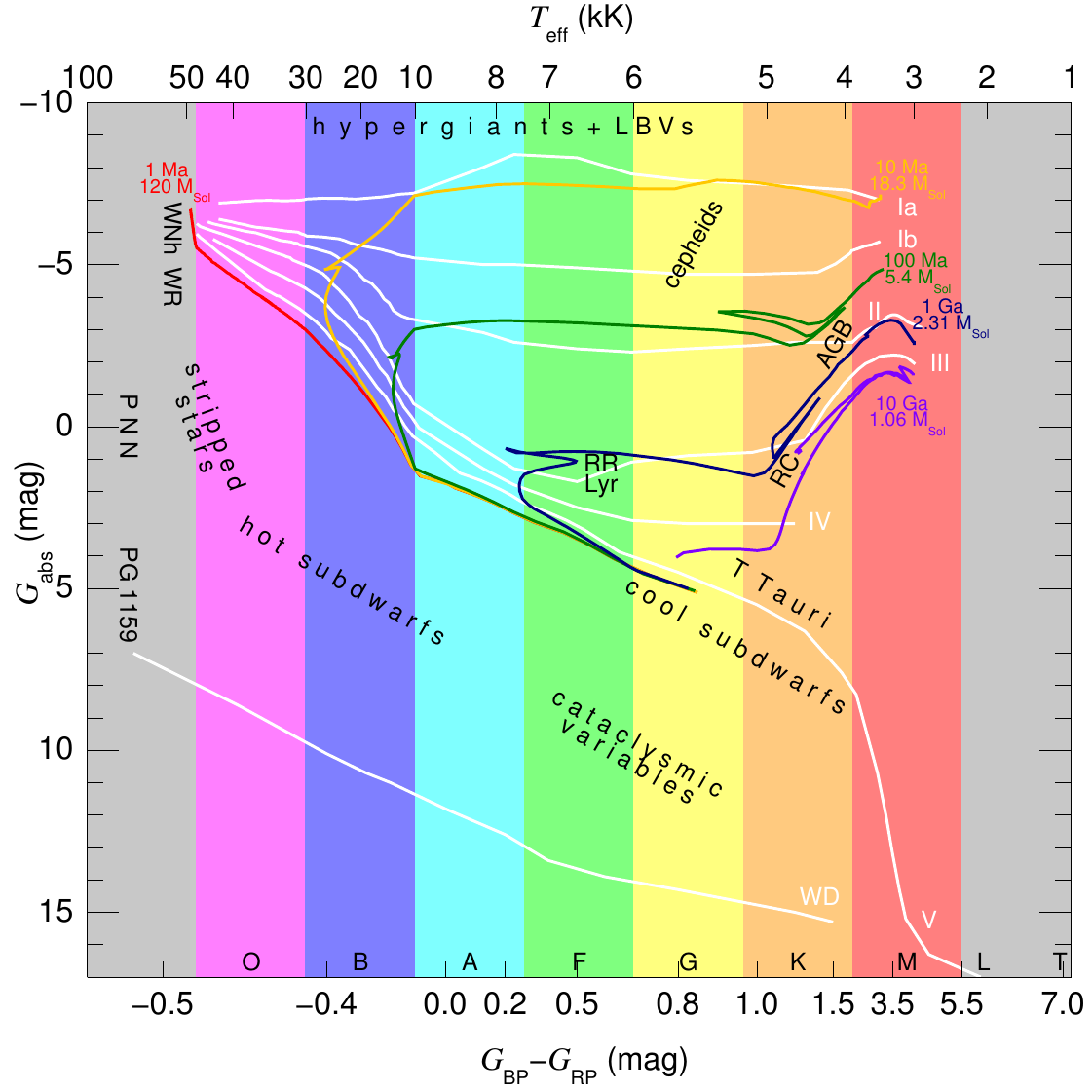} \
\includegraphics[width=0.533\linewidth]{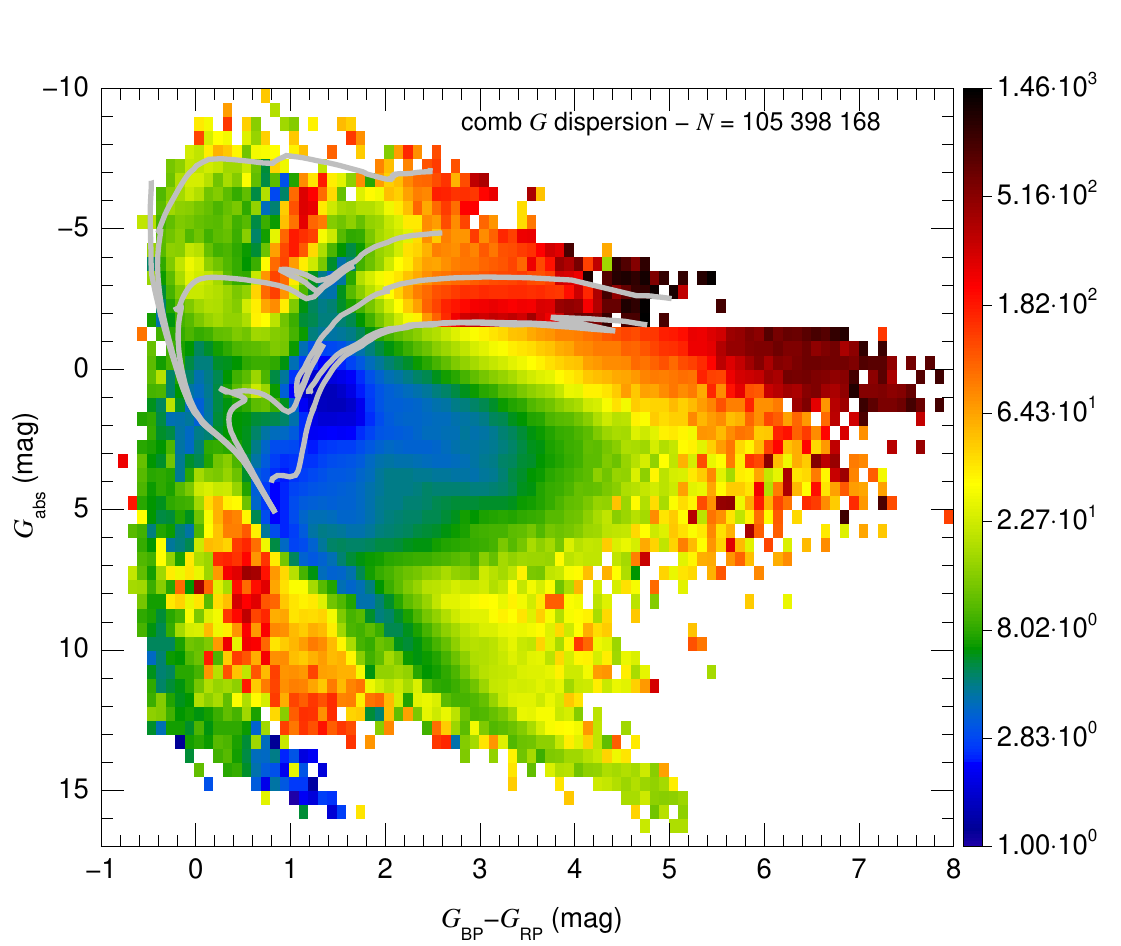} \\
\includegraphics[width=0.450\linewidth]{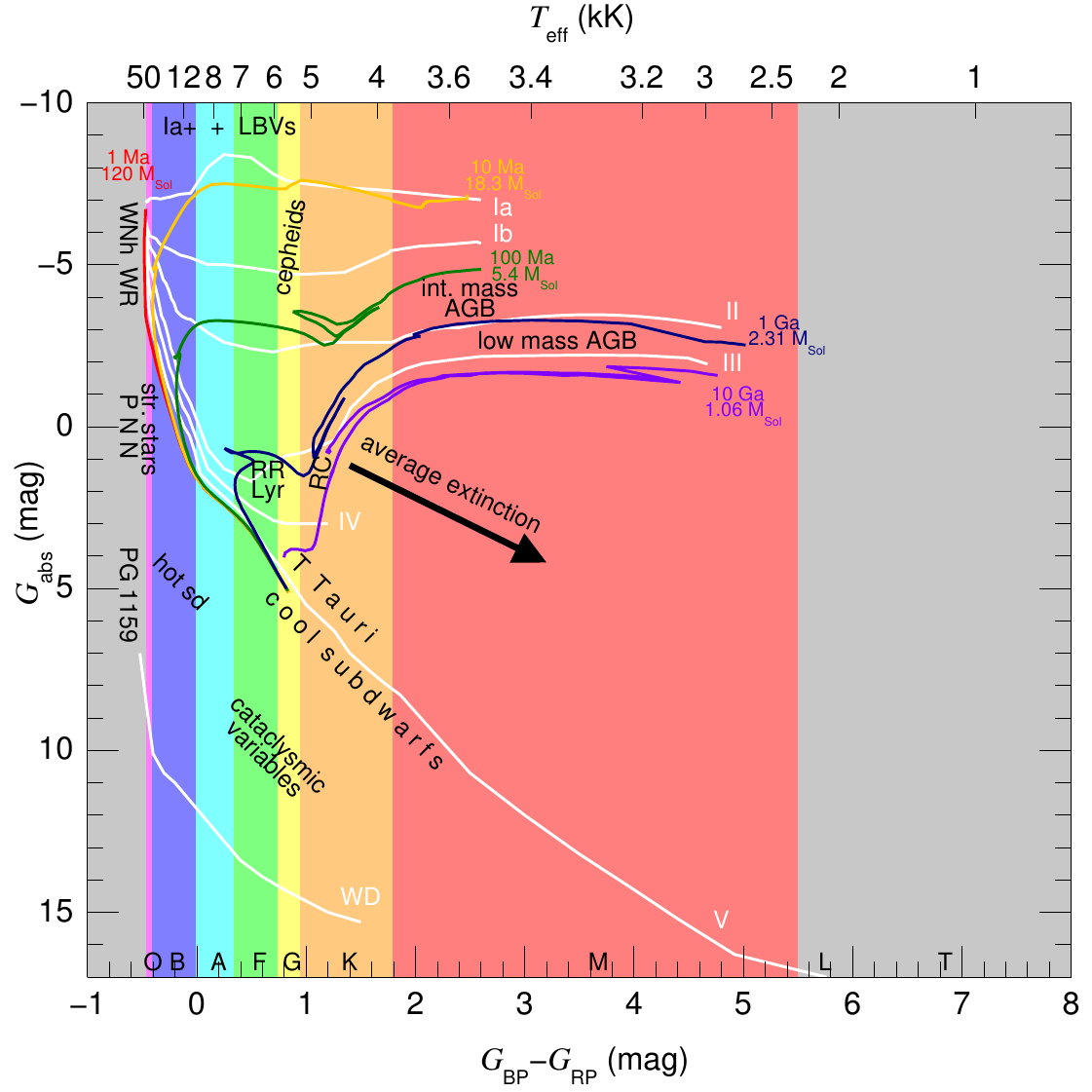} \
\includegraphics[width=0.533\linewidth]{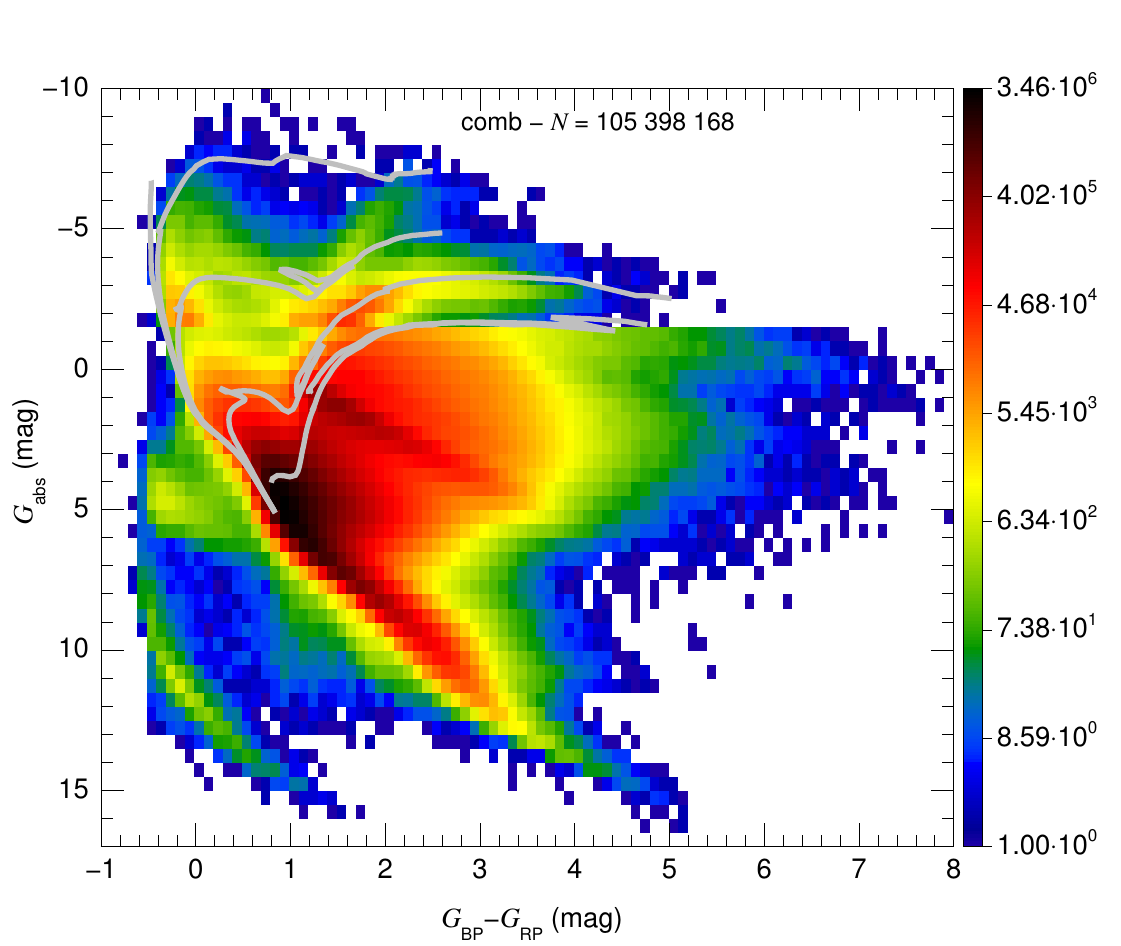}
\caption{Exploring the spectral classes. (Left column) HRD (top) and CAMD (bottom) with the sequences for different luminosity classes and white dwarfs (white), solar-metallicity isochrones of different ages (in color and labelled with their ages and maximum initial masses, all minimum initial masses are set to 1~\Msol), and different types of stars labelled (black). The two frames use common spectral type - \Teff\ - color scales that are only approximate, as they depend also on luminosity and metallicity. (Right column) \textit{Gaia} data from \citet{Maizetal23} to use as comparison, with objects brighter than $G_{\rm abs} = -1.5$~mag selected from the LMC sample (weak extinction) and objects fainter than that from the Milky Way sample (highly variable extinction, approximate direction marked with an arrow but see \citealt{Maiz24}). The top panel shows the average amplitude of the variability in the $G$ band in mmag and the bottom panel the source density, see \citet{Maizetal23} for details.}
\label{HRD}
\end{figure}

One particular mission that is likely to revolutionize spectral classification in the next decade is \textit{Gaia} \citep{Prusetal16}, as it has already obtained more optical spectra than the rest of the astronomical spectrographs in history combined. One instrument, the Radial Velocity Spectrometer \citep{Cropetal18,Recietal23}, is obtaining $\mathcal{R}$~=~\num[detect-all]{11500} spectroscopy in the 8470-8740~\AA\ range for unobscured stars as faint as $V = 16.5$ for spectral type G2~V (the $V$ limiting magnitude depends on spectral type and extinction) and can be used for standard spectral classification in the Ca triplet window. Alternatively, the data from the two low-resolution spectrographs can be combined to obtain $\mathcal{R}$~=30-100 spectra in the 3400-\num[detect-all]{10200}~\AA\ reaching even fainter stars \citep{DeAnetal23}. The low resolution implies that only strong lines are easily detected and the variable $\mathcal{R}$ across the range requires some special techniques to analyze the spectra \citep{Weiletal23} but the exceptional \textit{Gaia} flux calibration may compensate for it by e.g. measuring several Wolf-Rayet lines or the amplitude of the Balmer jump.

\subsection{A brief excursion through the spectral classes}

Before embarking on the spectral classification of different types of stars, we present Fig~\ref{HRD} as a tool to explore their different locations in an HRD or CAMD. The top left panel is a standard HRD with a linear scale in the seven spectral types OBAFGKM. As a means of comparing scales, we also show in the horizontal axes the \Teff\ (top) and \textit{Gaia} \BPRP~color (bottom, see \citealt{MaizWeil18} for band definitions), but note that those are approximate correspondences, as the \Teff\ and color of a given star also depend on luminosity class (or gravity) and metallicity. We plot the location in the HRD of different types of stars mentioned later in the text, as well as different luminosity classes (morphologically defined) and solar-metallicity isochrones (astrophysically calculated). 

The bottom left panel of Fig.~\ref{HRD} is the \textit{Gaia} CAMD equivalent to the HRD in the top left panel. It is presented as a way to directly compare observational CAMD diagrams (see description of right column panels) with the expected location for the different classes and types of stars presented below in the text. Two immediate observations can be made: [1] Late-type stars (and especially those of M type) cover a much larger dynamical range in \BPRP\ (and in most optical colors) than in spectral type. Note in particular how hot stars are hard to differentiate based on \BPRP\ alone. [2] The direction of an average extinction is marked by an arrow, but the direction of a given extinction can be somewhat different, as it also depends on the spectral type of the star and the amount and type of dust \citep{Maiz24}.

The bottom right panel of Fig.~\ref{HRD} is the density of \textit{Gaia} sources, with objects brighter than $G_{\rm abs} = -1.5$~mag selected from the LMC sample and objects fainter than that from the Milky Way sample. The reason for the division is that most of the LMC has a low extinction and a large population of luminous stars that is easy to associate with the different stellar types described in the text. \textit{Gaia} includes a much larger sample of Galactic stars but the highly variable degree of extinction blurs their location in the CAMD, something that is especially important for the most luminous stars due to their high average extinction in the Galactic sample. The most graphic tracer of the effect of extinction in this panel is how the Red Clump (RC) locus is transformed from a point into a (nearly straight but not perfectly so) line.

Finally, the top right panel of Fig.~\ref{HRD} is the average $G$ band photometric dispersion of the stars in the lower right panel (see \citealt{Maizetal23} for details). This panel is included to be used with the discussions below on the photometric variability of different types of stars.

\section{Spectral classification of hot stars}\label{sec4}

We define hot (or early-type) stars as those originally in Secchi's classes I and V, which in the standard spectral type sequence corresponds to O, B, A, and some F stars. The rest of the Secchi's classes correspond to cool (or late-type) stars and are analyzed in the next section, where we include all F stars (the transition spectral type) except RR Lyrae for simplicity. For an extensive description of the spectral classification of hot stars, the reader is referred to \citet{GrayCorb09}, including its chapter~3 on OB stars by Nolan R. Walborn. Most of \citet{GrayCorb09} remains up to date save for some significant modifications for O stars from the GOSSS project \citep{Maizetal11} and some minor ones for B stars \citep{Neguetal24} and WR stars \citep{CrowWalb11}, as described below. One important nomenclature issue (often missed in the literature) is that an OB star is not simply a star of either O or B spectral type, as one additional condition is required: that it is a massive star ($M \gtrsim 8$~\Msol). In spectral types terms, a working approximate definition is that it is of spectral subtype B2 or earlier if a dwarf, of B5 or earlier if a giant, and of any O or B type if a supergiant (\citealt{Morg51} and chapter by Walborn cited above). The distinction is important because most B stars are mid-to-late dwarfs, hence, of intermediate mass.

This section is divided into four parts: one each for ``normal'' stars of spectral types O, B, and A, and a last one for other hot stars, most of them highly evolved objects.

\subsection{O stars}

\begin{figure}[ht!]
\centerline{\includegraphics[width=1.00\linewidth]{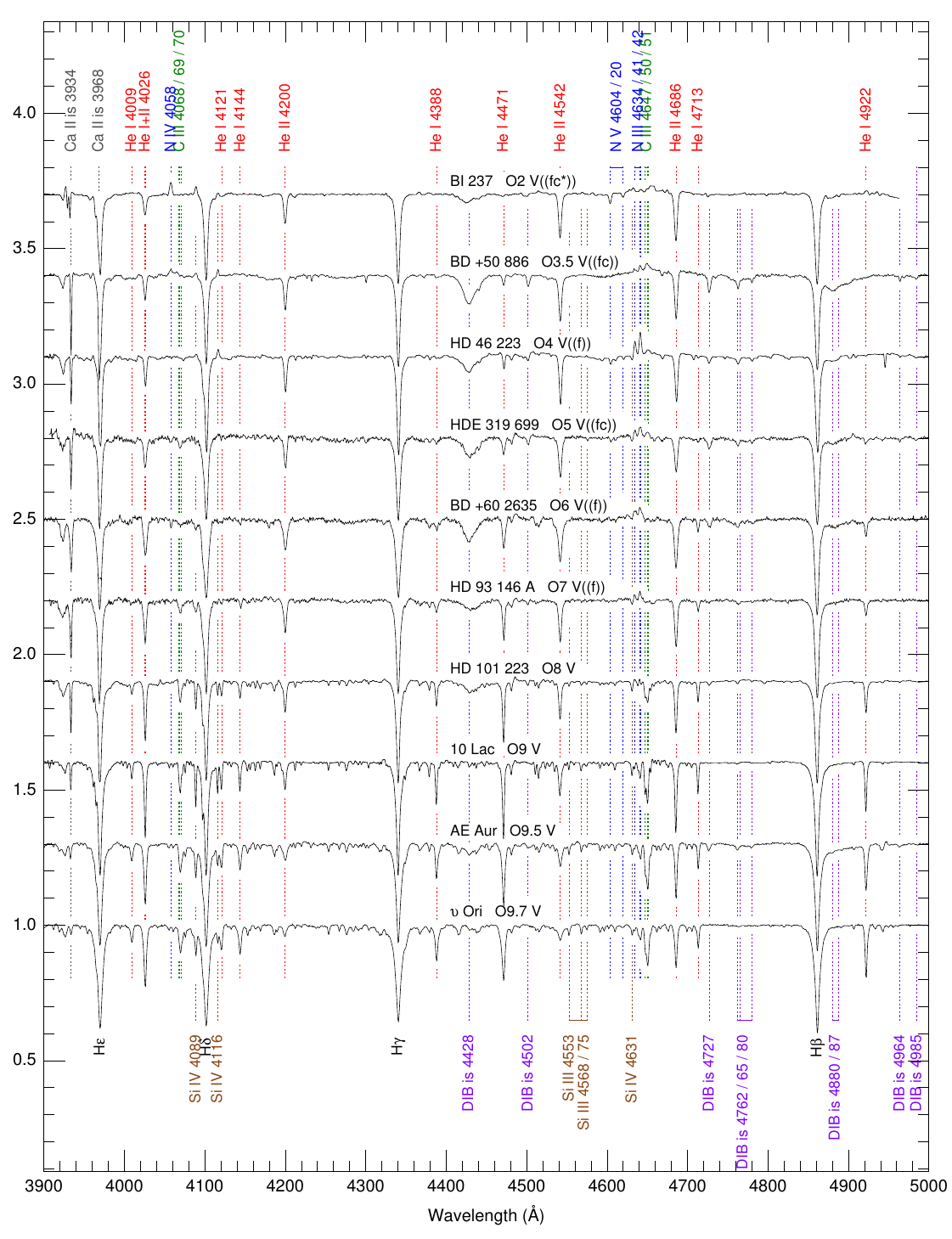}}
\caption{Spectral type sequence for O dwarfs at $\mathcal{R}\sim$~2500. Spectra are rectified, separated by 0.3 units, and from GOSSS or LiLiMaRlin.}
\label{OV}
\end{figure}

\begin{figure}[ht!]
\centerline{\includegraphics[width=1.06\linewidth]{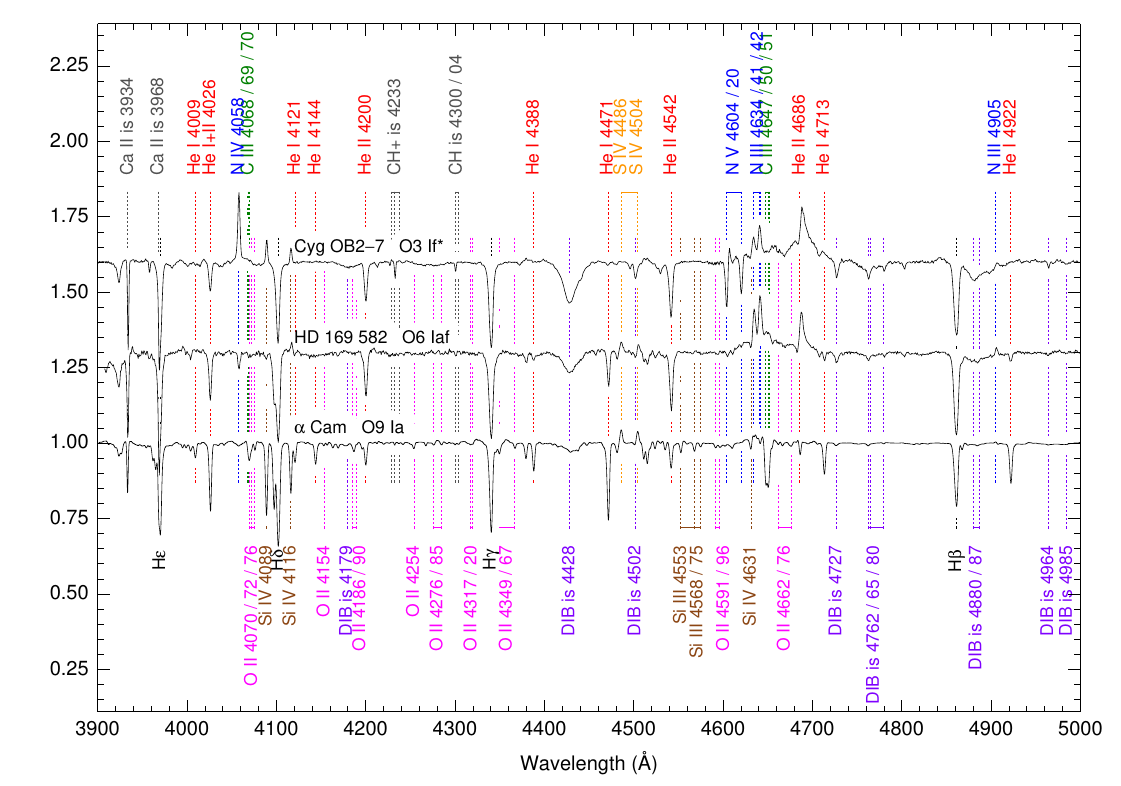}$\;\;\;\;\;\;\;\;\;\;$}
\caption{Examples of an early-, a mid-, and a late-type supergiants at $\mathcal{R}\sim$~2500. Spectra are rectified, separated by 0.3 units, and from GOSSS or LiLiMaRlin.}
\label{OI}
\end{figure}

O stars are the hot end of the spectral type sequence and, as such, were one of the last spectral types to have a fully developed classification grid. The original MKK atlas had O4 as the earliest spectral subtype and luminosity classes were only defined for O9 and O9.5 stars. Most of the subsequent work was done by Nolan R. Walborn or under his auspices from the 1970s to the 2010s\footnote{See also the series of papers by Peter Conti and collaborators starting with \citet{ContAlsc71}.}. He proposed luminosity classification criteria for the whole spectral type range \citep{Walb71a}, extended the spectral subtype range first to O3 \citep{Walb71b} and then to O2 \citep{Walbetal02b}, discovered the existence of carbon and nitrogen anomalies \citep{Walb71c,Walb76} and of the Onfp and Of?p phenomena described below \citep{Walb72}, and compiled an atlas of extreme stars of different OB types \citep{WalbFitz00}. He also helped to found the GOSSS project that has introduced further refinements to the spectral classification of O stars: the discovery of the Ofc phenomenon \citep{Walbetal10a}, the extension of the O9.7 subtype to all luminosity classes \citep{Sotaetal11a}, the definition of the O9.2 subtype \citep{Sotaetal14},  the extension of luminosity class IV to subtype O4 \citep{Maizetal16}, and the first large study of the OVz phenomenon \citep{Ariaetal16}. In a future paper with GOSSS data (Ma\'{\i}z Apell\'aniz et al. in preparation) the luminosity class II will be extended to the full O spectral type range.

\subsubsection{Classification criteria for O stars}

$\,\!$\indent \textbf{Spectral type.} The primary classification criterion for spectral subtype is the \HeII{4542}/\HeI{4471} ratio, which is unity at O7 (Figs.~\ref{nindex},~\ref{OV}), higher for earlier subtypes, and lower for later ones. Near both extremes of the sequence, additional criteria are used. For the earliest subtypes, the \NIV{4058}/\NIIIt{4634-40-42} ratio is used instead due to the common contamination of \HeI{4471} by hidden OB companions (Table~3 in \citealt{Walbetal02b}). For late-O stars, the additional criteria are \HeII{4542}/\HeI{4438} and \HeII{4200}/\HeI{4144} (both unity at O9) on the one hand and \SiIII{4552}/\HeII{4542} (unity at O9.7) on the other (Table~3 in \citealt{Sotaetal14}). Originally, O stars were defined as those having He\,{\sc ii} lines but, with modern digital spectra at good S/N and $R\approx$2500, \HeII{4542} is now detected until $\sim$B0.5 and \HeII{4686} until $\sim$B1.

\begin{figure}[ht!]
\centerline{\includegraphics[width=0.752\linewidth]{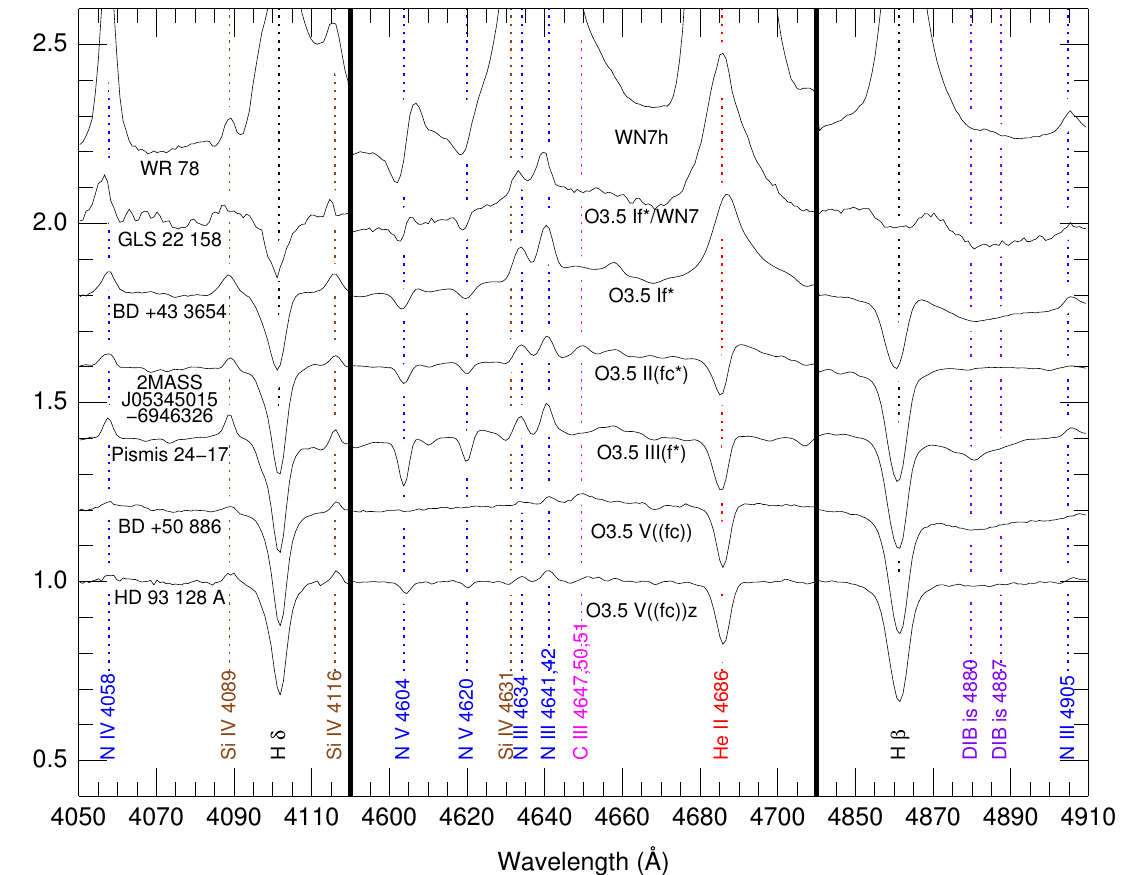}}
\centerline{\includegraphics[width=0.490\linewidth]{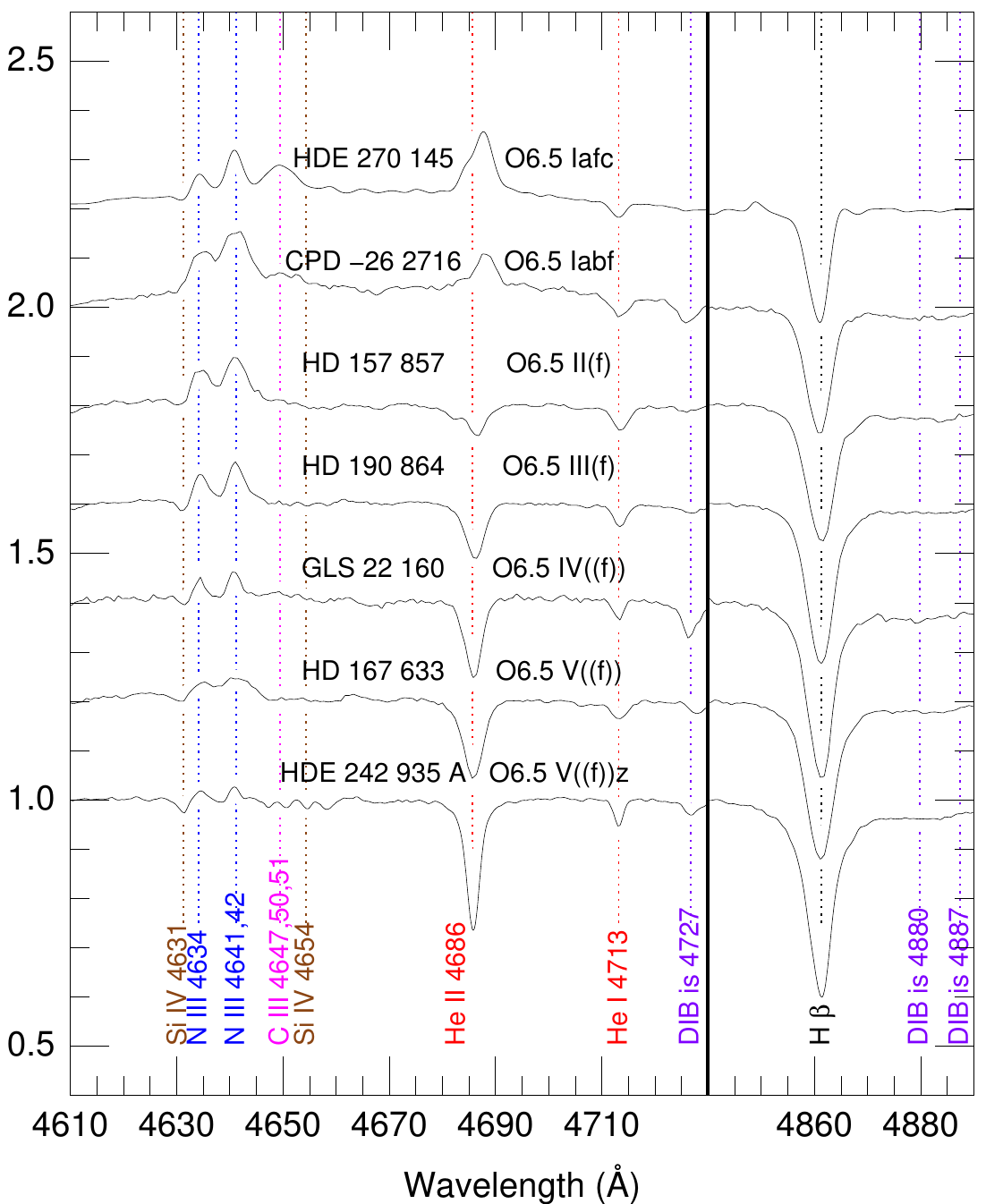}
            \includegraphics[width=0.490\linewidth]{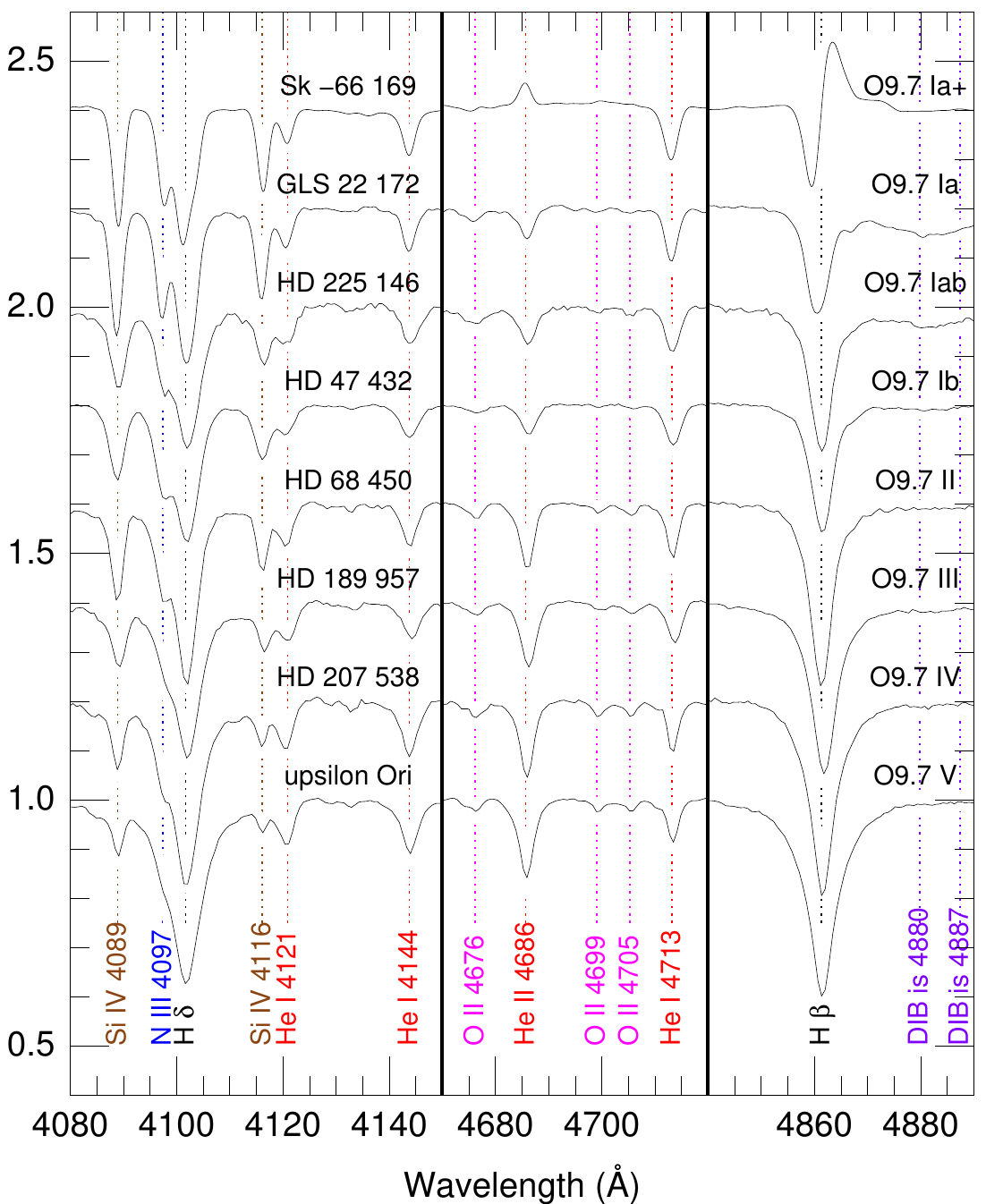}}
\caption{Luminosity class sequences for O3.5 (top), O6.5 (left), and O9.7 (right) at $\mathcal{R}\sim$~2500. Spectra are rectified, separated by 0.2 continuum units, and from GOSSS or LiLiMaRlin. The O3.5 sequence includes an Of/WN star (GLS~\num[detect-all]{22158}) and a WN star (WR~78), with part of the intense emission lines of the second object outside the frame.}
\label{OLC}
\end{figure}

\begin{figure}[ht!]
\centerline{\includegraphics[width=1.00\linewidth]{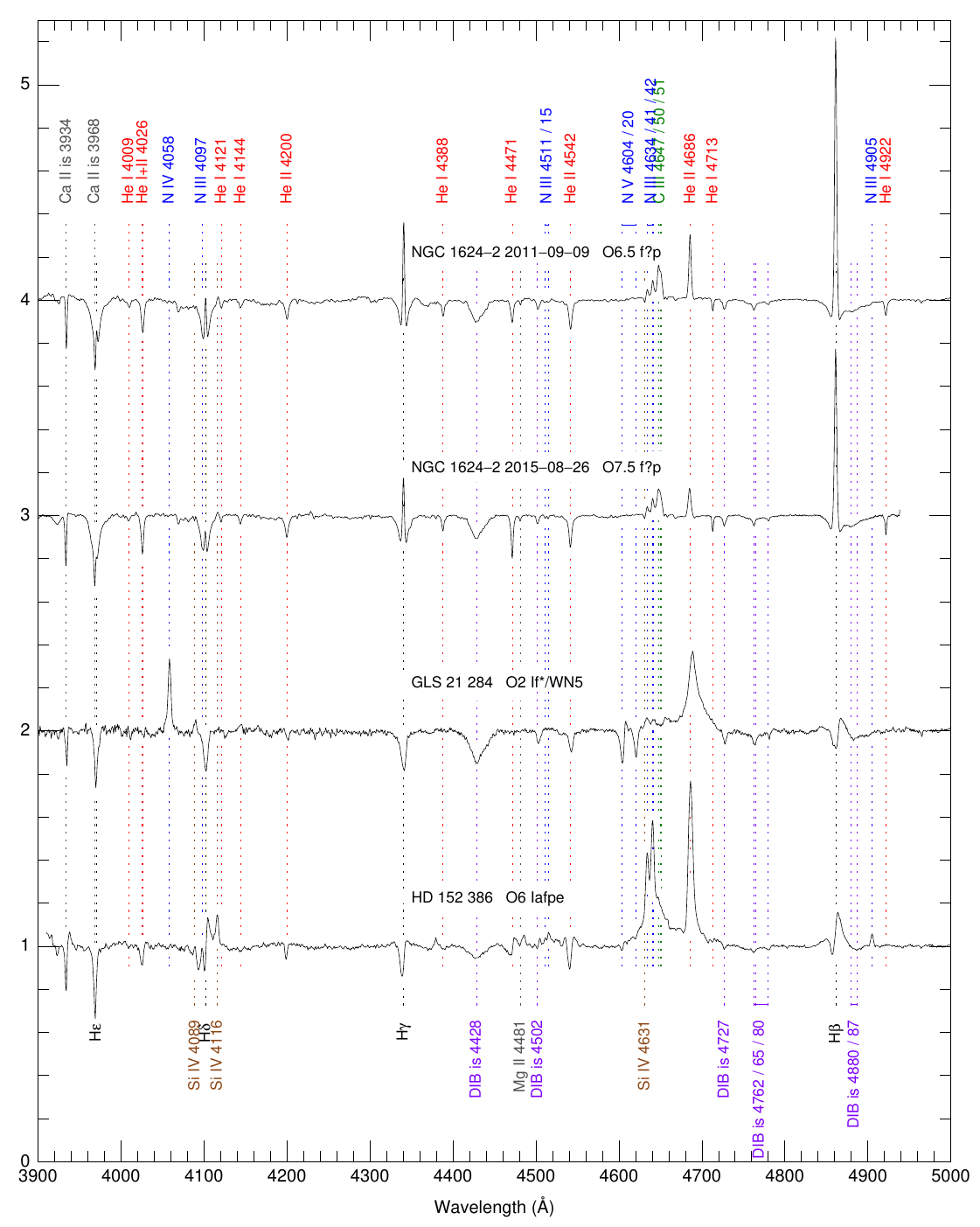}}
\caption{Examples of an Of?p, an Of/WN, and an O Iafpe stars at $\mathcal{R}\sim$~2500. For the Of?p star, two epochs with different spectral types due to the infilling of \HeI{4471} are shown. Spectra are rectified, separated by one continuum unit, and from GOSSS or LiLiMaRlin.}
\label{Opec}
\end{figure}

\textbf{Luminosity class.} The primary luminosity criterion for O stars is the Of phenomenon \citep{Walb71a}, by which (a) \HeII{4686} goes from strong absorption in dwarfs to emission in supergiants (a negative luminosity effect from the point of view of absorption) and (b) \NIIIt{4634-40-42} in emission increases in intensity along the same sequence. The Of phenomenon is expressed using the qualifiers in Table~\ref{qualifiers}. \HeII{4686} and \NIIIt{4634-40-42} are selective emission lines, that is, they can be in emission while other lines of the same ion remain in absorption. The OVz phenomenon (\citealt{Ariaetal16}, see criterion in Table~\ref{qualifiers}) is the opposite of the Of effect: OVz stars correspond to objects near the ZAMS and have \HeII{4686} in absorption stronger than normal class V stars. The location of the normal V sequence above the ZAMS can be seen in the top left panel of Fig.~\ref{HRD} by comparing it with the 1~Ma isochrone, which is a reasonable approximation to the latter.

The Of and OVz phenomena are not caused directly by luminosity but indirectly: they represent the infilling of absorption lines by the increase in the strength of the wind that accompanies an increase in luminosity for a constant \Teff. The Of phenomenon appears in early- and mid-O supergiants and is stronger for the earliest subtypes (Fig.~\ref{OI}). The standard Of phenomenon is shown in the left panel of Fig.~\ref{OLC} but for early types the phenomenon is even stronger and involves other selective emission lines such as \NIV{4068},  \SiIV{4089} and \SiIV{4116} (top panel of Fig.~\ref{OLC}). For late-O stars (right panel of Fig.~\ref{OLC}) the OVz phenomenon disappears and only a weak version of the Of phenomenon is present, not affecting \NIIIt{4634-40-42} and only showing up with a progressive infilling of \HeII{4686}.

The decay of the Of effect with increasing spectral subtype leads to the development of other luminosity criteria for late-O stars (which extend into the B stars, as we will see below). The most commonly used one is the ratio of either \SiIV{4089} or \SiIV{4116} to \HeI{4026} or \HeI{4144} (right panel of Fig.~\ref{OLC}), with the Si absorption lines showing a positive luminosity effect. Note that the use of a Si/He line ratio may be sensitive to metallicity effects while some He lines experience complex behaviors when originating from unresolved binary systems composed of a late-O and an early-B star (which are quite common, see \citealt{SimDetal15a}), and so luminosity discrepancies between different criteria in the spectral subtype range are common \citep{Walbetal14}.

We also point out that one of the luminosity criteria for B and A stars, the width of the Balmer lines, may also be used to some degree for O stars. The negative luminosity effect due to the Stark effect is clear for late-O types (right panel of Fig.~\ref{OLC}) and is combined with a wind infilling analogous to the Of phenomenon that is strong enough to create P-Cygni profiles in H$\beta$ for O9.7 hypergiants and Of/WN stars (top panel of Fig.~\ref{OLC}, be careful with the broad double DIB to the right of H$\beta$, \citealt{Maizetal14b}). The wind infilling effect is even more noticeable for H$\alpha$, outside the blue-violet range, but is weak or disappears for H$\gamma$ and higher-order Balmer lines.

For O stars, one should not equate the main sequence (MS) phase of stellar evolution with luminosity class V. O-type giants and even supergiants are still burning hydrogen in their cores \citep{MartPala17}. For most B and A stars, the transition between core and shell hydrogen burning happens during or near the end of their giant phase. Some early-B supergiants may also be burning hydrogen in their cores, but this issue is still controversial \citep{deBuetal24}. Therefore, in general the main sequence extends beyond luminosity class V for early-type stars.

\subsubsection{Peculiar O stars}

Here we briefly describe the peculiar categories for O stars.

\begin{itemize}
 \item \textbf{Early Of/WN.} The Of phenomenon reaches its extreme with WNh stars (see below): high-luminosity, narrow-line, late-type WN stars \citep{Walb74}, which are early O supergiants on steroids in terms of wind strength. An intermediate class between those WN stars and early-O supergiants is the early Of/WN (or ``hot slash'') category \citep{Walb82b,Sotaetal14}. The dividing criterion is set by H$\beta$: It is in absorption for supergiants, has a P-Cyg profile for ``hot slash'' stars, and is in emission for WN stars (\citealt{CrowWalb11}, top panel of Fig.~\ref{OLC} and Fig.~\ref{Opec}).
 \item \textbf{O Iafpe.} Another class of extreme Of stars are the O~Iafpe objects, which have a complex identification history. They were originally known as Ofpe/WN9, late Of/WN, or ``cool slash'' \citep{Walb82c} but were subsequently reclassified as a cool extension to the late WN sequence (WN9-11, \citealt{CrowSmit97}). Some authors prefer their division into either O or WR stars \citep{CrowBoha97} while others group them in a single one as O Iafpe \citep{Sotaetal14} because the category now includes stars as early as O4.5 and because they are not an intermediate stage between O supergiants and WN stars. Instead, they are objects that can be characterized as either. They are defined by having \HeI{4471} P-Cyg profiles and a complex morphology in the H$\delta$ region (Fig.~\ref{Opec}).
 \item \textbf{Of?p.} Of?p stars \citep{Walb72} are magnetic oblique rotators \citep{Munoetal20} that are characterized by strong and variable \CIIIt{4647-50-52} emission accompanied by variable H and He lines that can appear in absorption, emission, or as P-Cyg profiles \citep{Walbetal10a}. The variability is produced by the change in the direction from which the magnetosphere is observed and the variations in the infilling of \HeI{4471} cause changes in the observed spectral type as a function of the rotational phase (Fig.~\ref{Opec}). There are currently six known examples in the Milky Way \citep{Maizetal19b} and six in the Magellanic Clouds \citep{Walbetal15b,Hubretal24}. Of?p stars are of mid-O spectral type and are isolated or in non-interacting binaries, but  magnetism can occur in O stars under different circumstances and exhibiting different spectral phenomena \citep{Nazeetal16a,Nazeetal17b,Wadeetal20}.
 \item \textbf{Ofc.} The Ofc phenomenon was discovered by \citet{Walbetal10a} by analyzing the large sample of GOSSS stars with good-quality digital data. It takes place at early- and mid-O types and is characterized by \CIIIt{4647-50-52} stable emission comparable to that of \NIIIt{4634-40-42}, sometimes with the addition of \CIV{4658} for the earliest types. Hence, it is integrated in the same nomenclature as the Of phenomenon adding a c (Table~\ref{qualifiers}).
 \item \textbf{Onfp.} Onfp stars were defined by \citet{Walb72} as Of spectra with \HeII{4686} emission with a central reversal. Most of them are rapid rotators, hence the Onfp denomination. 
 \item \textbf{ON/OC.} \citet{Walb71c,Walb76} discovered the existence of anomalous C and N line strengths with respect to the He+Si reference frame and devised a nomenclature for this class in the form of suffixes: N$\rightarrow$N str$\rightarrow$normal$\rightarrow$N wk$\rightarrow$C in order of decreasing N and increasing C strengths (Table~\ref{qualifiers}). The surface composition of O stars may show signs of CNO processing in their interiors, having N enrichment and C depletion. Normal evolved O stars are N enriched with respect to their initial composition but some are more so (ON stars) and some less so (OC stars). The ON/OC phenomenon appears at late O and early B stars (though some very early O stars also show significant differences in the strength of N lines) and some ON giants are rapid rotators (ONn or ``double n'' stars, \citealt{Walbetal11}), pointing towards rotational mixing as a source of the additional enrichment. 
 \item \textbf{Oe.} Oe stars are the hotter equivalent of Be stars \citep{Neguetal04}, see below and Fig.~\ref{esuffix}.
\end{itemize}

\subsection{B stars}

\begin{figure}[t]
\centerline{\includegraphics[width=1.00\linewidth]{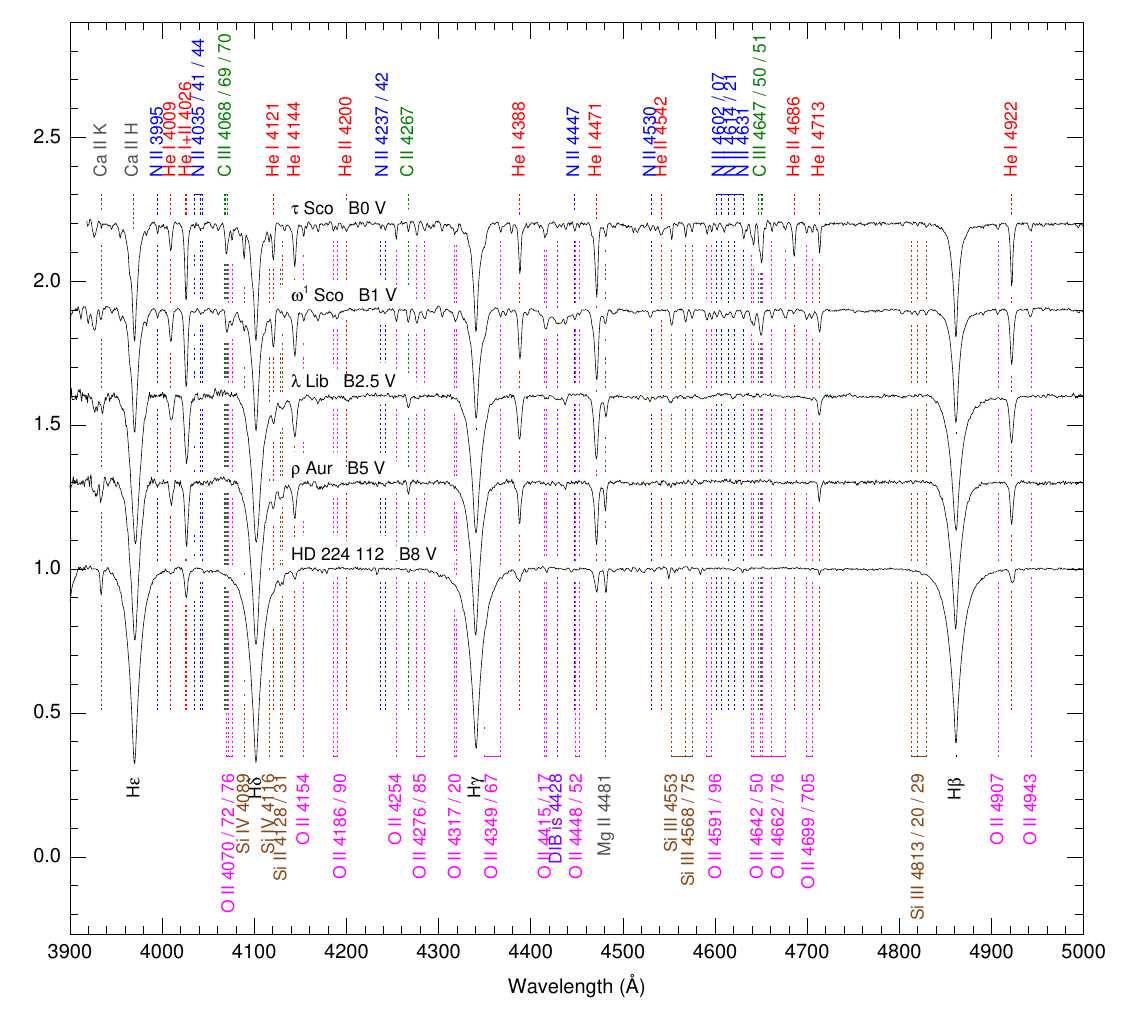}}
\caption{Spectral type sequence for B dwarfs at $\mathcal{R}\sim$~2500. Spectra are rectified, separated by 0.3 continuum units, and from GOSSS or LiLiMaRlin. The Ca\,{\sc ii}~H+K lines are of interstellar origin for the early-to-mid subtypes.}
\label{BV}
\end{figure}

\begin{figure}[t]
\centerline{\includegraphics[width=1.00\linewidth]{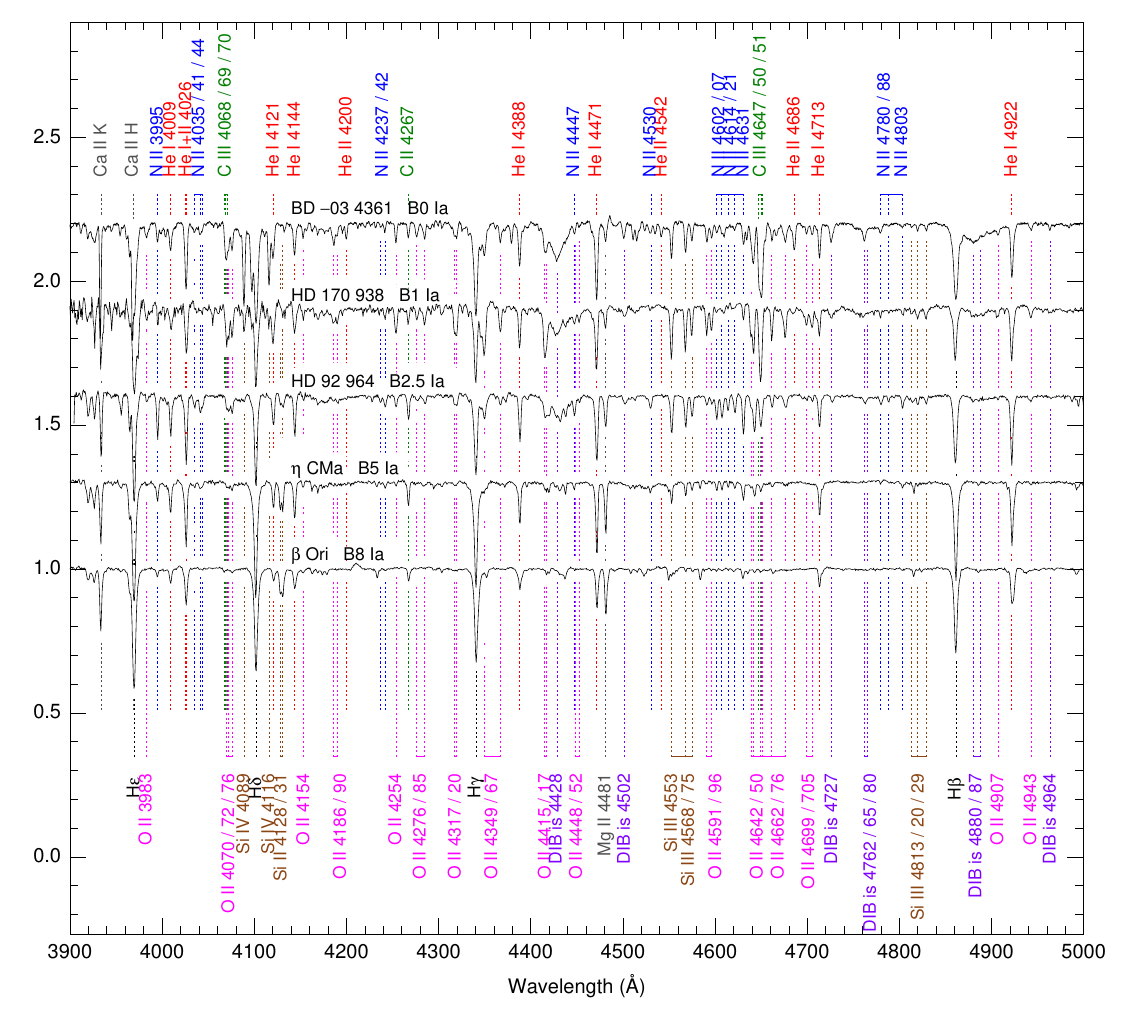}}
\caption{Spectral type sequence for B supergiants at $\mathcal{R}\sim$~2500. Spectra are rectified, separated by 0.3 continuum units, and from GOSSS or LiLiMaRlin. The Ca\,{\sc ii}~H+K lines are of interstellar origin for the early-to-mid subtypes.}
\label{BI}
\end{figure}

Starting with the B spectral class we find significant differences between the natures of dwarfs and supergiants, with the latter being massive stars (8~\Msol\ or more, so they are able to fuse elements beyond helium in their advanced stages) and most of the former being of intermediate mass (which only reach helium burning and do so without a flash). For the earliest B types (B0-2), the difference is not yet large, as their luminosity class V are still massive  and have MS lifetimes of 8-30~Ma. However, by the time we reach B9 V we find stars with masses around 3~\Msol\ and MS lifetimes around 300~Ma. Hence, early-B and late-B dwarfs are very different types of stars.

The whole range of the B spectral type was already covered in the MKK atlas, even though only the B0, B0.5, B1, B2, B3, B5, and B8 subtypes were included at first. By the time of \citet{JohnMorg53} the B1.5, B6, B7, and B9 subtypes had already been added. The list was completed with B2.5 and B4 by \citet{Lesh68} and with B0.2 and B0.7 by \citet{Walb71a}. \citet{Morgetal78} also introduced the B9.5 subtype but the recent analysis by \citet{Neguetal24} recommends not using it because it is difficult to differentiate from either B9 or A0. The reader is referred to \citet{Neguetal24} for details, noting that reference also suggests the possibility of dropping the B4 subtype and merging B6 and B7 into one.

\begin{figure}[t]
\centerline{\includegraphics[width=1.00\linewidth]{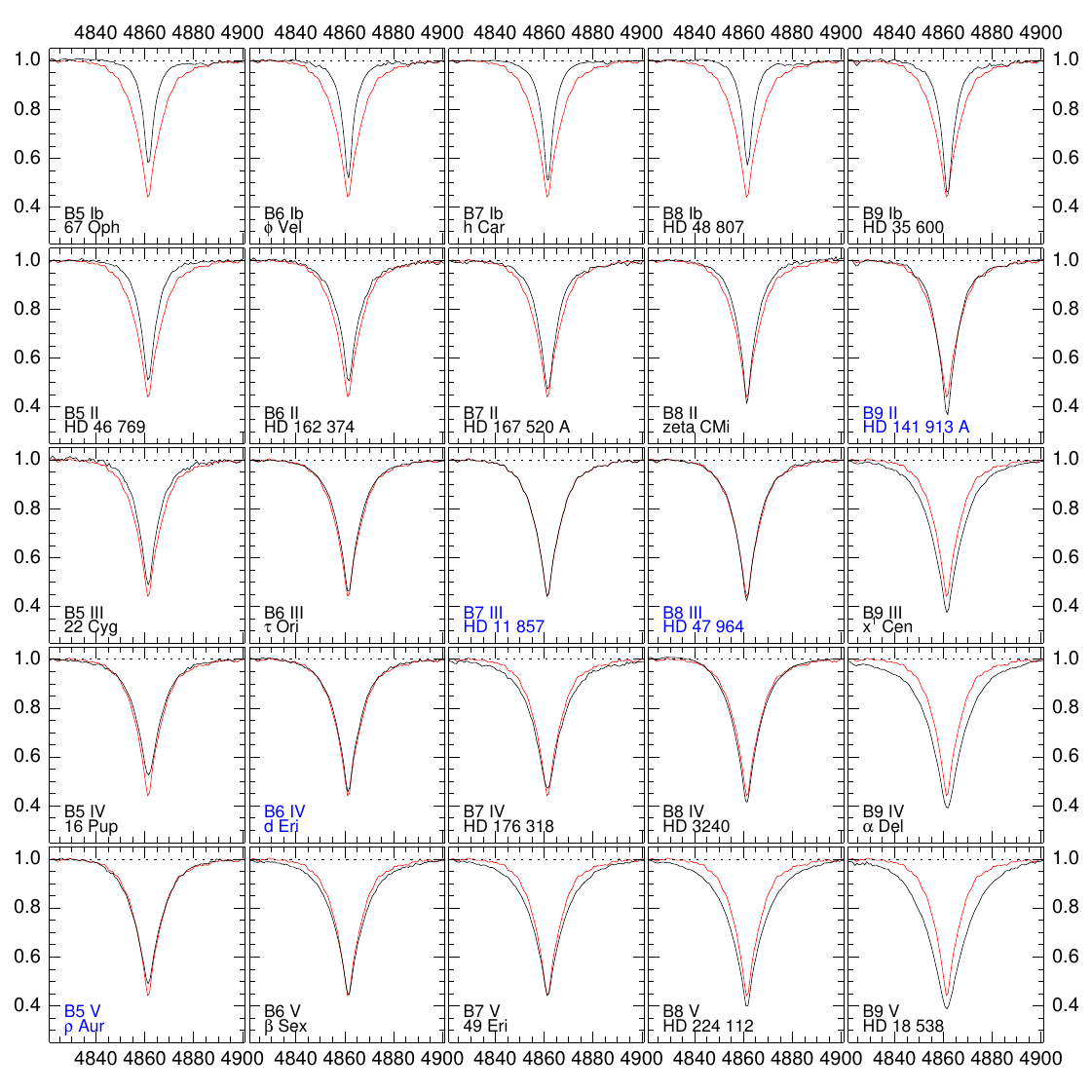}}
\caption{The diagonal effect as observed in H$\beta$ at $\mathcal{R}\sim$~2500, compare with \citet{GarrGray94} The grid of plots shows the rectified line for the spectral subtype range B5-B9 and the luminosity class range V-Ib. Stars with names in blue mark the diagonal where the H$\beta$ profile is very similar. The profile of the B7~III star
 (HD~\num[detect-all]{11857}) is shown in all plots in red for comparison. Profiles above the diagonal are narrower and of lower EW while those below it are broader and of higher EW. The horizontal axes are in \AA}
\label{diagonal}
\end{figure}


\subsubsection{Classification criteria for B stars}

$\,\!$\indent \textbf{Spectral type.} There are three criteria that can be used to obtain the spectral subtype for B stars. Each one has its pros and cons, so they should be used in combination if possible. The spectral sequences for dwarfs and supergiants are shown in Figs.~\ref{BV}~and~\ref{BI}, respectively. See \citet{GrayCorb09} and \citet{Neguetal24} for more details:

\begin{itemize}
 \item \textit{H and He.} Originally, B stars were defined as those with He\,{\sc i} and without He\,{\sc ii} lines, but that is no longer valid with modern digital data. On the one hand, \HeII{4542} can be detected until B0.5 and \HeII{4686} until B1. On the other, \HeI{4471} can be detected in early A stars, especially in supergiants. Besides the He\,{\sc ii} lines, the strength of the He\,{\sc i} lines changes along the B spectral range and Balmer lines become increasingly stronger for later types. The combination of H and He lines can be used to determine or constrain the horizontal classification, especially for early subtypes, with two caveats: [a] the additional dependence of both of luminosity (Fig.~\ref{diagonal}) and [b] the possibility of He abundance anomalies, see section below on peculiar B stars.
 \item \textit{Si}. The preferred criterion for obtaining the spectral subtype are the line ratios of Si lines: \SiIV{4089}/\SiIII{4552} for early subtypes and \SiIII{4552}/\SiIId{4128,32} for mid and late subtypes. The main advantage is its independence of composition anomalies or metallicities. The main disadvantage is that the strength of the metal lines is luminosity dependent, making this criterion difficult to apply in mid and late dwarfs and late giants. This is worsened by the significant fraction of fast rotators among B stars, which can make weak Si lines more difficult to detect. 
 \item \textit{Mg/He and others.} The difficulty detecting weak Si lines may force the use of other criteria. \MgII{4481}/\HeI{4471} is the most important one, increasing with subtype until it becomes unity at B8. Two caveats are in order for this criterion: [a] \HeI{4471} is intrinsically broader due to the Stark effect, so line strengths should be used, not depth (see line broadening above) and [b] the ratio is metallicity dependent, leading to different spectral subtypes than those from Si ratios at low metallicities \citep{Lenn97}. In addition to Mg, other metallic lines can be used to constrain the spectral subtype. For example, \CII{4267} reaches its maximum strength at B2 for class V.
\end{itemize}

\textbf{Luminosity class.} As with the spectral type, there are also three luminosity classification criteria for B stars. Also, due to the effect of composition anomalies and metallicity effects, if possible they should be used simultaneously.

\begin{itemize}
 \item \textit{Si/He.} In general, metal lines show a positive luminosity effect and, of those, the most used ratios are \SiIV{4116}/\HeI{4121} for early-B subtypes and \SiIII{4552}/\HeI{4387} or \SiIII{4552}/\HeI{4471} for early- and mid-B subtypes. As those same Si lines are used to obtain the spectral subtype, both classifications have to be done simultaneously. The two objections to this criterion are its metallicity and He anomalies dependences and the weakness of the lines for late types.
 \item \textit{H.} The width and EW of the Balmer lines depend on both the spectral subtype and the luminosity class \citep{BaloCram74}, leading to the diagonal effect where stars in a grid diagonal have a similar profile (Fig.~\ref{diagonal}): both the width and the EW increase with spectral subtype and with luminosity. Therefore, using a combination of the spectral type criteria and the diagonal effect, a full classification can be obtained. Traditionally, this method was used mostly for mid- and late-subtypes but \citet{Neguetal24} have shown it can be used for all the B subtypes. For hypergiants (class Ia+) H$\beta$ shows a P Cyg or emission profile, sometimes with a broad pedestal \citep{Walbetal15a}.
 \item \textit{Other metallic lines.} Lines such as \OII{4070}, \OII{4416}, \OII{4348}, and \NII{3995} also show a positive luminosity effect and can be used for classification. However, some B stars are affected by anomalous C and N strengths in a similar way to some late O stars, so the luminosity classification should be confirmed by another criterion or, otherwise, the anomaly should be noted.
\end{itemize}

\subsubsection{Peculiar B stars}

Here we briefly describe the peculiar categories for B stars.

\begin{itemize}
 \item \textbf{He strong and He weak.} Some B stars have large He abundance anomalies. The cases among the earlier subtypes are enriched in He (e.g. $\sigma$~Ori~E) and the ones among the later ones are He poor (e.g. 3~Cen~A). Most of the examples in this category have been found to have magnetic fields.  
 \item \textbf{Bp.} In addition to He, B stars can have anomalous abundances of other elements. Some of them are the hot versions of the classical magnetic Ap stars described below. Others are enriched in singly-ionized mercury and manganese and are accordingly called HgMn stars \citep{Paunetal21}. They have spectral subtypes between B7 and B9 and luminosity classes between V and III.
 \item \textbf{Be and B shell.} Be stars are the most common type of peculiar B stars, to the point that already Secchi created a class for them (V, see above). They are characterized by having some (at least H$\alpha$) or all of the Balmer lines in emission but with an extreme variety in intensity (see Fig.~\ref{esuffix}) and in the lines affected, as objects with an intense Be phenomenon also have other lines (e.g. from He\,{\sc i} and Fe\,{\sc ii}) in emission. Be stars are highly variable, both in the emission lines themselves \citep{Dimietal18} and photometrically in general \citep{Maizetal23}. Indeed, a Be star can go through a non-Be phase where its emission lines disappear but even then it is considered a Be star i.e. ``once a Be, always a Be'' \citep{PortRivi03}. The Be phenomenon includes a diversity of objects, but the most common ones are classical Be stars, which are of luminosity classes V to III (and sometimes II) and for which the emission lines are produced in a disk around it that is tied up to pulsations in the star. Be stars with weak emission lines are easy to classify following the same criteria as normal B stars, but those with strong emission lines can be notoriously difficult. Some Be stars can experience a B shell phase, where narrow cores in the Balmer lines and absorption lines of singly ionized metals appear. See the A-star section for the Herbig Be stars.
 \item \textbf{B[e].} B[e] stars are objects that, in addition to strong Balmer emission, show emission in forbidden metal lines, mostly of [Fe\,{\sc ii}] and [O\,{\sc i}]. The B[e] phenomenon has diverse origins \citep{Lameetal98}: supergiant stars in one of the LBV phases (\citealt{WalbFitz00}, see below), PMS stars (see below for Herbig AeBe stars), symbiotic stars with a cool and a hot component, and even some for which the origin is currently unknown.
 \item \textbf{B Iape.} These objects are the early-B extension of the O Iafpe stars \citep{WalbFitz00}.
 \item \textbf{BN/BC.} BN and BC stars are the cooler equivalents of ON and OC stars, see above.
\end{itemize}

\subsection{A stars}

The A stars are the first spectral class where the population is divided into three distinct types according to their luminosity. Most dwarfs are low-mass stars\footnote{They only reach helium burning and ignite it with a flash.} and are the majority of the population while supergiants are evolved massive stars in a brief stage of their evolution, making them rare in absolute numbers but easy to observe as they are some of the most luminous stars in the optical due to their small bolometric correction compared to earlier or later spectral types. In between, we find different types of objects, from low-mass stars around the end of their hydrogen core burning to intermediate-mass stars evolving to the right or the left of the HRD and low-metallicity horizontal branch (HB) stars burning helium in their cores.

\subsubsection{Classification criteria for A stars}

\begin{figure}[t]
\centerline{\includegraphics[width=0.490\linewidth]{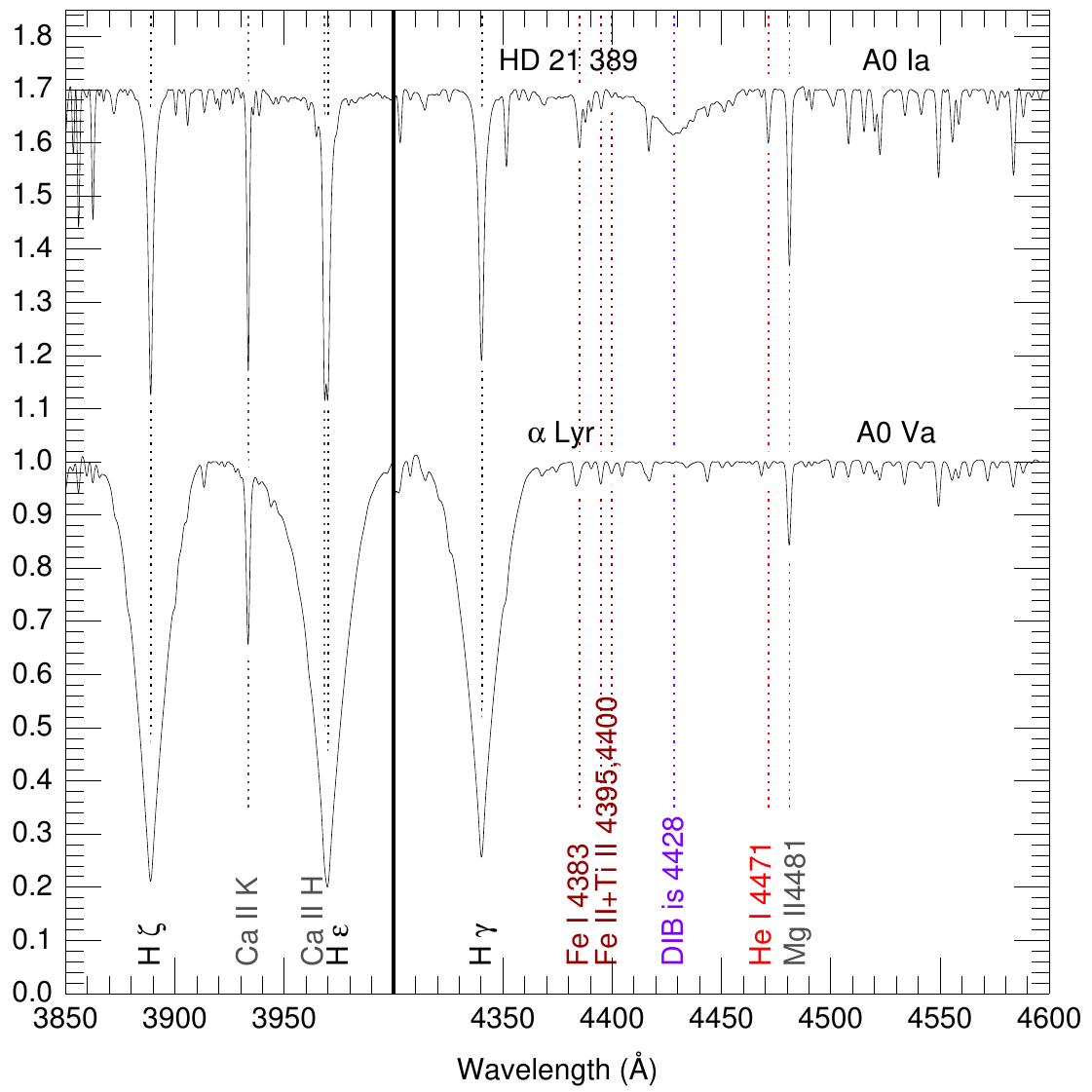}
            \includegraphics[width=0.490\linewidth]{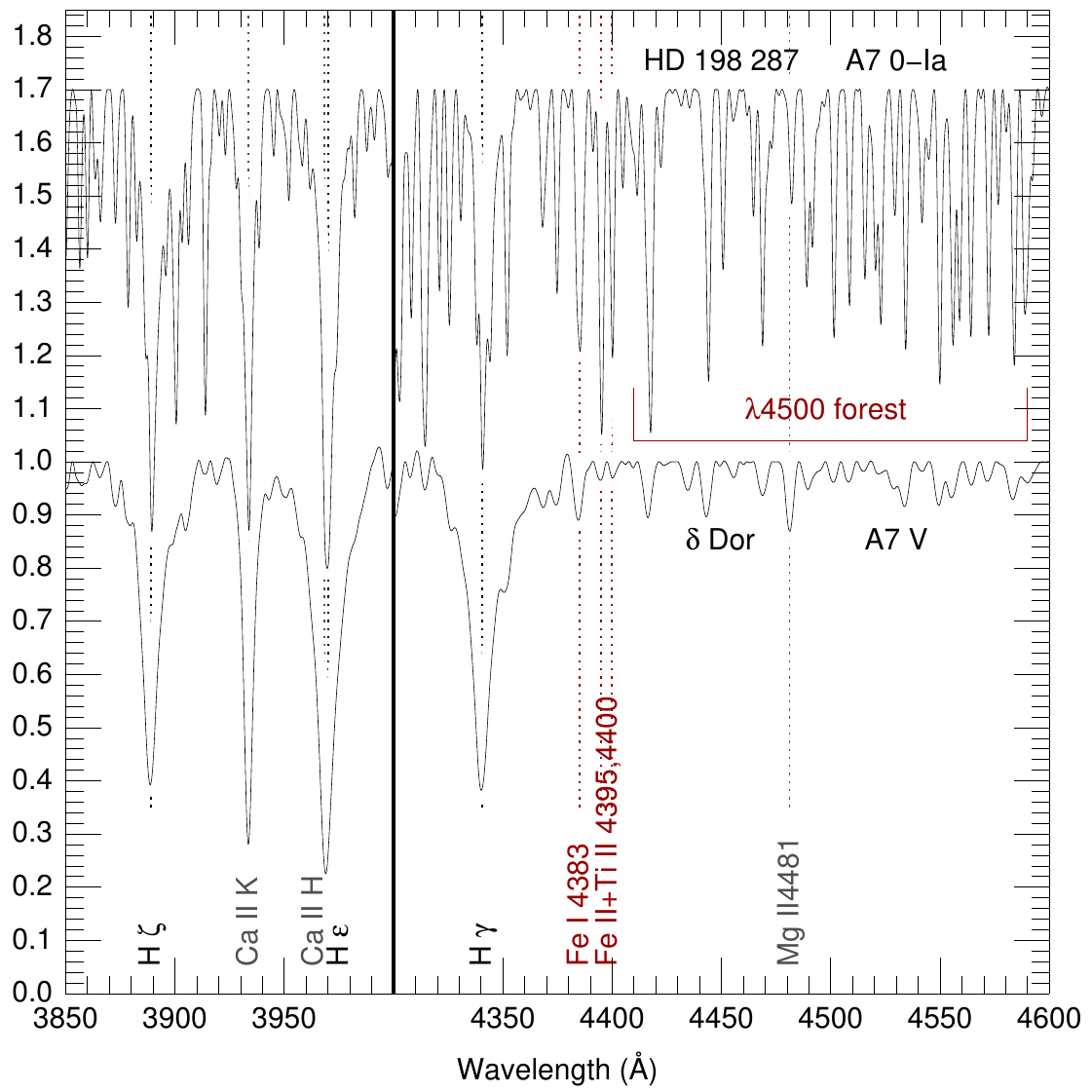}}
\caption{Selected spectral regions (3850--4000~$\AA$ and 4300--4600~$\AA$) of A0 (left) and A7 (right) dwarfs (bottom) and supergiants (top) at $\mathcal{R}\sim$~2500. Spectra are rectified, separated by 0.7 continuum units, and from LiLiMaRlin.}
\label{A}
\end{figure}

$\,\!$\indent \textbf{Spectral type.} Originally, A0 stars were defined as those where He\,{\sc i} lines were no longer visible, but with modern data that is no longer the case. There are three criteria used to determine the spectral type \citep{GrayGarr87,GrayGarr89b}, which are showcased in Fig.~\ref{A}:

\begin{itemize}
 \item \textit{H.} Balmer lines increase from A0 to A2 and decrease afterwards.
 \item \textit{Ca}. \CaII{3934} (or Ca K), already detectable for late-B stars, greatly increases with subtype for A stars. Two notes are made in this respect: for A stars \CaII{3968} (or Ca H)  is mixed with H$\epsilon$ (dominated by the latter at A0 but comparable for late subtypes) and both  Ca K+H can be significantly contaminated by interstellar component(s). 
 \item \textit{Metallic lines.} The overall metallic contribution increases with subtype, with a nearly featureless spectrum for early subtypes (much as for late-B stars) to a large number of lines for late subtypes (mostly weak at this stage, as the transition from Secchi's class I to II takes place for F stars, see below). Among the useful lines we find \CaI{4226}, \FeI{4271}, and the \MnI{4030} blend. 
\end{itemize}

\textbf{Luminosity class.} The primary luminosity criterion for A stars is the profile (width and EW) of the Balmer lines. As it happens for the B stars, both have a negative luminosity effect, but they also depend on the spectral subtype. The effect is especially strong for the early subtypes (Fig.~\ref{A}), where some authors divide the V class into Va (normal MS) and Vb (ZAMS) and even introduce Va$^-$ for the intermediate class between the two and Va$^+$ for the intermediate one between Va and IV-V. This can be done because of the significant separation in the HRD between the normal V sequence and the ZAMS for this spectral type, see Fig.~\ref{HRD}. For the late subtypes the Balmer lines become less sensitive to luminosity and metallic lines with a positive luminosity effect have to be used instead, in particular different blends of Fe\,{\sc ii} and Ti\,{\sc ii} at 4172-4179~\AA, 4395~\AA, 4400~\AA, and a forest of lines around 4500~\AA\ (Fig.~\ref{A}).

Two complications arise with the spectral classification of A stars. The first is the significant number of fast rotators, something that was also seen for O and B stars, which hampers the classification by making lines less prominent (Fig.~\ref{nindex}). The second is the large fraction of A stars with chemical peculiarities, already present in B stars but more important for A stars. As there are three spectral classification criteria for A stars, it is possible to derive three different spectral subtypes, leading to a notation such as kA2hA9mF2. This corresponds to A2 for the Ca K-line subtype (k), A9 for the hydrogen-line type (h), and F2 for the metallic-line type (m). This notation is used for some of the peculiar A stars and for HB stars (population II objects).

\subsubsection{Peculiar A stars}

Here we briefly describe the peculiar categories for A stars.
\begin{itemize}
 \item \textbf{Am.} Metallic-line or Am stars have different spectral classifications according to the K-line and the metallic-line criteria, with the first being earlier and the second later, possibly as much as F, as in the example given in the previous section. The origin of Am stars is the chemical separation in their interiors caused by their slow rotation.
 \item \textbf{Ap.} The Ap stars are the extension of the He strong, He weak, and Bp stars to the B-star regime, that is, objects in which only a few selected heavy elements are enhanced in the stellar atmosphere (as opposed to most heavy elements for Am stars). In terms of notation, the enhanced elements are listed behind the p suffix. As with their B-type counterparts, the chemical peculiarities are associated with magnetic fields \citep{Hummetal20}. Many Ap stars are spectroscopic variables. 
 \item \textbf{$\lambda$ Bootis stars.} They are metal-poor (population-II) A-type stars. The k-h-m notation described above is used followed by $\lambda$~Boo.
 \item \textbf{Herbig Ae/Be stars.} These objects are PMS stars that are the hotter equivalent to T~Tauri stars, discussed below. They are usually associated with interstellar clouds detected as a thermal IR excess, dark cloud, and/or reflection nebula, all of them produced by cool and/or hot dust. They are photometrically variable and their light is polarized. 
 \item \textbf{A-type shell stars.} A-type shell stars have most spectral features of late-B to early-F stars of luminosity classes V to III but their Fe\,{\sc ii} and Ti\,{\sc ii} lines are enhanced, as it would happen in A supergiants.
\end{itemize}

\subsection{Other hot stars}

In this section we discuss early-type stars in advanced stages of stellar evolution and/or those whose spectra are radically different from the previous ``normal'' stars. Two differences between the stars here and the previous ones are a more diverse range of masses, with some objects having initial values lower than those of a mid-F star in the MS, and a more disorganized distribution in the HRD, as some of these stars occupy locations unexplored until now or can be found in the same locations as ``normal'' stars but with quite different initial masses.

\subsubsection{Wolf-Rayet stars, Luminous Blue Variables, and iron stars}

Massive stars, especially the most massive ones, can experience phases where their winds are so strong and/or their variability is so high as to defy the traditional assignment of fixed spectral classes through the analysis of absorption lines. There is a whole zoo of such stars \citep{WalbFitz00}, of which we have already seen some examples. In this subsubsection, we analyze the two most extreme cases, Wolf-Rayet stars and Luminous Blue Variables (LBVs).

As already mentioned, WR stars were one of the first types of peculiar stars to be identified \citep{WolfRaye67}. The spectra of WR stars are dominated by broad emission lines of H, He, C, N, and/or O, with few or no absorption lines (intrinsic or caused by companions). Depending on the dominant CNO species in the emission spectrum, they are divided into two classes with significant membership, WN and WC, and one with a few known examples, WO. Sequences for each of them based on the ratios of lines from different ionization states can be used to provide spectral subtypes, see \citet{Smitetal96} for WN stars and \citet{Crowetal98} for WC and WO stars, and Table~\ref{qualifiers} for qualifiers.

The defining characteristic for WR stars is their strong wind responsible for the emission lines. This can happen in four ways:

\begin{itemize}
 \item \textbf{Initial stage of very massive stars.} The most massive stars of all ($\sim 100$~\Msol\ or more) have very strong winds from the start of their lives and are classified as WNh stars, with the h indicating a strong hydrogen component in the Balmer lines (instead of pure He\,{\sc ii} lines, see Table~\ref{qualifiers}). Some examples are the central stars at the core of NGC~3603 \citep{Drisetal95} and R136 \citep{Crowetal10,Bestetal20}. They are located at the hot extreme end of the sequence for normal O stars in the HRD (Fig.~\ref{HRD}).
 \item \textbf{Final stage of massive stars.} The only stars of O type in the ZAMS that become red supergiants are those of the latest subtypes\footnote{The precise subtype may depend on metallicity.}. The rest lose enough mass after they leave the MS to change the direction of their evolution in the HRD (\citealt{HumpDavi79}, see LBVs below) and become WR stars before their core collapses. This is called the Conti's scenario for WR production \citep{Maed96b} and is likely to be the primary one, at least at solar metallicities. These WR stars are hotter than O stars but are less luminous than the WRs that represent the initial stage of very massive stars (Fig.~\ref{HRD}).
 \item \textbf{End product of binary evolution.} As an alternative to the Conti's scenario, a massive star may also lose enough mass through mass transfer to a binary companion to become a WR star \citep{DeGr96}. It is still unclear whether the majority of WR stars are formed through this channel or through the previous one i.e. whether binary mass transfer or stellar winds are the dominant effect. The results of both channels are called classical Wolf-Rayet stars.
 \item \textbf{Planetary Nebula Nuclei (PNNi).} Some low-mass stars can have stellar winds at the end of their lives when at the planetary nebula stage (see below) strong enough to have a WR-like spectrum. They are designated as [WR] stars \citep{GornStas95}.
\end{itemize}

Luminous Blue Variables (also known as S Dor stars and Hubble-Sandage Variables) are evolved very luminous blue stars (Fig.~\ref{HRD}) that experience large photometric and spectroscopic variations \citep{Cont84}. The instability is caused by their proximity to the Eddington limit. Their spectroscopic variations are large enough to make them move horizontally across the top left of the HRD from a quiescent state with a hot \Teff\ to an active phase with a cool \Teff\ in time scales of years, with very different spectral types in each (see \citealt{Walbetal17} for examples). They can also experience eruptions, the most famous of which took place in the mid-19th century,  causing $\eta$~Car to become the second brightest star in the sky \citep{SmitFrew11,Smit11a}. During some of its phases, the spectrum of an LBV is dominated by Fe\,{\sc ii} emission lines, in which case it is called an iron star \citep{WalbFitz00}. 

\subsubsection{Horizontal Branch stars and RR Lyrae}

At the stage where population-I low-mass stars burn helium in their cores they form a structure in the HRD called the Red Clump (RC) and they are classified as late-G or early-K giants. Their population II counterparts, however, form a structure at similar luminosities called the Horizontal Branch (HB) and most of its members are classified as B, A, and F stars, with its left end below the MS (see below for hot subdwarfs) and its right one above it. 

The (low-metallicity) HB is divided into red and blue parts. In the gap in between we find the RR Lyrae, which are A and F giants with variable spectral types (Fig.~\ref{HRD}). This is a class of variable stars presenting cyclical changes in their brightness. The length of these cycles is proportional to the intrinsic brightness of the star, as in classical cepheids (discussed below), and RR~Lyrae are their low-luminosity equivalents on the same instability strip. Their location is marked by a photometric variability maximum in the top right panel of (Fig.~\ref{HRD}) even with the blurring introduced by Galactic extinction. Their spectral types are variable through their cycles, but most of the time they appear as F-type giants with weak metallic lines (see below for metallicity effects on F and later-type stars). 

\subsubsection{Hot subdwarfs, stripped stars, white dwarfs, and cataclysmic variables}

The term subdwarf originally described only (late-type) population-II MS stars which, due to their low metallicity, were located below\footnote{Actually, to the left, as those stars are mostly bluer, not fainter, than their high-metallicity counterparts (Fig.~\ref{HRD})}. the luminosity class V sequence in the HRD. However, a population of hot subluminous stars (hot or OBA sudwarfs) also exists (Fig.~\ref{HRD}). Most of them are extreme HB stars (core He burning, see above) that are the result of binary evolution but some may have been formed by other mechanisms such as two white dwarfs merging \citep{Hebe16}. \citet{Driletal13} have developed a classification system to integrate hot subdwarfs into the MK system.

Hot subdwarfs are the low-mass equivalent of the massive Wolf-Rayet stars formed through the binary channel. But what about the masses in between the two ranges? Those objects are called stripped stars (Fig.~\ref{HRD}); they have been only recently studied in detail, and they have spectral characteristics that are intermediate between those of sdOBA and WR stars \citep{Gotbetal18,Gotbetal23}.

\begin{table}[t]
\caption{Spectral classification symbols for white dwarfs from \citet{GrayCorb09}.}
\label{WD}
\centering
\begin{tabular}{ll}
\\
\hline
Symbol & Definition  \\
\hline
\multicolumn{2}{l}{Main symbols} \\
\hline
DA     & Only Balmer lines, no He\,{\sc i} or metallic lines present \\
DB     & Only He\,{\sc i} lines, no Balmer or metallic lines present \\
DC     & Continuous spectrum, no lines deeper than 5\% in the whole electromagnetic spectrum \\
DO     & He\,{\sc ii} strong, He\,{\sc i} or Balmer lines present \\
DZ     & Metallic lines only, no Balmer or He\,{\sc i} lines present \\
DQ     & Carbon (atomic or molecular) features present somewhere in the electromagnetic spectrum \\
\hline
\multicolumn{2}{l}{Additional symbols} \\
\hline
P      & Magnetic WDs with detectable polarization \\
H      & Magnetic WDs without detectable polarization \\
X      & Peculiar or unclassifiable spectrum \\
E      & Emission lines present \\
? or : & Uncertain spectral classification \\
V      & Variable \\
d      & Circumstellar dust present \\
C\,{\sc i}, C\,{\sc ii}, O\,{\sc i}, O\,{\sc ii} & Added within parentheses for hot DQ types to indicate presence of those species \\ 
\hline
\end{tabular}
\end{table}

White Dwarfs (WDs) are the end point of stellar evolution for most\footnote{We say ``most'' instead of ``all' because, once binaries are included, other scenarios are possible.} low- and intermediate-mass stars. Their spectral classification follows a different scheme from the MK system, with the symbols described in Table~\ref{WD} and indicating the different types of lines detected in the spectrum. The symbols are followed by a numerical index from .1 to 13 i.e. DA.1\ldots DA.9, DA1, DA1.5, DA2\ldots DA13 that represents 50.4~kK/\Teff, so the sequence above runs from a \Teff\ of 504~kK to one of 3.9~kK. They are located at the bottom of the HRD (Fig.~\ref{HRD}).

Cataclysmic variables are binary systems composed of a white-dwarf primary and a mass-donor secondary, with an accretion disk usually involved in the process. They come in different varieties, are usually located in the region between the class~V and WD sequences, and their membership includes some of the most variable stars known (top right panel of Fig.~\ref{HRD}).

\subsubsection{Post-Asymptotic Giant Branch stars, Planetary Nebula Nuclei, and PG 1159 stars}

At the end of their lives, low-mass stars cross the HRD from the upper right (Asymptotic Giant Branch or AGB phase) to the lower left (WD phase). They do so by quickly moving first near-horizontally to the left until they reach extremely hot \Teff\ values, and then downwards. 

The first phase corresponds to post-AGB (pAGB) stars. These objects can be classified as peculiar supergiants of K to O type (depending on the evolutionary moment at which they are caught). They have peculiar abundances due to their strong mass loss rates, infrared excesses from the recently ejected dust, and a mismatch between their spectroscopic luminosity classification as supergiants and their location in the HRD somewhat below (as their masses are lower than those of the usual denizens of the region, their gravity is lower than that expected at their luminosity). 

Once the extremely hot ($\Teff\sim$200~kK, {Fig.~\ref{HRD}) degenerate core is exposed, the ejected layers become a planetary nebula and the core, a planetary nebula nucleus (PNN), some of which have WR-like features (see [WR] stars above). This phase lasts only a few tens of ka, with the star moving downwards in the HRD on its way to become a WD. In between the two phases, the object is known as a PG~1159 star (after the prototype PG~1159$-$035), which resembles a hot DQ spectrum but is even hotter (Fig.~\ref{HRD}). 

\section{Spectral classification of cool stars}\label{sec5}

As previously mentioned, we define cool (or late-type) stars as those with low values of \Teff\ and, more specifically, as those that would correspond to the original II, III, and IV Secchi's classes. We have divided this section in five parts:

\begin{itemize}
 \item \textbf{F stars.} They are the transition type between hot and cool stars.
 \item \textbf{G+K stars.} We group them together because of their similarities.
 \item \textbf{Cool luminous stars.} They are the FGKM giants and supergiants, which we group together due to their significantly different nature or evolutionary state compared to cool dwarfs.
 \item \textbf{M dwarfs.} They constitute the most numerous of all the stellar categories in this paper and represent the lowest end of the stellar mass function.
 \item \textbf{Ultracool dwarfs.} They are the most recent addition to the spectral classes and are objects of substellar mass (but see below).
\end{itemize}

\subsection{F stars}
In many senses, F-type stars represent a transition. Early-F stars still have a radiative envelope, with only superficial convection zones, but later types have fully-developed convective envelopes. Beyond this transition, magnetic braking is effective, and fast rotators are rare, except among very young stars, an effect that can be directly observed when comparing spectra \citep{Kraf67,BeyeWhit24}. Chemical peculiarities are not observed in main-sequence stars with convective envelopes, except for some rare cases of contamination by a companion. It must be noted, however, that Am stars are observed with  spectral types as late as F2.

In the earliest F subtypes, Balmer lines have an important negative luminosity effect, as seen in A-type stars. The Balmer lines, however, weaken significantly with increasing subtype, and by late F, they have become weak and not responsive to luminosity. Conversely, the overall metallic spectrum strengthens both with decreasing temperature and increasing luminosity. These combined effects mean that F-type stars may present very different appearances. An extreme case is shown in Fig.~\ref{fig:fstars}.  The spectrum of a fast-rotating early-F star is still dominated by the Balmer lines, and its overall appearance is that of an early-type star. Conversely, a late-F star looks decidedly like a cool star.

\subsubsection{Classification criteria for F stars}

$\,\!$\indent \textbf{Spectral type.} At solar metallicity, the main criterion to classify F subtypes is the increasing strength of metallic lines when compared to Balmer lines \citep{GrayGarr89a}. The decreasing strength of the Balmer lines on its own gives a very good approximation to the subtype, thus providing a useful tool to classify low metallicity stars (more on this below). A number of metallic lines that are not sensitive to luminosity, such as \CaI{4226} or \FeI{4046,4383}, can be used as comparators to the Balmer lines. An important feature for classification is Fraunhofer's G band. First seen before F5, this is a broad feature due to the diatomic molecule CH, which becomes prominent in late-F, G, and early-K stars (Fig.~\ref{OBAFGKM}). Centred around $\lambda$4307, it can extend over about 20~\AA\ and become strengthened by several Fe\,{\sc i} and Ca\,{\sc i} lines.

\textbf{Luminosity class.} The luminosity class of stars up to F5 is primarily determined by the strength of Fe\,{\sc ii} and Ti\,{\sc ii} lines compared to metallic lines insensitive to luminosity, such as those mentioned in the previous paragraph. The lines of ionized Fe and Ti appear mostly in blends, giving rise to strong features at $\lambda\lambda$4172--8, or $\lambda\lambda$4395--4400. These lines become weaker with decreasing temperature, and they are thus not very useful for classification of late-F stars, except for the supergiants. In late subtypes, the Sr\,{\sc ii} lines at $\lambda$4077 and $\lambda$4216 compared to Fe\,{\sc i} lines provide the main criterion.

\textbf{The issue of metallicity.} Hot stars, with a few exceptions such as HB, pAGB, sdOBA stars, or WDs (see above), are relatively young objects, belonging to Population~I. When we reach mid-F types, however, we start to find Population~II stars. The oldest stars known to be still on the main sequence, such as those in globular clusters or the halo ``turn-of'', have temperatures corresponding to F types, but very weak metallic spectra, more typical of the A-type dwarf standards. Phillip Keenan tried to develop a classification system to take into account differing metallicities by adding a qualifier Fe followed by a numerical index (for example, Fe-3 for a star with very weak metallic lines). It is, however, very difficult to find relevant matches between stars with really low metallicities and MK standards, all of which have approximately solar metallicity.  

\begin{figure}[t]
\vspace{-4.0mm}
\centerline{\includegraphics[width=1.00\linewidth]{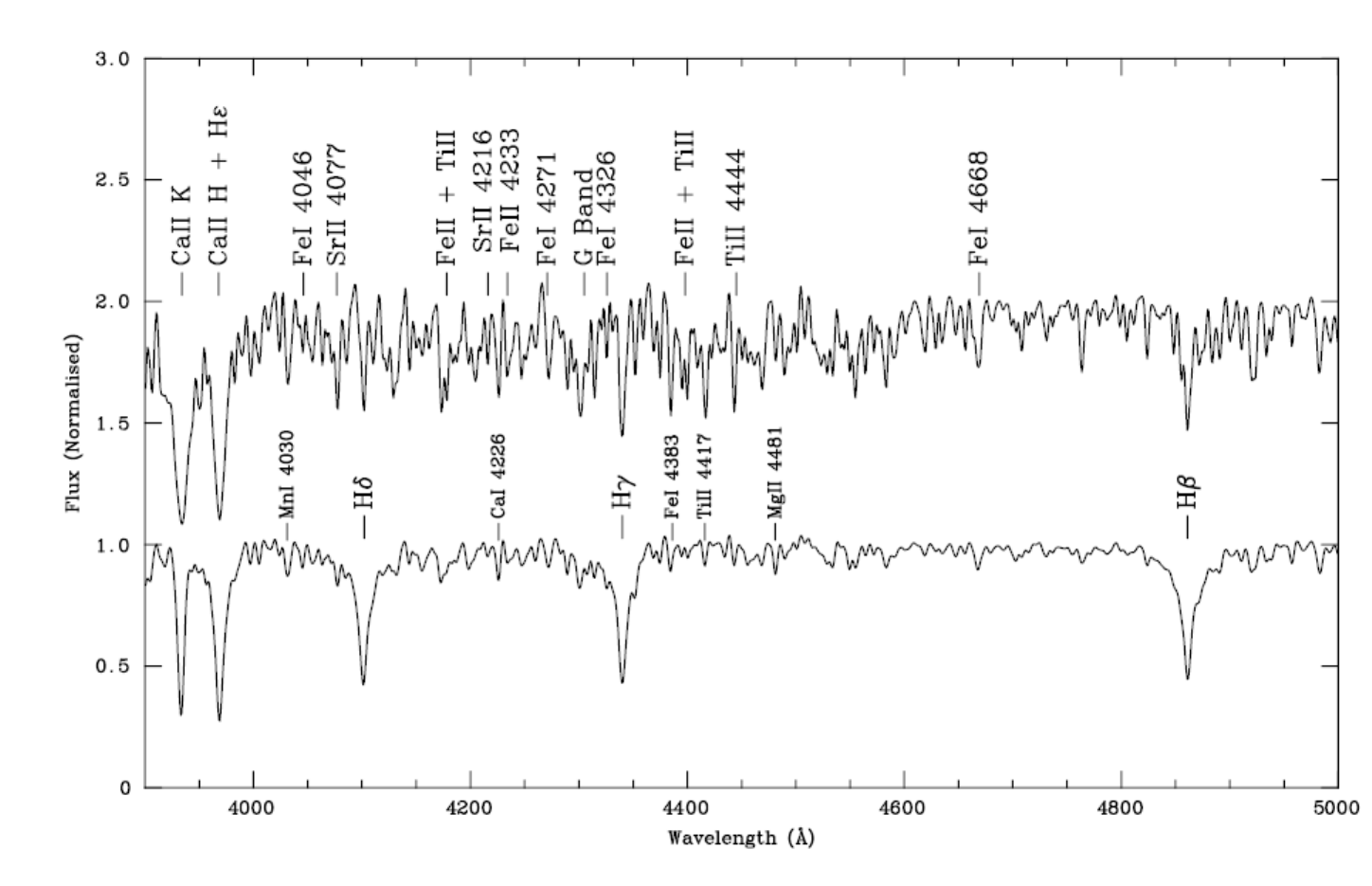}}
\caption{These two extreme spectra illustrate the very different morphologies that can correspond to F-type stars. The bottom spectrum is that of the fast-rotating (\vsini~$\approx$~160~km~s$^{-1}$; \citealt{Royeetal07}) F0~V star 89~Tau (HD~\num[detect-all]{29375}), which looks decidedly like a hot star. The top spectrum corresponds to the F8~Ia supergiant $\delta$~CMa (HD~\num[detect-all]{54605}), which would have gone into Secchi's class II. The spectra have been adapted from the {\sc miles} spectral library \citep{Falcetal11}. Lines useful for classification are indicated. }
\label{fig:fstars}
\end{figure}

Naturally, this issue extends to later spectral types.
In recent years, stars with metallicities as low as [Fe/H] $< -5$ have been found. These objects present ratios between chemical elements that are extremely different from those of MK standards (for example, they are very rich in C in comparison to other elements; e.g. \citealt{Aguaetal18}). Such objects cannot be accurately classified within the system without an extension to further dimensions.

\subsubsection{Peculiar F stars}

A peculiar type of F dwarfs (see below for peculiar luminous F stars) are barium dwarfs. These are late-F main-sequence stars that display enhanced lines of $s$-process elements. This is mainly evident in the very strong \SrII{4077} and \SrII{4216} lines. Since these two lines are used in primary luminosity classification criteria, the main-sequence nature of these star must be established from other ratios. They are called barium dwarfs in analogy with the barium giants (discussed below). The heavy-element contamination is believed to be due to a binary companion, which produced it during its AGB phase and is now a white dwarf.

\subsection{G+K stars}

There is a smooth transition between late-F and early-G stars (compare $\alpha$~Cen~A to $\alpha$~CMi in Fig.~\ref{OBAFGKM}), and FGK stars of low luminosity are sometimes grouped together as cool dwarfs.
The most salient characteristic of G-type stars in the classification region is Fraunhofer's G band, which extends around $\lambda$4307 (Fig.~\ref{OBAFGKM}). At longer wavelengths, the \MgIt{5167,72,84} triplet (Fraunhofer's b band) increases in strength with decreasing temperature, but also responds to luminosity. The Balmer lines have become weak, and lost all sensitivity to gravity. They gradually weaken with decreasing temperature, and their ratio to the G band or metallic lines can be used as a quick estimator of spectral type for G and K types.

The G band increases in strength until about K2 (it is very strong in the spectrum of $\alpha$~Cen B, K1~V, in Fig.~\ref{OBAFGKM}), and then fades in later K stars, where two MgH bands become noticeable.  The \CaI{4226} line is very strong in dwarfs, especially in mid-K types, but it has a very important negative luminosity effect (which can be easily noticed in Fig.~\ref{fig:M1s}). The spectra of M-type stars are dominated by TiO bands, which grow in strength as temperature decreases, until most of the blue continuum has been eroded away (see the spectrum of $\alpha$~Cen C in Fig.~\ref{OBAFGKM}). For this reason, it is preferable to use spectra extending into the green (or even the red) to classify M-type stars, as done in Figures~\ref{fig:lategiants}~and~\ref{fig:M1s}.

\subsubsection{Classification criteria for G and K stars}

The same characteristics that we saw in late-F stars continue into later types. Balmer lines become progressively weaker, with H$\gamma$ being of about the same depth as the G band at G0, but becoming inconspicuous when compared to neighbouring metallic lines by late G. Similarly, H$\beta$ is a salient feature in early G stars, but is lost in the metallic forest by early K. \CaI{4226} increases in strength with decreasing temperature and becomes very prominent in mid-K dwarfs. The same ratios of neutral \FeI{} lines to Balmer lines used for F stars can be applied to classify G and K stars at solar metallicity. In stars with lower metal content, ratios between Cr\,{\sc i} and \FeI{} lines are used, notably \CrI{4254}/\FeI{4250} and \CrI{4290}/\FeI{4326}. These ratios can also be used at solar metallicity.

In the green region, the \MgIt{5167,72,84} triplet is a very salient feature that increases in strength with decreasing temperature, although it is also sensitive to luminosity. This feature increases in strength with decreasing temperature in G dwarfs. 

The original system \citep[e.g.][]{JohnMorg53} did not contain subtypes K6, K8 or K9. In later years, Keenan introduced fractional subtypes for K stars (i.e.\ K0.5, K1.5, and so on) and occasionally used the K6 subtype (e.g.\ Gl~529 is K6~Va in \citealt{KeenMcNe89}). There is also an example of K8 (HD~\num[detect-all]{142574} is K8~IIIb) and Gl~638 is even given as K7.5~Ve in \citet{KeenMcNe89}. Nevertheless, he never considered these fine classifications to represent full subtypes. The only true subtypes admitted in \citet{Keen84} are G0, G2, G5, G8, K0, K1, K2, K3, K4, K5 and M0 (followed by all the integer M subtypes). The reasoning behind this choice is that stars must be evenly distributed into subtypes, while possible subdivisions are utilized only for interpolation whenever the quality of the data renders their use reasonable. In any event, as can be seen in Fig.~\ref{fig:lategiants}, the differences between mid-K and early-M giants are marginal, and the use of these late-K types is therefore uncommon.

\subsubsection{Luminosity criteria for G and K stars}
In stars of higher luminosity, molecular bands strongly affect the shape of the continuum. The violet system of CN bands is hardly seen in dwarfs, but generates deep depressions in the spectra of giants (and more luminous stars) between G5 and early K. The $\lambda$4215 band creates a shallow depression between $\sim\lambda$4120 and $\lambda$4215 in the spectra of giants, which is deeper in supergiants. These features are also very strongly dependent on metallicity, and Phillip Keenan developed a qualifier CN to indicate their strength, in an analogous manner to the Fe qualifier.

\begin{figure}[t]
\vspace{-4.0mm}
\centerline{\includegraphics[width=1.00\linewidth]{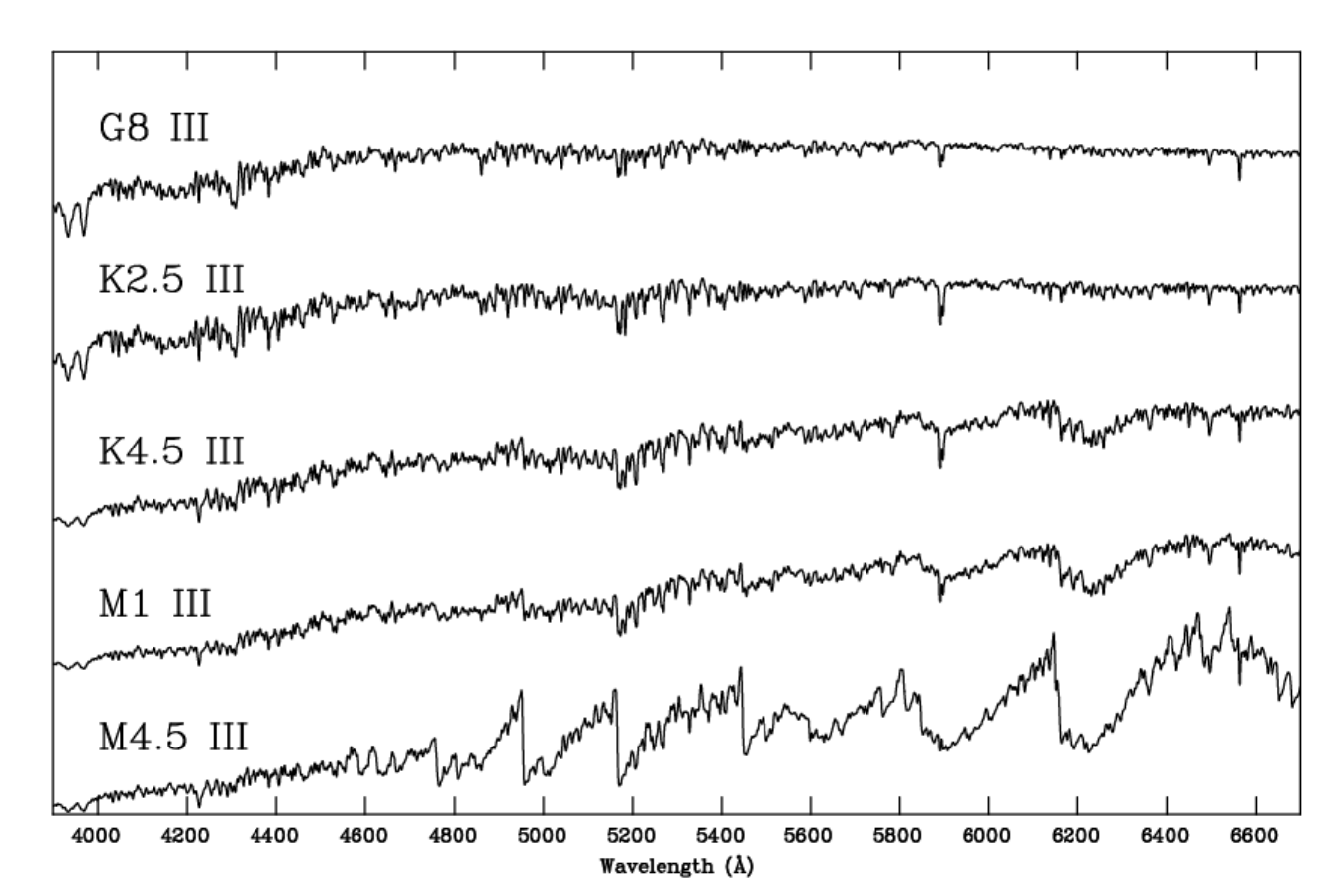}}
\caption{Temperature sequence for late-type giants. The stars displayed are, from top to bottom, HD~\num[detect-all]{10761}, HD~\num[detect-all]{216640}, HD~\num[detect-all]{99167}, HD~\num[detect-all]{102212}, and HD~\num[detect-all]{123657}. The spectra have been adapted from the Indo-US spectral library \citep{Valdetal04}, and spectral types are from \citet{KeenMcNe89}. See Fig.~\ref{fig:M1s} for the most relevant spectral features. Note the high similarity between the K4.5~III and M1~III stars, which explains the lack of definition for late-K types.}
\label{fig:lategiants}
\end{figure}

At lower temperatures, TiO bands start to be visible. They are not noticeable in dwarfs before type M, but they can be clearly seen in mid-K giants (note, for example, the prominent bandhead at $\lambda$6158 and incipient bandheads at  $\lambda$4954, 5446 in the spectrum of the K4.5~III star HD~\num[detect-all]{99167} in Fig.~\ref{fig:lategiants}). Conversely, the MgH band around $\lambda$5200 produces a broad depression between $\lambda$5000 and $\lambda$5240 in mid-K dwarfs, but is hardly seen in high luminosity stars. A second MgH band, around $\lambda$4780, gives rise to a much narrower feature, which also shows a marked negative luminosity effect. The G band also shows a negative luminosity effect, although it is still moderately strong in supergiants. 

For accurate luminosity classification, the ratios of \SrII{4077} and \SrII{4216} to neighbouring Fe\,{\sc i} lines can be used. The best criterion in the blue part of the spectrum is the ratio of \YII{4367} to \FeI{4383}.

\subsubsection{Peculiar G+K stars}

A peculiar type of low-luminosity G+K stars (see below for peculiar luminous G+K stars)  with high astrophysical relevance are T~Tauri stars.
 These young stars are divided into classical T Tauri stars (CTTSs) and weak-lined T Tauri stars (WTTSs). 
The former are very young stars that retain a remnant of their natal envelope, resulting in the presence of emission lines, frequently strong, most notably H$\alpha$ and other Balmer lines, but also others from Ca\,{\sc ii} and Fe\,{\sc ii}.  
These emission lines originate from infalling channels of plasma and neutral gas from circumstellar protoplanetary disks, or from the disks themselves. CTTSs sometimes present P~Cygni profiles or forbidden emission lines.
Their underlying spectrum, which may be veiled, normally corresponds to luminosity classes III to V, with the latter being more frequent (see their typical location in the HRD in Fig.~\ref{HRD}). 
The nature of T~Tauri stars can be confirmed through other observational characteristics, such as photometric variability, infrared excess, or X-ray emission. Their spectra experience ``veiling'', a general loss of contrast, accompanied by blurring of atomic lines \citep{Fangetal20}, as Mira variables also do (see below).
CTTSs stars have G to M types, though some F-type T~Tauri stars are known, but not so well studied. 
It has also been suggested that the T Tauri phase extends to brown dwarfs, as there are very young M-type objects with masses below the hydrogen-burning limit in star-forming regions exhibiting very similar characteristics to CTTSs. 
While CTTSs are the low-mass counterpart of Herbig Ae/Be stars, WTTSs represent the low-activity counterparts of CTTSs, having lost most of their natal envelopes. The peculiar characteristics in their spectra are thus weak.

\subsection{Cool luminous stars}

 Cool stars of high luminosity, i.e. FGKM giants and supergiants, despite the overall resemblance of their spectra, include a wide variety of stellar types. While a star classified as O9~III and a star classified as O9~Ia share most characteristics and are essentially the same sort of object, a K0~III star is likely a He-burning low-mass star in the RC, but a K0~Ia star is likely a massive star in a rare, transitional phase. Nevertheless, differences between their spectra are moderately subtle, when compared -- for example -- with a main sequence star of the same spectral type. It is important to remark once again that spectral classification is morphological; the fact that a star is classified as a supergiant does not necessarily imply that it is a massive star, as some have been found to be of intermediate mass. This is in contrast to OBA stars, where normal (non-pAGB) supergiants are always massive stars. 

There are very few high-luminosity supergiants of G and K type at solar metallicity, to the point that the original list of MK standards by \citet{JohnMorg53} only contains the K3~Iab standard $o^{1}$~CMa (later given as K2.5~Iab in \cite{MorgKeen73} and K2~Iab in \citealt{KeenMcNe89}) and no Ia standards between G0 and M0.
The few objects observed tend to be very luminous and highly variable, such as the hypergiant $\rho$~Cas, which has been observed at spectral types between F and M. Contrarily, there are many supergiants of M type. These correspond mainly to the He-core burning phase of massive stars between $\sim10$ and $25$~\Msol, although some objects of luminosity class Ib can correspond to lower-mass giant branch or AGB stars. An example at hand is the MK standard $\alpha$~Her (Rasalgheti), M5~Ib--II, which is believed to be an AGB star of intermediate mass (2.3--3~\Msol). True supergiants tend to be weakly variable in spectral type but more significantly so from a photometric point of view. Most are of early-M type, with mid-M types generally observed amongst those believed to be more massive. It must be noted, however, that the properties of He-core burning massive stars are very strongly dependent on metallicity. In the SMC, there are few M-type supergiants; luminous red supergiants have types G and K and tend to show moderate spectral type variability \citep{Dordetal16b}.

Given the variety of masses and evolutionary phases found among cool luminous stars, there are significant differences in their photometric variability. For example, studies of open clusters show that low-mass RC stars typically have spectral types in the G6~III--K2~III range \citep[e.g.][]{MorgHilt65,MermMayo89} and recent \textit{Gaia} analyses have shown those to be quite stable from the photometric point of view \citep{Maizetal23}, see top right panel of Fig.~\ref{HRD}. On the other hand, stars with masses between 4~\Msol\ and 8~\Msol\ in the He-core burning phase appear as late-G or K-type stars of luminosity class II or Ib \citep[e.g.][]{Alonetal19,Alonetal20}. He-core burning stars with masses higher than $\sim5\:M_{\odot}$ can loop blueward, and then they will appear as yellow supergiants, with spectral type F and luminosity class Ib or II. Some of these objects are periodically variable in spectral type and luminosity, the classical cepheids. Many giants and supergiants of M type are photometrically variable \citep{Maizetal23}. Variations can be irregular, semi-regular and (quasi-)periodic, many of which are classified as long period variables (LPVs). When a period can be defined, stars can be sorted into families according to their period/luminosity relation \citep[e.g.][]{Lebzetal19}. Most LPVs are AGB stars, the He shell burning for stars of intermediate and low-mass, with the two types forming two branches in the \textit{Gaia} CAMD (Fig.~\ref{HRD}) and the latter reaching a significantly redder location in \BPRP\ than the former. M-type LPVs constitute a major subclass of Mira variables (or Miras, for short), which show very large photometric amplitudes and large changes in their spectra, attributed to thermal pulsations.  

A modern guide to classification of cool luminous stars, which gathers information spread over many classical references, including the use of features in the green region of the spectrum, can be found in \citet{Dordetal18b}.

\subsubsection{M giants and supergiants}

\begin{figure}[t]
\vspace{-4.0mm}
\centerline{\includegraphics[width=1.00\linewidth]{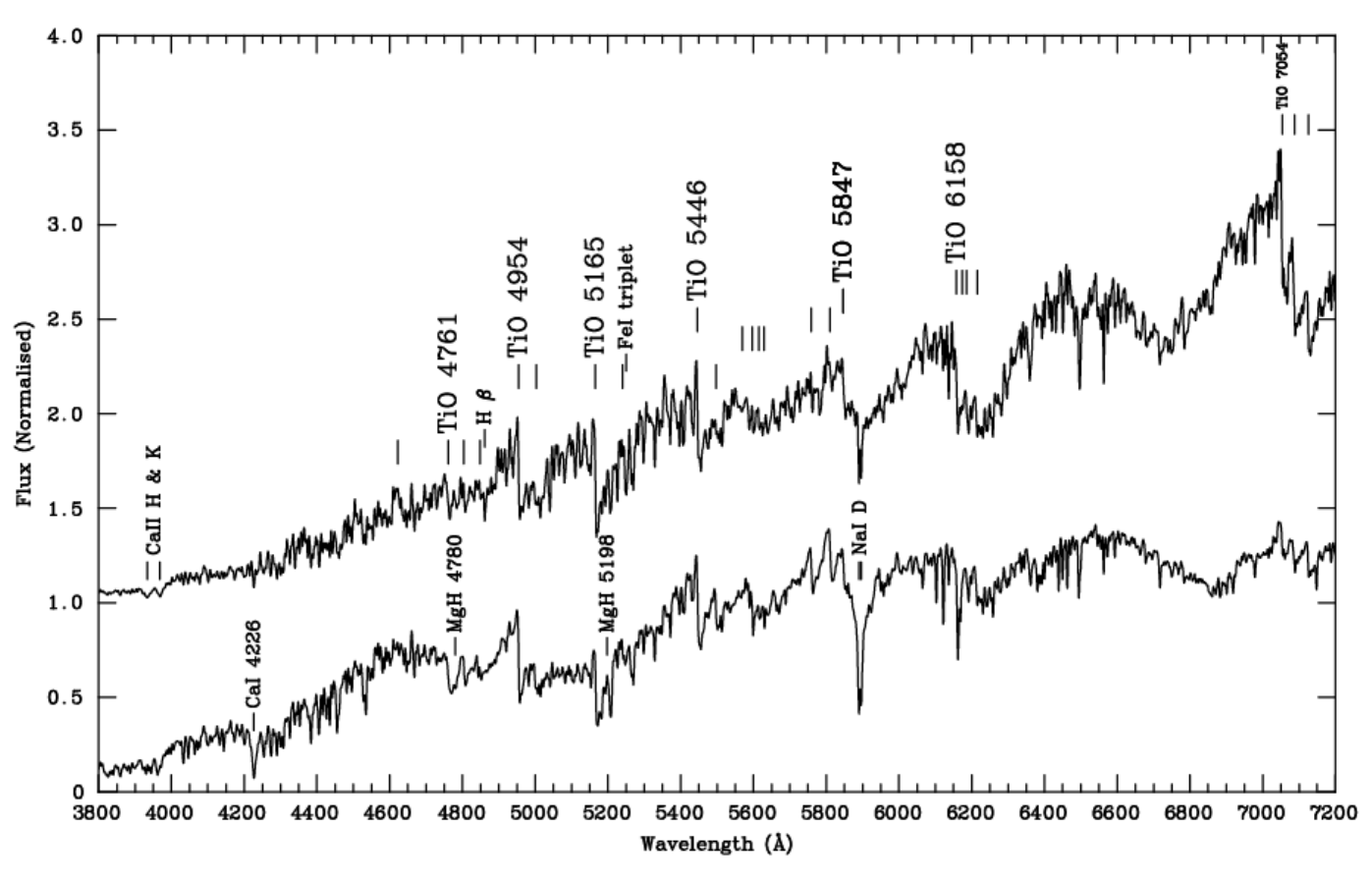}}
\caption{Comparison of a red dwarf (HD~\num[detect-all]{147379}; M1~V, bottom) to a red supergiant of very similar temperature (6~Gem; M1--M2~Ia-Iab). The spectra have been adapted from the Indo-US spectral library \citep{Valdetal04}, and spectral types are from \citet{KeenMcNe89}.
The spectral type of the planet-host star HD~\num[detect-all]{147379} has been revised with newer redder data to M0.0~V \citep{Alonetal15,Reinetal18}.}
\label{fig:M1s}
\end{figure}

The vast majority of type M stars are dwarfs.
For example, about 70~\% of the almost 500 stars in the immediate vicinity of the Sun at $d <$ 10~pc are M dwarfs \citep{Reyletal21}, 
while there are only a dozen giants (luminosity class III) in the 8$\times$ larger volume of 20~pc \citep{Kirketal24}.  
Nevertheless, M-type giants have radii thousands of times larger than those of M dwarfs, and are thus many orders of magnitude brighter.
As a result, while the brightest M dwarf, Lacaille~8760 (AX~Mic), is at $V \sim$ 6.7~mag slightly fainter than the naked-eye limit, there are six M-type giants and supergiants with apparent magnitudes $V <$ 2.5~mag among the 100 brightest stars in both hemispheres: Betelgeuse, Antares, Gacrux, Mirach, $\beta$~Gru, and $\beta$~Peg.
For this obvious reason, the first spectral classification studies of M stars focused on M giants and supergiants, starting with Secchi and his ``class III'' (orange to red stars with complex band spectra) and in subsequent pioneering work (e.g. \citealt{Wild36,Keen42}). 

The spectral type of M-type giants and supergiants on photographic plates was determined by the progressive appearance of different TiO bandheads in the blue part of the spectrum \citep{Keenetal74}. These bands are much stronger than in dwarfs of the same type. On modern spectrograms, ratios of TiO bands are preferable for a number of reasons, among which the possibility of ``veiling'' in Mira variables \citep{CrowGarr88}. Strong bands reach saturation and stop growing, while weaker bands become visible at later types. Some of the line ratios that are used for K-type stars can still be used in early-M giants, but the loss of the continuum makes them disappear at later types. A remarkable exception is the \CaI{4226} line. This line, which is weak in luminous early-M stars because of a marked luminosity effect (see Fig.~\ref{fig:M1s}) becomes very strong in giants of types M5 and later (compare the M4.5~III star HD~\num[detect-all]{123657} with the earlier giants in Fig.~\ref{fig:lategiants}). 

Nevertheless, the spectral classification of luminous late-M stars is complex. Most of them are variable in spectral type, and thus not suitable as standards. Moreover, when we reach mid-M spectral types, the strength of the TiO band systems has increased so much that there is no continuum left anywhere in the classical classification region. As a consequence, the subsequent increase of the band strengths with decreasing temperature does not result in large changes in the spectrum. It becomes mandatory to resort to redder regions of the spectrum \citep{Solf78}. Inclusion of the visual region allows the use of VO bands to help defining spectral types. Starting at M5, the VO bandheads can be seen as relatively weak, narrow features appearing within the dominant TiO bands.

Although the latest reference spectra given by \citet{KeenMcNe89} are M8, it is possible to extrapolate the even later subtypes M9 and M10, which are only observed in LPVs, such as Miras. The extreme object VX~Sgr, which has luminosity class Ia, and could be either a very peculiar supergiant or an extremely luminous AGB star \citep{Tabeetal21}, was used by \citet{Solf78} to define M10, but many Miras reach M9 during their cycles. The use of M10 indicates that spectral type L is not considered for luminous stars (type M11 has been employed in some references).

Similarly, exact luminosity classification is tricky. Some criteria based on line ratios that are used for K-type stars can also be used in the M type, until they lose applicability because the lines involved disappear within growing TiO band strength \citep{Dordetal18b}. The criterion reaching later types is the ratio of \YII{4367} to \FeI{4383}, which can be used effectively up till $\sim$M5. For later types, the general strength of the metallic line spectrum is a useful guide, but must be employed with caution, because of ``veiling'' in Mira variables. 

Most M supergiants are mildly variable, with \citet{KeenMcNe89} giving ranges of spectral types for a significant number of them. For example, TV~Gem (HD~\num[detect-all]{42475}) is listed as M0--M1.5~Iab. The supergiant with the latest spectral type listed as a reference is EV~Car (HD~\num[detect-all]{89845}), at M4.5~Ia. There are a few supergiants of later type (e.g.\ MY~Cep in NGC~7419; \citealt{MarcNegu13}), but their exact luminosity class cannot be given. Late-M giants are very luminous, as bright as the fainter supergiants, but their luminosity class is difficult to determine. See \citet{Neguetal12} for an attempt at finding criteria in the nearest infrared, where some strong metallic lines are visible at classification resolution, and references to classical sources.

\subsubsection{Peculiar cool luminous stars}

In a sense, all cool luminous stars are peculiar, with the possible exception of normal clump stars, as they are subject to photometric and, in many cases, spectroscopic variability. Here we list some of the most relevant cases. 

\begin{itemize}
\item \textbf{Cepheids.} Classical cepheids are variable stars whose intrinsic magnitude undergoes cyclical variability. The recognition of a direct relation between the length of their period and their intrinsic brightness \citep{LeavPick12} represents the first step in the distance scale and the foundation of modern cosmology. Classical cepheids are He-core burning stars of between $\sim5$~\Msol\ and $\sim12$~\Msol\ that are passing through the instability strip (clearly seen in the top right panel of Fig.~\ref{HRD}) as part of their blue loop evolution \citep[e.g.][]{Andeetal14b}. Along their cycle, they also display spectral variability. All cepheids have a base spectral type at maximum light not far from F8~Ib (some can be as early as F5 and the less luminous examples will have luminosity class II). Throughout their photometric cycle, they will move to later spectral types, with the most luminous (and, hence, massive) cepheids reaching into the K type \citep[e.g.][]{Claretal15b}.
\item \textbf{CN-strong stars.} Some giants have stronger CN (and C$_2$) bands than normal stars of the same spectral type. This anomaly is also marked with the CN qualifier, this time followed by a positive numerical index.
\item \textbf{Barium stars.} Some G and K giants display enhanced lines of $s$-process elements, specifically \BaII{4554} and the Sr\,{\sc ii} lines. The heavy-element contamination is caused by a binary companion, which produced it during its AGB phase and is now a white dwarf.
\item \textbf{Mira variables.} Mira-type variables are intrinsically very red LPV stars with some reaching photometric amplitudes larger than $\sim2.5$~mag in the optical bands (top right panel of Fig.~\ref{HRD}) and $\gtrsim1$~mag in the near infrared. Their photometric (quasi)periods are longer than $\sim150$~d. Physically, they are stars that have completed He core burning and are now on the second phase of the  AGB, characterized by alternating episodes of H-shell and He-shell burning. This irregular energy injection leads to very large pulsations of the whole envelope (the thermal pulses), which result in the observed variability.

There are three main types of Miras: O-rich (which appear as M giants), S stars (characterized by strong bands of ZrO) and C-rich (carbon stars, Secchi's group IV). Lower-mass stars (up to a limit of 3--4~\Msol, depending on metallicity) undergo evolution M\,$\rightarrow$\,S\,$\rightarrow$\,C (intermediate stages are known). This is understood as a sequence of progressive enrichment of the atmosphere with C generated during the He-shell burning flashes, due to the third dredge-up. More massive stars do not raise so much  C to the atmosphere and remain O-rich (see \citealt{Herw05} for an overview of the AGB). In consequence, the Miras with the longest periods ($\gtrsim1000$~d; and, hence, the highest luminosities) are expected to be very late-M giants (although some are S stars; e.g.\ \citealt{Smitetal95}). However, many of these high-luminosity Miras are enshrouded in thick dust shells, due to their very heavy mass loss, which reprocess their emitted light into infrared wavelengths \citep[e.g.][]{GarHetal07}.

Miras show a very high degree of spectral variability. Their spectral type changes through the photometric cycle, normally by several subtypes. When close to maximum light, they generally display emission lines, with very strong Balmer emission, and weaker lines of Fe\,{\sc i} (and sometimes Fe\,{\sc ii}) and other metals. In addition, they suffer from ``veiling'' \citep{CrowGarr88}. These changes are not consistent, in the sense that they do not always happen at a given phase of the photometric cycle. 

Carbon stars are generally classified in two (parallel) temperature sequences, which correspond to the old R and N spectral types of Cannon, although much more complex classifications have been devised \citep[and references]{Keen93}. The S stars have subtypes that intend to mirror those of M-type giants, together with a C/O index that runs from 1 to 10, where stars with index 7 to 10 are considered intermediate SC stars that merge into the C-N sequence.

\item \textbf{Symbiotic stars.} A symbiotic binary (or symbiotic star) is a binary system whose spectrum shows very different components. In general, symbiotic stars show the spectrum of a cool giant superimposed with strong nebular emission, similar to a planetary nebula. The hot component is, in most cases, a white dwarf, which can be seen in UV spectra. The high excitation lines generate in an extended envelope or circumstellar disk created by material lost from the red giant, as it is illuminated by energetic photons coming from the white dwarf. See \citet{Muna19} for a review of related phenomena.
\end{itemize}

\subsection{M dwarfs}

M dwarfs are among the coolest and smallest stars, preceded only by old early L-type stars (see below for further details). 
The study of M dwarfs has been revived because of their importance in the last decade for low-mass exoplanet detection and characterization (e.g., \citealt{Bonfetal13,Ribaetal23}). 
As introduced above, their spectra were first described in the 19th century by \citet{Secc66}, although he referred to M giants. 
Until the first parallax measurements of such faint objects, in the first two decades of the 20th century, the high proper motion of M dwarfs with respect to those of M giants and the subtle variations in their spectra, afterwards ascribed mainly to very different $\log{g}$, were the only observable differences between M dwarfs and giants.
With new parallaxes in hand, \citet{AdamJoy22b} and \citet{Luyt22} were the first to prepare M-dwarf catalogs.
Furthermore, \citet{AdamJoy22b} assigned spectral type M (actually, Ma and Mb) to stars previously classified as K8, which inaugurated the curious ``gap'' in spectral type between K7 and M0.0 stars.

\begin{figure}[t]
\centerline{\includegraphics[width=1.00\linewidth]{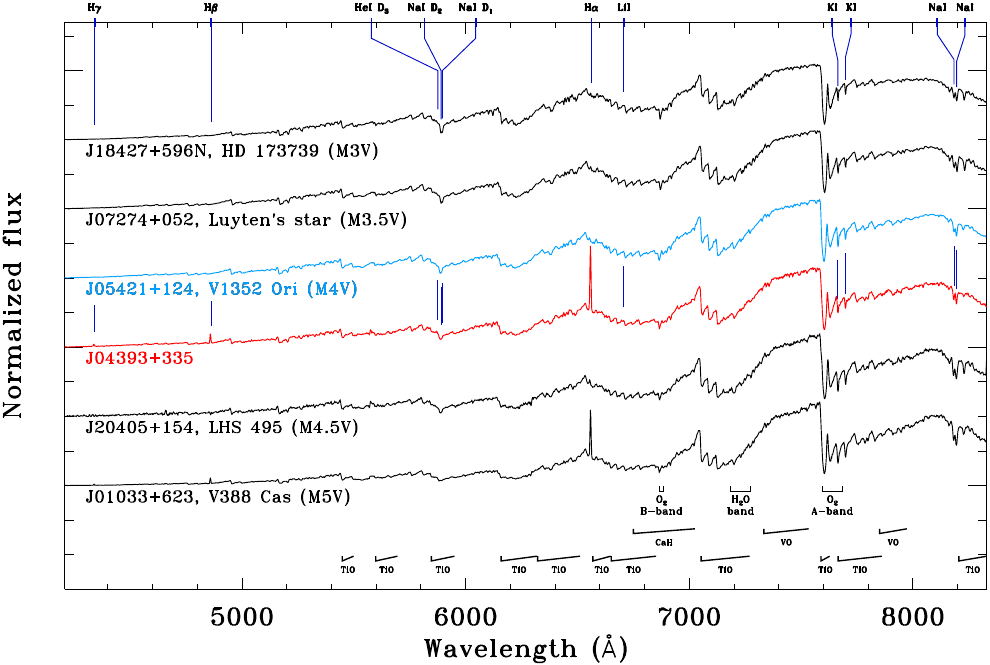}}
\caption{Six representative spectra of mid-M dwarfs from \citet{Alonetal15}. 
From top to bottom, spectra of standard stars with spectral type 1.0 and 0.5 subtypes earlier than the
target (black), standard star with the same spectral type as the target (cyan), the target star (red; in this case, V583~Aur~B, M4.0 V), and standard stars with spectral type 0.5 and 1.0 subtypes later than the target (black). We mark activity-, gravity-, and youth-sensitive lines and doublets at the top of the figure (H$\gamma$, H$\beta$, \HeI{5876}, \NaId{5890,6}, H$\alpha$, \LiI{6708}, \KId{7665,99}, and \NaId{8183,95}, from left to right) and molecular absorption bands at the bottom. Note the three first lines of the Balmer series in emission in the spectrum of the target star.
All the spectra were collected with CAFOS/2.2~m Calar Alto and the same instrumental configuration.}
\label{fig:AF15-1}
\end{figure}

The main feature of the M spectral type is the prevalence of molecular bands at optical wavelengths (Figs.~\ref{fig:AF15-1}~and~\ref{fig:AF15-2}), especially the ones produced by titanium oxide (TiO). 
These bands have been used to establish subtypes since \citet{Morg38b}  
described the M0, M1 and M2 subtypes, followed by the studies of \citet{Kuip42}, \cite{Joy47}, and \citet{KeenSchr52},  
who expanded the classification to mid- and late-M dwarfs based on the strength of the TiO molecular bands. 
However, it was not until \citet{JoyAbt74} and \citet{Boes76} 
when the first standard classification of M subtypes from M0 to M6 was established over the optical wavelength region (4400--6800~{\AA}), which was further extended to M7, M8 and M9 subtypes in later publications \citep{Lee84,Bide85,BoesTyso85}.  
This primordial classification was based on the strength of TiO bands for early M dwarfs, and on the ratio between the molecular bands of vanadium oxide (VO) at 5736~{\AA} and TiO at 5759~{\AA}, together with the appearance of the calcium hydroxide (CaOH) molecular band at 5530~{\AA}, for later M subtypes. 
However, longer wavelengths become more important for the spectral classification according to the advance of the spectral type sequence towards cooler subtypes, showing other easily recognisable molecular bands, such as additional VO and CaOH molecular features. 
The strength of TiO bands are distinguishable from mid-K to M6, after which the VO bands and the sharper slope of the pseudocontinuum stand out (e.g., VO bandhead at 7300~{\AA}). 
To the contrary of stars of earlier spectral types, the spectral continuum in M dwarfs cannot be defined because of the virtual omnipresence of molecular absorption bands; however, a pseudocontinuum can indeed be defined, at least locally.

It was \citet{Kirketal91} who reanalyzed the entire M-dwarf sequence using CCDs and longer wavelengths in the far red (from 6300 to 9000~{\AA}) and provided a new, more complete, spectral-type classification in agreement with previous ones (except for an offset of up to one subtype in early M dwarfs with respect to MK).
Kirkpatrick et al. (1991) identified a list of M-dwarf reference stars and of M giants for calibration purposes, and compiled a number of molecular bands (TiO, VO, and CaOH, but also CaH, CN) and neutral and ionized atomic lines (Na~{\sc i}, Mg~{\sc i}, K~{\sc i}, Ti~{\sc i}, Fe~{\sc i}, Ba~{\sc ii}) in absorption used for spectral classification. 
Some atomic lines (e.g. H$\alpha$, Ca~{\sc ii}) are also sensitive to $T_{\rm eff}$, although they are so much more to chromospheric activity and, therefore, to rotation and age.
The Li~{\sc i} $\lambda$6708~{\AA} line (actually a triplet) is a more reliable age indicator, since the original lithium is destroyed in a few tens of megayears in the interior of M dwarfs, which have extended convective envelopes (spectral type $\lesssim$ M4~V) or are even completely convective (spectral type $\gtrsim$ M4~V).

Following \citet{Kirketal91}, there have been a number of remarkable studies on spectral classification of M dwarfs, such as those under the umbrella of the Palomar/Michigan State University nearby star spectroscopic survey (PMSU: \citealt{Reidetal95,Reidetal02,Hawletal96,Gizietal02})
or those by \citet{Riazetal06}, \citet{LepiGaid11}, and \citet{Lepietal13}.  
To date, \citet{Alonetal15}  have provided the most exhaustive study of M-dwarf spectral classification, including an updated list of reference stars from M0.0~V to M8.0~V, a comparison with previous works, and a comprehensive summary of spectral indices previously used in the literature.
They employed two independent methods for spectral typing: $\chi^2$ minimization of the difference between target and standard spectra, and spectral indices.
The spectral indices methodology for spectral typing is based on computing flux ratios at certain wavelength intervals in low-resolution spectra.
\citet{Alonetal15} compiled virtually all indices used in spectral classification of M dwarfs in the last decades 
\citep{Kirketal91,Kirketal95,Reidetal95,MartKun96,Martetal96,Martetal99,Hawletal02,Lepietal03,Wilketal05,Shkoetal09}.
Of the 31 tabulated indices, nine were related to TiO features, seven to VO, six to CaH, three to the pseudocontinuum (i.e., relative absence of features), and the rest to H and neutral metallic lines (Na~{\sc i}, Ti~{\sc i}).
Some indices are sensitive not only to spectral type (i.e., $T_{\rm eff}$), but also to $\log{g}$, metallicity, or even activity.
A few more spectral indices, mostly related to the TiO bands, may be defined in the blue region, but in most occasions the low observed flux of M dwarfs at these wavelengths would prevent their use.
There are five spectral indices that have the widest range of application and least scatter in the dataset of \citet{Alonetal15}, who used CAFOS at the 2.2~m Calar Alto telescope with spectral resolution $\mathcal{R} \sim$ 1500 and wavelength range 4200--8300~{\AA}.
The indices are TiO~2 and TiO~5 \citep{Reidetal95}, PC1 \citep{Martetal96}, VO-7912 \citep{Martetal99}, and Color-M \citep{Lepietal03}. 

\begin{SCfigure}
\centering
\includegraphics[width=0.40\textwidth]{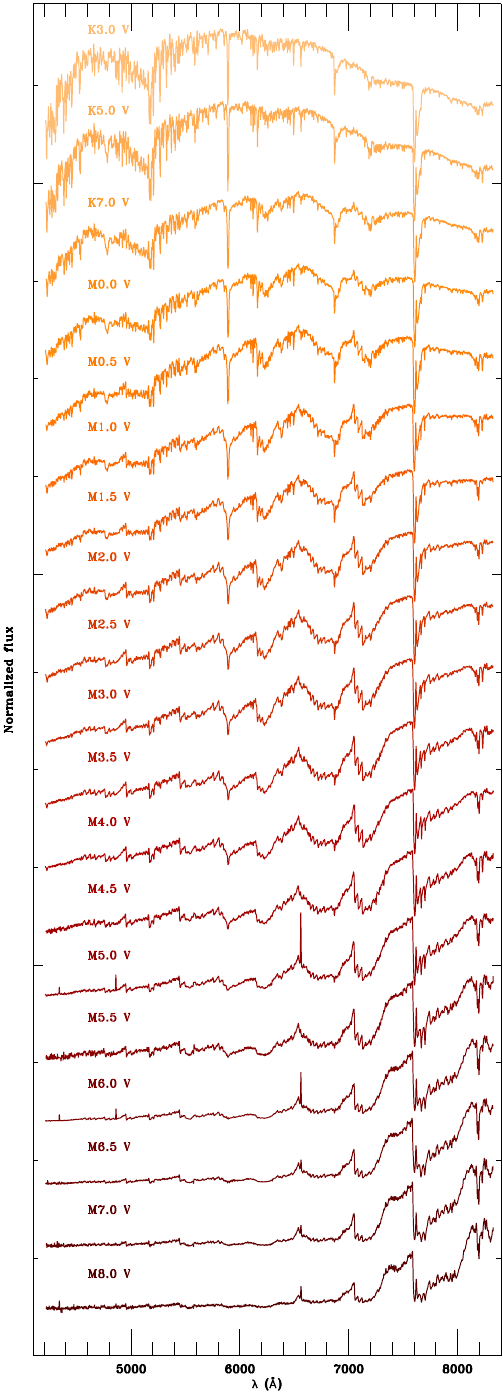} $\;\;$
\caption{CAFOS spectra of the prototype stars in \citet{Alonetal15}. From top to bottom, K3~V, K5~V, K7~V, M0.0--7.0~V in steps of 0.5 subtypes, and M8.0~V.
Note the strengthening of molecular absorption bands, the weakening of the Na~{\sc i} D line, and the decrease of flux at the bluest wavelengths with the spectral sequence.
Although standard M dwarfs were selected to be quiet, some of them have the Balmer series in emission (here H$\alpha$, H$\beta$, H$\gamma$).}
\label{fig:AF15-2}
\end{SCfigure}

In spite of the huge effort for over a century to determine the best spectral types in M dwarfs, recent papers on low-mass transiting planets around nearby stars tend to forget or ignore previous spectral-type determinations (and even classical star designations; e.g., \citealt{Daietal24}).  
Instead, some modern authors prefer to use quantitative spectroscopy only to (re-)determine their own model-dependent $T_{\rm eff}$, which are subject to a number of unsettled uncertainties that come from gravity and metallicity effects, the data themselves, and different methodologies, grids of models, and line libraries \citep{Marfetal21}. See the text in the introduction on the need for spectral classification as a prior step for quantitative spectroscopy, which is opposed to the modern suggestion that in most cases a combination of information on stellar luminosity (from the integration of multi-wavelength spectral energy distribution and precise \textit{Gaia} parallaxes; \citealt{Cifuetal20}) and on metallicity either from high-resolution spectroscopy \citep{Passetal22} or photometry \citep{Duquetal23} can replace most of the information provided by spectral typing in M dwarfs.
For those who still believe in the usefulness of simple, model-free spectral classification of M dwarfs, here we report a few final hints:

\begin{itemize}
    \item {\bf Subtype.}
    The uncertainty of the adopted spectral types can be down to 0.5 subtype. 
    As a result, in M dwarfs the zero at the right of the dot should be written when the spectral type is derived with 0.5 subtype accuracy (e.g. ``M3.0~V'' instead of ``M3~V'').
    More precise classifications should rather carefully consider gravity and metallicity effects and avoid using more than one digit after the decimal point.
    \item {\bf K--M and M--L boundaries.}
    For continuity with the original HD scale, possible spectral types of late K and early M dwarfs in the MK, \citet{Kirketal91} and \citet{Alonetal15} classification schemes are ``K5, K7, M0.0, M0.5, M1.0, M1.5...''.
    As a result, there are no K6, K8 or K9 dwarfs and the gap between K7 and M0.0 remains. 
    On the other end, the latest M dwarf has spectral type M9.5, the next cooler spectral type being L0.0 \citep{Kirk05}.
    \item {\bf Configuration.}  
    For internal consistency of derived spectral types, both standard and target stars in a homogeneous data set must be observed with the same instrumental configuration and reduced with the same pipeline (especially instrumental response correction). If spectra with different instrumental configurations are compared, precise spectral types cannot be determined either from $\chi^2$ fitting or from spectral indices.
    \item \textbf{Emission.}
    Related to the suffixes/qualifiers used for spectral classification in Table~\ref{qualifiers}, the qualifier ``e'' for emission lines was extensively used in the astronomical literature of M dwarfs in the 1980s and 1990s.
    Such emission lines are in general the Balmer series and the Ca~{\sc ii} H \& K doublet and infrared triplet.
    However, a large portion of mid- and late-M dwarfs, if not all during a significant part of their lives, have such chromospheric lines in emission (\citealt{Jeffetal18} and references therein). 
    This chromospheric emission must be differentiated from the much more intense emission from accretion from protoplanetary disks \citep{WhitBasr03,BarrMart03}.  
    As a result, the ``e'' qualifier should not be used for field M dwarfs, but for really active, very young, M-type T~Tauri stars displaying emission lines from accretion.
    If not the case, virtually all M dwarfs except the oldest ones would also carry the ``e'' qualifier.
    \item \textbf{Subdwarfs.}
    In rare occasions, a qualifier ``pec'' is also used for peculiar M dwarfs (that is, identical to ``p'').
    The usual suspect for this peculiarity, once the close M+WD binarity scenario is discarded, is very low metallicity.
    Actually, very low metallicity in M dwarfs can be detected in general only on high-S/N, high-resolution spectra and with specific software that simultaneously compares the EWs of a large number of faint Fe~{\sc i}, Fe~{\sc ii}, and Ti~{\sc i} lines with those from a grid of synthetic spectra.
    Just in those rare occasions, the metallicity is so low ([Fe/H] $\ll$ --1) that its effect is visible in low-resolution spectra.
    When the peculiarity is confirmed to be ascribed to low metallicity (i.e., old age) based on the spectra themselves, kinematics compatible with the Galactic thick disk or the halo, and colours often much bluer than stars of the same \Teff, the  prefix ``sd'' for subdwarfs is used instead of the class \citep{Jaoetal08}.
    The qualitative nomenclature ``esd'' and ``usd'' has also been proposed for extreme and ultra subdwarfs, respectively, with far lower metallicity than ``sd'' \citep{Gizi97,Lepietal03,Lepietal07,Burgetal09}. 
    Low-metallicity subdwarfs are core-hydrogen burning stars (hence, in the main sequence from an evolutionary point of view) but not class V objects, as they are located below (or, more precisely, to the left of) the dwarf sequence in an H-R diagram.
    \item \textbf{Pre-main sequence stars (and brown dwarfs).} 
    On the other side of the subdwarfs, there are other ``M dwarfs'' that have not arrived to the main sequence yet.
    Stars with masses less than about 0.6~$M_\odot$ and ages of a few tens of megayears are in the $L$ vs. \Teff\ diagram in the vertical Hayashi tracks above the zero-age main sequence.
    These young M stars are overluminous with respect to M dwarfs of the same $T_{\rm eff}$ because of their larger radius.
    Spectroscopically, they are characterized by weaker gravity-sensitive alkali lines, as in the case of M giants.
    Besides, and perhaps more importantly, most young stars rotate fast and, therefore, develop strong magnetic fields and have intense chromospheric emission lines in their spectra; some M stars are so young that even have accretion from a disk (see above).
    Since they have not had time to fully contract and reach the typical $\log{g}$ of 4.5--5.5 of M dwarfs, this nomenclature or the class V should not be used for M stars in very young nearby kinematic groups, open clusters, and associations.
    Furthermore, some very late, very young M-type objects have masses below the hydrogen burning mass limit and, therefore, never become stars.
    They are instead brown dwarfs.
\end{itemize}

\subsection{L, T and Y ultracool dwarfs}

\begin{figure}[t]
\centerline{\includegraphics[width=1.00\linewidth]{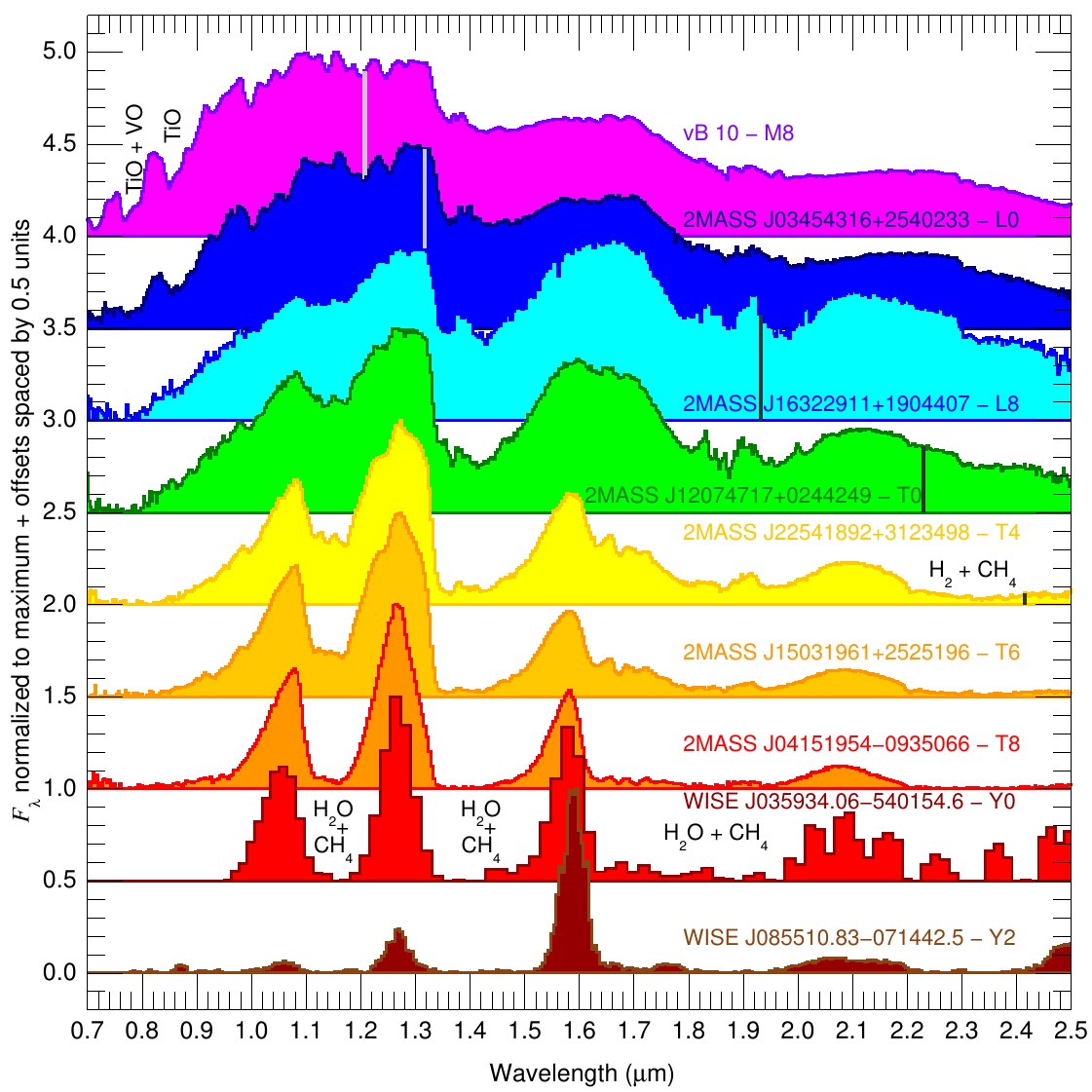}}
\caption{NIR spectral sequence in normalized $F_\lambda$ for the M8-Y2 spectral range. For the top five stars, the wavelength at which the peak of a black body of the same \Teff\ would be located is marked with a vertical line (for the bottom four, it would be outside the frame to the right). Some relevant absorption bands are marked.
Note that for T and Y stars the flux is concentrated in four windows similar to the $YJHK$ terrestrial ones, mostly due to the effect of H$_2$O and CH$_4$ bands.
All spectra were taken from the SpeX Prism Spectral Libraries, maintained by Adam Burgasser at \url{http://www.browndwarfs.org/spexprism}, 
except for the Y dwarfs, which were taken from \citet{Beiletal23} and \citet{Luhmetal24}.}

\label{MLTY}
\end{figure}

There have been numerous theoretical and observational studies dedicated to clarifying the existence of objects with masses intermediate between those of stars and planets.
More than sixty years ago, \citet{Kuma63} and \citet{HayaNaka63} 
proposed that stable thermonuclear fusion of the lightest isotope of hydrogen ($^1$H) in a stellar core would not take place below a certain mass, beyond which electron degeneration pressure would compensate for gravitational collapse to maintain hydrostatic balance.
These types of objects were named with the term brown dwarfs by \citet{Tart76}.  
Brown dwarfs, along with giant exoplanets, whose mass is insufficient even for stable deuterium ($^2$H) combustion, are the two types of substellar objects.

Observational efforts to detect these bodies took more than a quarter of a century to bear fruit. In the same year that the first giant exoplanet around a solar-like star was discovered with the radial velocity method, 51~Pegasi~b \citep{MayoQuel95}, the brown dwarfs were unveiled at the cosmic fauna zoo: Teide~1 in the Pleiades cluster \citep{Reboetal95} and GJ~229~B orbiting at about 45~au from an M2.0~V star \citep{Nakaetal95}.  
Despite its relatively recent discovery, the number of candidates for substellar objects  already amounts to several thousands and grows month by month.

The effective temperatures that have been determined in substellar objects vary between about \num[detect-all]{3000}~K for the most massive brown dwarfs in very  young star-forming regions and less than 300~K (a typical room temperature!) for old Y dwarfs at less than 10~pc from the Sun, not counting planetary companions in wide orbit to stars. 
Due to their low surface temperature, substellar objects emit most of their energy in the near and medium infrared (1--10~$\mu$m), with hardly any emission in the optical. 
The classic spectral series O, B, A, F, G, K, M had to be extended to encompass such cool bodies with the new spectral types L and T, first \citep{Martetal99,Kirketal99,Burgetal02b,Gebaetal02},  
and Y afterwards \citep{Deloetal08,Cushetal11,Kirketal12,Kirketal21}, as illustrated by Fig.~\ref{MLTY}.
As in earlier spectral types, spectral classification is carried out based on the intensity of some easily identifiable features in the photospheric spectra (absorption lines of alkaline elements, bands of metal oxides, water vapor, methane, or ammonia, slope of the pseudocontinuum).
There is a mix of spectral classifications and standard ultracool dwarfs in the far red (only for L dwarfs) and in the near infrared (for L, T, and Y) that do not agree completely, yet.
In any case, all of those classifications are in general performed with very low spectral resolutions, down to $\mathcal{R} \sim$ 100 or even less, as their spectra are very faint and dominated by broad bands with a poorly defined continuum, as for M dwarfs.

One must not mix up the terms substellar object, brown dwarf, ultracool dwarf, and the different late spectral types. 
Substellar objects can be split into brown dwarfs ($M \sim$ 0.072--0.013~M$_\odot$; they do not burn $^1$H but burn $^2$H) and ``planetary-mass objects'' ($M \lesssim$ 0.013~M$_\odot$; they burn neither $^1$H nor $^2$H).
In isolation, the later seem to originate from the star and brown-dwarf formation process extrapolated to very low masses \citep{Caba18},  
while in orbit to stars they seem to originate in protoplanetary disks, as the planets of our Solar System.
There are extremely young objects with late M spectral types that are brown dwarfs (e.g. in the star-forming regions of Orion, Taurus, or Upper Scorpius), while there are very old objects with early L spectral types that are stars (e.g. in the solar neighbourhood).
As a result, one cannot assign a stellar or substellar mass to a late-M- or an early-L-type object unless its age is determined.
Moreover, there can be extremely young T- and moderately-young Y-type objects that are not brown dwarfs, but planetary-mass objects beyond the deuterium-burning limit.
For convention, in the field the term ``ultracool'' is used to denominate any object with spectral type M7.0 or later, regardless of its mass and age.
The wide use of the term ``dwarf'' for such ultracool mixture of stars at the bottom of the main sequence, brown dwarfs, and giant exoplanets is only historical: while one cannot certify whether some of them have fusion reactions in their interiors, at least one can be sure of their extreme faintness, which prevented their discovery for decades.
Such faintness of ultracool dwarfs has also led to difficulties in their spectral characterization.

\subsubsection{L dwarfs}

The first discovered L dwarf was GD~165~B, which orbits a white dwarf \citep{BeckZuck88}.  
Its spectrum was nevertheless so different from what was known at that time that a decade had to elapse for it to be recognised as an L dwarf \citep{Kirketal99}, 
especially after the first DENIS and Kelu-1 discoveries \citep{Delfetal97,Martetal97,Ruizetal97}.   
The atmospheres of L dwarfs, as the spectral sequence progresses, are characterized by the increasing weakening and disappearance of the absorption bands of metal oxides (TiO and VO, especially), due to their condensation in solid grains of dust of refractory species (such as perovskite –-CaTiO$_3$–- and corundum –-Al$_2$O$_3$–-). 
The intensity of the molecular bands of metal hydrides (CrH, FeH, CaH), which replace the oxides, and water vapor (H$_2$O), grows along the L sequence. 
In turn, the absorption of alkali metal neutrals (Li~{\sc i}, Na~{\sc i}, K~{\sc i}, Rb~{\sc i}, Cs~{\sc i}) in the optical also increases considerably as the effective temperature decreases. 
On the other hand, as there are fewer donor electrons, the opacities due to H$^-$ and H$^-_2$ ions decrease. 
However, these opacities are still decisive when it comes to depressing the continuum, as is the collision-induced absorption (often named CIA) between H$_2$ and He molecules. 
Atmospheric dust masks some of the weakest spectral features and carbon is mainly found in the form of monoxide, CO \citep{BurrShar99,Kirketal99,Martetal99,Allaetal01}.   
Li~{\sc i} is also sensitive to $T_{\rm eff}$ because the alkali atom bonds to hydrogen to form lithium hydride (LiH).

The Greek letters $\alpha$, $\beta$, and $\gamma$ are used as a proxy of the gravity instead of the luminosity class, which must not be used in brown dwarfs (and, by extension, LTY ultracool dwarfs).
As an example, an L3$\beta$ has the same $T_{\rm eff}$ but lower $\log{g}$ than an L3$\alpha$, and the same $T_{\rm eff}$ but greater $\log{g}$ than an L3$\gamma$.
The $\alpha$ letter is often dropped.
Low-gravity L dwarfs have weaker alkali lines and fainter VO bands than their regular counterparts.
Their spectra may also display a triangular shape of the pseudocontinuum in the $H$ band.
A Greek letter $\delta$ has been proposed for extremely low-gravity L dwarfs, too, but is not as extensively used.
Other used qualifiers are in common to M dwarfs, such as ``sd'' and ``pec'', although ``red'' and ``blue'' (related to metallicity and unresolved multiplicity) are also found.

\subsubsection{T dwarfs}

Decreasing even further in temperature, the sequence of spectral type T dwarfs begins with the appearance of the methane (CH$_4$) bands, and continues with the strengthening of the H$_2$O and methane bands in the infrared, especially in the $H$ and $K$ bands, and the condensation of alkaline species in the form of chlorides, which makes the resonance doublet of Li~{\sc i}~$\lambda$6707.8 disappear in the coldest T dwarfs \citep{BurrShar99,Burgetal02b,Gebaetal02}. 
The rich chemical diversity presented above is subject to significant changes in the physical conditions of the atmosphere, which cause temporal and spatial variations in the concentration of molecular species, gravitational sedimentation, formation of dust cloud layers (above or below the photosphere) or opening of holes in the cloud cover (which would allow additional flow to emerge from warmer inner layers). These phenomena are especially important in the transition between L and T dwarfs and  explain the overluminosity of early T dwarfs in the $J$ band \citep{AckeMarl01,Burgetal02b}
and enhanced photometric variability at the L--T boundary \citep{Artietal09}.  
The first T dwarf, GJ~229~B, was discovered in 1995, and it took a further decade to discover the first Y dwarfs.

\subsubsection{Y dwarfs}

While the M--L and L--T boundaries are at about 2200~K and 1300~K, respectively, the letter ``Y'' was already identified by \citet{Kirketal99}
as a potential spectral type cooler than the coolest T dwarfs at about 700~K.
With the advent of the Wide-field Infrared Survey Explorer (WISE, \citealt{Wrigetal10b}),  
which operated in the bands centred at 3.4, 4.6, 12, and 22~$\mu$m,
the first Y dwarfs were discovered.
They no longer have photospheric CH$_4$ clouds, but mainly H$_2$O vapour and ammonia (NH$_3$).
The carbon is instead bonded to oxygen in CO and CO$_2$ molecules, although traces of CH$_4$ remain in early Ys.
At the moment of writing this lines (September 2024), there are only about 50 Y dwarfs known \citep{Kirketal24,Luhmetal24}. 
All of them are within 20~pc, have masses near the deuterium burning limit (i.e. brown dwarf-planet boundary), and display spectral types Y0--2.
There are, however, a few Y dwarfs that may be later, such as WISE J053516.80--750024.9, WISE J182831.08+265037.7, or, especially, WISE J085510.83--071442.5, with about 285~K \citep{Luhmetal24,Martetal24}.  
Phosphine (PH$_3$) and H$_2$O ice clouds are now being looked for, and the metallicity, fast rotation, impact on its atmosphere of such clouds are being investigated. The future of Y-dwarf classification will have to be supported by mid-infrared observations, with e.g. \textit{James Webb} Space Telescope.
In ten years, this paragraph on Y-dwarf spectral classification will be wildly outdated.
In a futuristic (or, rather, sci-fi) future, it may happen that Jupiter, with an equilibrium temperature of 88~K, is given the spectral type H0, just slightly cooler than the Y9 given to \textit{Nemesis}, a hypothetical ultracool dwarf at a distance closer than Proxima Centauri.

\section{Outlook}

Despite claims to the contrary, spectral classification is not obsolete. It is needed to provide an image of nature as an intermediate step prior to its understanding by a
subsequent analysis via quantitative spectroscopy. The phenomenology of stellar spectra is too complex and our knowledge of stellar atmospheres still needs improvement, so
the measurements provided by quantitative spectroscopy are subject to modification and to possibly miss important aspects about the nature of the object of study. In that
sea of change, spectral classification provides an island of stability.

Indeed, the future could be very bright for spectral classification with the avalanche of spectra that new surveys are providing and which without a doubt would have
impressed Secchi, Cannon, Morgan, or Walborn with their potential. They can possibly extend the sample of spectral classifications two or three orders of magnitude beyond what was
done by the Harvard group a century ago. But to do that, it is important to follow the procedures described in this entry: establish a grid of standards in common with
the one of the primary MK system in the blue-violet region; determine the characteristics that are useful to obtain the spectral subtype and luminosity class for each type; derive spectral
classifications by overall comparisons between the sample and the standards instead of restricting oneself to a single line strength or ratio; maintain a
consistent notation; and be clear about what special characteristics such as line broadening or composition anomalies can be measured and with which (un)certainty with the
data at hand, using the appropriate suffixes. If that is not done, the classifier is bound to add more noise than signal to our knowledge, a peril that has frequently affected
the field. 

\begin{ack}[Acknowledgments]

This article would not have been possible without the knowledge of spectral classification that Nolan R. Walborn passed on to us. We thank Paul Crowther, Danny Lennon, Brian Skiff, Roberto Gamen, Sophie Rosu, and Fabian Schneider for helpful suggestions and/or corrections.
We acknowledge support from the Spanish Government Ministerio de Ciencia e Innovaci\'on and Agencia Estatal de Investigaci\'on (\num{10.13039}/\num{501100011033}) through grants PID2022-\num{136640}~NB-C22 (J.M.A.),
PID2021-\num{122397}~NB-C22 (I.N.), and  PID2022-\num{137241}NB-C42 (J.A.C.).
\end{ack}


\bibliographystyle{Harvard}
\bibliography{general}

\begin{thebibliography*}{220}
\providecommand{\bibtype}[1]{}
\providecommand{\natexlab}[1]{#1}
{\catcode`\|=0\catcode`\#=12\catcode`\@=11\catcode`\\=12
|immediate|write|@auxout{\expandafter\ifx\csname
  natexlab\endcsname\relax\gdef\natexlab#1{#1}\fi}}
\renewcommand{\url}[1]{{\tt #1}}
\providecommand{\urlprefix}{URL }
\expandafter\ifx\csname urlstyle\endcsname\relax
  \providecommand{\doi}[1]{doi:\discretionary{}{}{}#1}\else
  \providecommand{\doi}{doi:\discretionary{}{}{}\begingroup
  \urlstyle{rm}\Url}\fi
\providecommand{\bibinfo}[2]{#2}
\providecommand{\eprint}[2][]{\url{#2}}

\bibtype{Article}%
\bibitem[Ackerman and Marley(2001)]{AckeMarl01}
\bibinfo{author}{Ackerman AS} and  \bibinfo{author}{Marley MS}
  (\bibinfo{year}{2001}).
\bibinfo{title}{Precipitating condensation clouds in substellar atmospheres}.
\bibinfo{journal}{{\em ApJ}} \bibinfo{volume}{556} (\bibinfo{number}{2}):
  \bibinfo{pages}{872--884}.

\bibtype{Article}%
\bibitem[Adams and Joy(1922{\natexlab{a}})]{AdamJoy22b}
\bibinfo{author}{Adams WS} and  \bibinfo{author}{Joy AH}
  (\bibinfo{year}{1922}{\natexlab{a}}).
\bibinfo{title}{A list of dwarf \uppercase{M}-type stars}.
\bibinfo{journal}{{\em PASP}} \bibinfo{volume}{34} (\bibinfo{number}{199}):
  \bibinfo{pages}{174}.

\bibtype{Article}%
\bibitem[Adams and Joy(1922{\natexlab{b}})]{AdamJoy22a}
\bibinfo{author}{Adams WS} and  \bibinfo{author}{Joy AH}
  (\bibinfo{year}{1922}{\natexlab{b}}).
\bibinfo{title}{A spectroscopic method of determining the absolute magnitudes
  of \uppercase{A}-type stars and the parallaxes of 544 stars}.
\bibinfo{journal}{{\em ApJ}} \bibinfo{volume}{56}: \bibinfo{pages}{242}.

\bibtype{Article}%
\bibitem[Aguado et al.(2018)]{Aguaetal18}
\bibinfo{author}{Aguado DS}, \bibinfo{author}{Allende~Prieto C},
  \bibinfo{author}{Gonz{\'a}lez~Hern{\'a}ndez JI} and  \bibinfo{author}{Rebolo
  R} (\bibinfo{year}{2018}).
\bibinfo{title}{J0023+0307: \uppercase{A} mega metal-poor dwarf star from
  \uppercase{SDSS/BOSS}}.
\bibinfo{journal}{{\em ApJL}} \bibinfo{volume}{854} (\bibinfo{number}{2}),
  \bibinfo{eid}{L34}.

\bibtype{Article}%
\bibitem[Allard et al.(2001)]{Allaetal01}
\bibinfo{author}{Allard F}, \bibinfo{author}{Hauschildt PH},
  \bibinfo{author}{Alexander DR}, \bibinfo{author}{Tamanai A} and
  \bibinfo{author}{Schweitzer A} (\bibinfo{year}{2001}).
\bibinfo{title}{The limiting effects of dust in brown dwarf model atmospheres}.
\bibinfo{journal}{{\em ApJ}} \bibinfo{volume}{556} (\bibinfo{number}{1}):
  \bibinfo{pages}{357--372}.

\bibtype{Article}%
\bibitem[Alonso-Floriano et al.(2015)]{Alonetal15}
\bibinfo{author}{Alonso-Floriano FJ}, \bibinfo{author}{Morales JC},
  \bibinfo{author}{Caballero JA}, \bibinfo{author}{Montes D},
  \bibinfo{author}{Klutsch A}, \bibinfo{author}{Mundt R},
  \bibinfo{author}{Cort{\'e}s-Contreras M}, \bibinfo{author}{Ribas I},
  \bibinfo{author}{Reiners A}, \bibinfo{author}{Amado PJ},
  \bibinfo{author}{Quirrenbach A} and  \bibinfo{author}{Jeffers SV}
  (\bibinfo{year}{2015}).
\bibinfo{title}{\uppercase{CARMENES} input catalogue of \uppercase{M} dwarfs.
  \uppercase{I. L}ow-resolution spectroscopy with \uppercase{CAFOS}}.
\bibinfo{journal}{{\em A\&A}} \bibinfo{volume}{577}, \bibinfo{eid}{A128}.

\bibtype{Article}%
\bibitem[Alonso-Santiago et al.(2019)]{Alonetal19}
\bibinfo{author}{Alonso-Santiago J}, \bibinfo{author}{Negueruela I},
  \bibinfo{author}{Marco A}, \bibinfo{author}{Tabernero HM},
  \bibinfo{author}{Gonz{\'a}lez-Fern{\'a}ndez C} and  \bibinfo{author}{Castro
  N} (\bibinfo{year}{2019}).
\bibinfo{title}{A comprehensive study of \uppercase{NGC} 2345, a young open
  cluster with a low metallicity}.
\bibinfo{journal}{{\em A\&A}} \bibinfo{volume}{631}, \bibinfo{eid}{A124}.

\bibtype{Article}%
\bibitem[Alonso-Santiago et al.(2020)]{Alonetal20}
\bibinfo{author}{Alonso-Santiago J}, \bibinfo{author}{Negueruela I},
  \bibinfo{author}{Marco A}, \bibinfo{author}{Tabernero HM} and
  \bibinfo{author}{Castro N} (\bibinfo{year}{2020}).
\bibinfo{title}{Three open clusters containing cepheids: \uppercase{NGC} 6649,
  \uppercase{NGC} 6664, and \uppercase{B}erkeley 55}.
\bibinfo{journal}{{\em A\&A}} \bibinfo{volume}{644}, \bibinfo{eid}{A136}.

\bibtype{Article}%
\bibitem[Anderson et al.(2014)]{Andeetal14b}
\bibinfo{author}{Anderson RI}, \bibinfo{author}{Ekstr{\"o}m S},
  \bibinfo{author}{Georgy C}, \bibinfo{author}{Meynet G},
  \bibinfo{author}{Mowlavi N} and  \bibinfo{author}{Eyer L}
  (\bibinfo{year}{2014}).
\bibinfo{title}{On the effect of rotation on populations of classical cepheids.
  \uppercase{I. P}redictions at solar metallicity}.
\bibinfo{journal}{{\em A\&A}} \bibinfo{volume}{564}, \bibinfo{eid}{A100}.

\bibtype{Article}%
\bibitem[Arias et al.(2016)]{Ariaetal16}
\bibinfo{author}{Arias JI}, \bibinfo{author}{Walborn NR},
  \bibinfo{author}{Sim{\'o}n~D{\'\i}az S}, \bibinfo{author}{Barb{\'a} RH},
  \bibinfo{author}{Ma{\'\i}z~Apell{\'a}niz J},
  \bibinfo{author}{Sab{\'\i}n-Sanjuli{\'a}n C}, \bibinfo{author}{Gamen RC},
  \bibinfo{author}{Morrell NI}, \bibinfo{author}{Sota A},
  \bibinfo{author}{Marco A}, \bibinfo{author}{Negueruela I},
  \bibinfo{author}{Le{\~a}o JRS}, \bibinfo{author}{Herrero A} and
  \bibinfo{author}{Alfaro EJ} (\bibinfo{year}{2016}).
\bibinfo{title}{Spectral classification and properties of the \uppercase{OV}z
  stars in the \uppercase{G}alactic \uppercase{O-S}tar
  \uppercase{S}pectroscopic \uppercase{S}urvey (\uppercase{GOSSS})}.
\bibinfo{journal}{{\em AJ}} \bibinfo{volume}{152}, \bibinfo{eid}{31}.

\bibtype{Article}%
\bibitem[Artigau et al.(2009)]{Artietal09}
\bibinfo{author}{Artigau {\'E}}, \bibinfo{author}{Bouchard S},
  \bibinfo{author}{Doyon R} and  \bibinfo{author}{Lafreni{\`e}re D}
  (\bibinfo{year}{2009}).
\bibinfo{title}{Photometric variability of the \uppercase{T}2.5 brown dwarf
  simp \uppercase{J013656.5+093347: E}vidence for evolving weather patterns}.
\bibinfo{journal}{{\em ApJ}} \bibinfo{volume}{701} (\bibinfo{number}{2}):
  \bibinfo{pages}{1534--1539}.

\bibtype{Article}%
\bibitem[Balona and Crampton(1974)]{BaloCram74}
\bibinfo{author}{Balona L} and  \bibinfo{author}{Crampton D}
  (\bibinfo{year}{1974}).
\bibinfo{title}{The \uppercase{H}$\gamma$-absolute magnitude calibration.}
\bibinfo{journal}{{\em MNRAS}} \bibinfo{volume}{166}:
  \bibinfo{pages}{203--217}.

\bibtype{Article}%
\bibitem[Barrado~y Navascu{\'e}s and Mart{\'\i}n(2003)]{BarrMart03}
\bibinfo{author}{Barrado~y Navascu{\'e}s D} and  \bibinfo{author}{Mart{\'\i}n
  EL} (\bibinfo{year}{2003}).
\bibinfo{title}{An empirical criterion to classify \uppercase{T T}auri stars
  and substellar analogs using low-resolution optical spectroscopy}.
\bibinfo{journal}{{\em AJ}} \bibinfo{volume}{126} (\bibinfo{number}{6}):
  \bibinfo{pages}{2997--3006}.

\bibtype{Article}%
\bibitem[Becklin and Zuckerman(1988)]{BeckZuck88}
\bibinfo{author}{Becklin EE} and  \bibinfo{author}{Zuckerman B}
  (\bibinfo{year}{1988}).
\bibinfo{title}{A low-temperature companion to a white dwarf star}.
\bibinfo{journal}{{\em Nature}} \bibinfo{volume}{336} (\bibinfo{number}{6200}):
  \bibinfo{pages}{656--658}.

\bibtype{Article}%
\bibitem[Beiler et al.(2023)]{Beiletal23}
\bibinfo{author}{Beiler SA}, \bibinfo{author}{Cushing MC},
  \bibinfo{author}{Kirkpatrick JD}, \bibinfo{author}{Schneider AC},
  \bibinfo{author}{Mukherjee S} and  \bibinfo{author}{Marley MS}
  (\bibinfo{year}{2023}).
\bibinfo{title}{The first \uppercase{JWST} spectral energy distribution of a
  \uppercase{Y} dwarf}.
\bibinfo{journal}{{\em ApJL}} \bibinfo{volume}{951} (\bibinfo{number}{2}),
  \bibinfo{eid}{L48}.

\bibtype{Article}%
\bibitem[Bestenlehner et al.(2020)]{Bestetal20}
\bibinfo{author}{Bestenlehner JM}, \bibinfo{author}{Crowther PA},
  \bibinfo{author}{Caballero-Nieves SM}, \bibinfo{author}{Schneider FRN},
  \bibinfo{author}{Sim{\'o}n-D{\'\i}az S}, \bibinfo{author}{Brands SA},
  \bibinfo{author}{de~Koter A}, \bibinfo{author}{Gr{\"a}fener G},
  \bibinfo{author}{Herrero A}, \bibinfo{author}{Langer N},
  \bibinfo{author}{Lennon DJ}, \bibinfo{author}{Ma{\'\i}z~Apell{\'a}niz J},
  \bibinfo{author}{Puls J} and  \bibinfo{author}{Vink JS}
  (\bibinfo{year}{2020}).
\bibinfo{title}{The \uppercase{R136} star cluster dissected with
  \uppercase{H}ubble \uppercase{S}pace \uppercase{T}elescope/\uppercase{STIS -
  II. P}hysical properties of the most massive stars in \uppercase{R136}}.
\bibinfo{journal}{{\em MNRAS}} \bibinfo{volume}{499} (\bibinfo{number}{2}):
  \bibinfo{pages}{1918--1936}.

\bibtype{Article}%
\bibitem[Beyer and White(2024)]{BeyeWhit24}
\bibinfo{author}{Beyer AC} and  \bibinfo{author}{White RJ}
  (\bibinfo{year}{2024}).
\bibinfo{title}{The \uppercase{K}raft break sharply divides low mass and
  intermediate mass stars} \eprint{arXiv:2408.02638}.

\bibtype{Article}%
\bibitem[Bidelman(1985)]{Bide85}
\bibinfo{author}{Bidelman WP} (\bibinfo{year}{1985}).
\bibinfo{title}{\uppercase{G.P. K}uiper's spectral classifications of
  proper-motion stars.}
\bibinfo{journal}{{\em ApJS}} \bibinfo{volume}{59}: \bibinfo{pages}{197--227}.

\bibtype{Phdthesis}%
\bibitem[Boeshaar(1976)]{Boes76}
\bibinfo{author}{Boeshaar PC} (\bibinfo{year}{1976}).
\bibinfo{title}{The Spectral Classification of M-Dwarf Stars.}
\bibinfo{comment}{Ph.D. thesis}, \bibinfo{school}{The Ohio State University}.

\bibtype{Article}%
\bibitem[Boeshaar and Tyson(1985)]{BoesTyso85}
\bibinfo{author}{Boeshaar PC} and  \bibinfo{author}{Tyson JA}
  (\bibinfo{year}{1985}).
\bibinfo{title}{New limits on the surface density of \uppercase{M} dwarfs.
  \uppercase{I. P}hotographic survey and preliminary \uppercase{CCD} data.}
\bibinfo{journal}{{\em AJ}} \bibinfo{volume}{90}: \bibinfo{pages}{817--822}.

\bibtype{Article}%
\bibitem[Bonfils et al.(2013)]{Bonfetal13}
\bibinfo{author}{Bonfils X}, \bibinfo{author}{Lo~Curto G},
  \bibinfo{author}{Correia ACM}, \bibinfo{author}{Laskar J},
  \bibinfo{author}{Udry S}, \bibinfo{author}{Delfosse X},
  \bibinfo{author}{Forveille T}, \bibinfo{author}{Astudillo-Defru N},
  \bibinfo{author}{Benz W}, \bibinfo{author}{Bouchy F}, \bibinfo{author}{Gillon
  M}, \bibinfo{author}{H{\'e}brard G}, \bibinfo{author}{Lovis C},
  \bibinfo{author}{Mayor M}, \bibinfo{author}{Moutou C}, \bibinfo{author}{Naef
  D}, \bibinfo{author}{Neves V}, \bibinfo{author}{Pepe F},
  \bibinfo{author}{Perrier C}, \bibinfo{author}{Queloz D},
  \bibinfo{author}{Santos NC} and  \bibinfo{author}{S{\'e}gransan D}
  (\bibinfo{year}{2013}).
\bibinfo{title}{The \uppercase{HARPS} search for southern extra-solar planets.
  \uppercase{XXXIV. A} planetary system around the nearby \uppercase{M} dwarf
  \uppercase{GJ} 163, with a super-\uppercase{E}arth possibly in the habitable
  zone}.
\bibinfo{journal}{{\em A\&A}} \bibinfo{volume}{556}, \bibinfo{eid}{A110}.

\bibtype{Article}%
\bibitem[Burgasser et al.(2002)]{Burgetal02b}
\bibinfo{author}{Burgasser AJ}, \bibinfo{author}{Kirkpatrick JD},
  \bibinfo{author}{Brown ME}, \bibinfo{author}{Reid IN},
  \bibinfo{author}{Burrows A}, \bibinfo{author}{Liebert J},
  \bibinfo{author}{Matthews K}, \bibinfo{author}{Gizis JE},
  \bibinfo{author}{Dahn CC}, \bibinfo{author}{Monet DG}, \bibinfo{author}{Cutri
  RM} and  \bibinfo{author}{Skrutskie MF} (\bibinfo{year}{2002}).
\bibinfo{title}{The spectra of \uppercase{T} dwarfs. \uppercase{I.
  N}ear-infrared data and spectral classification}.
\bibinfo{journal}{{\em ApJ}} \bibinfo{volume}{564} (\bibinfo{number}{1}):
  \bibinfo{pages}{421--451}.

\bibtype{Article}%
\bibitem[Burgasser et al.(2009)]{Burgetal09}
\bibinfo{author}{Burgasser AJ}, \bibinfo{author}{Witte S},
  \bibinfo{author}{Helling C}, \bibinfo{author}{Sanderson RE},
  \bibinfo{author}{Bochanski JJ} and  \bibinfo{author}{Hauschildt PH}
  (\bibinfo{year}{2009}).
\bibinfo{title}{Optical and near-infrared spectroscopy of the \uppercase{L}
  subdwarf \uppercase{SDSS J125637.13-022452.4}}.
\bibinfo{journal}{{\em ApJ}} \bibinfo{volume}{697} (\bibinfo{number}{1}):
  \bibinfo{pages}{148--159}.

\bibtype{Article}%
\bibitem[Burrows and Sharp(1999)]{BurrShar99}
\bibinfo{author}{Burrows A} and  \bibinfo{author}{Sharp CM}
  (\bibinfo{year}{1999}).
\bibinfo{title}{Chemical equilibrium abundances in brown dwarf and extrasolar
  giant planet atmospheres}.
\bibinfo{journal}{{\em ApJ}} \bibinfo{volume}{512} (\bibinfo{number}{2}):
  \bibinfo{pages}{843--863}.

\bibtype{Article}%
\bibitem[Caballero(2018)]{Caba18}
\bibinfo{author}{Caballero JA} (\bibinfo{year}{2018}).
\bibinfo{title}{A review on substellar objects below the deuterium burning mass
  limit: \uppercase{P}lanets, brown dwarfs or what?}
\bibinfo{journal}{{\em Geosciences}} \bibinfo{volume}{8}
  (\bibinfo{number}{10}): \bibinfo{pages}{362}.

\bibtype{Article}%
\bibitem[Cannon and Mayall(1949)]{CannMaya49}
\bibinfo{author}{Cannon AJ} and  \bibinfo{author}{Mayall MW}
  (\bibinfo{year}{1949}).
\bibinfo{title}{The \uppercase{H}enry \uppercase{D}raper extension.
  \uppercase{II}.}
\bibinfo{journal}{{\em AnHar}} \bibinfo{volume}{112}: \bibinfo{pages}{1--295}.

\bibtype{Article}%
\bibitem[Cannon and Pickering(1901)]{CannPick01}
\bibinfo{author}{Cannon AJ} and  \bibinfo{author}{Pickering EC}
  (\bibinfo{year}{1901}).
\bibinfo{title}{Spectra of bright southern stars photographed with the 13-inch
  \uppercase{B}oyden telescope as part of the \uppercase{H}enry
  \uppercase{D}raper \uppercase{M}emorial}.
\bibinfo{journal}{{\em Annals of Harvard College Observatory}}
  \bibinfo{volume}{28}: \bibinfo{pages}{129--P.6}.

\bibtype{Article}%
\bibitem[Cifuentes et al.(2020)]{Cifuetal20}
\bibinfo{author}{Cifuentes C}, \bibinfo{author}{Caballero JA},
  \bibinfo{author}{Cort{\'e}s-Contreras M}, \bibinfo{author}{Montes D},
  \bibinfo{author}{Abell{\'a}n FJ}, \bibinfo{author}{Dorda R},
  \bibinfo{author}{Holgado G}, \bibinfo{author}{Zapatero~Osorio MR},
  \bibinfo{author}{Morales JC}, \bibinfo{author}{Amado PJ},
  \bibinfo{author}{Passegger VM}, \bibinfo{author}{Quirrenbach A},
  \bibinfo{author}{Reiners A}, \bibinfo{author}{Ribas I},
  \bibinfo{author}{Sanz-Forcada J}, \bibinfo{author}{Schweitzer A},
  \bibinfo{author}{Seifert W} and  \bibinfo{author}{Solano E}
  (\bibinfo{year}{2020}).
\bibinfo{title}{\uppercase{CARMENES} input catalogue of \uppercase{M} dwarfs.
  \uppercase{V. L}uminosities, colours, and spectral energy distributions}.
\bibinfo{journal}{{\em A\&A}} \bibinfo{volume}{642}, \bibinfo{eid}{A115}.

\bibtype{Article}%
\bibitem[Clark et al.(2015)]{Claretal15b}
\bibinfo{author}{Clark JS}, \bibinfo{author}{Negueruela I},
  \bibinfo{author}{Lohr ME}, \bibinfo{author}{Dorda R},
  \bibinfo{author}{Gonz{\'a}lez-Fern{\'a}ndez C}, \bibinfo{author}{Lewis F} and
   \bibinfo{author}{Roche P} (\bibinfo{year}{2015}).
\bibinfo{title}{A long-period cepheid variable in the starburst cluster
  \uppercase{V}d\uppercase{BH} 222}.
\bibinfo{journal}{{\em A\&A}} \bibinfo{volume}{584}, \bibinfo{eid}{L12}.

\bibtype{Inproceedings}%
\bibitem[Conti(1984)]{Cont84}
\bibinfo{author}{Conti PS} (\bibinfo{year}{1984}), \bibinfo{title}{Basic
  observational constraints on the evolution of massive stars},
  \bibinfo{editor}{Maeder A} and  \bibinfo{editor}{Renzini A}, (Eds.),
  \bibinfo{booktitle}{Observational Tests of the Stellar Evolution Theory},
  \bibinfo{series}{IAU Symposium}, \bibinfo{volume}{105}, pp.
  \bibinfo{pages}{233}.

\bibtype{Article}%
\bibitem[Conti and Alschuler(1971)]{ContAlsc71}
\bibinfo{author}{Conti PS} and  \bibinfo{author}{Alschuler WR}
  (\bibinfo{year}{1971}).
\bibinfo{title}{Spectroscopic studies of \uppercase{O}-type stars.
  \uppercase{I. C}lassification and absolute magnitudes}.
\bibinfo{journal}{{\em ApJ}} \bibinfo{volume}{170}: \bibinfo{pages}{325}.

\bibtype{Article}%
\bibitem[Cropper~et al.(2018)]{Cropetal18}
\bibinfo{author}{Cropper~et al. } (\bibinfo{year}{2018}).
\bibinfo{title}{\uppercase{G}aia \uppercase{D}ata \uppercase{R}elease 2.
  \uppercase{G}aia \uppercase{R}adial \uppercase{V}elocity
  \uppercase{S}pectrometer}.
\bibinfo{journal}{{\em A\&A}} \bibinfo{volume}{616}, \bibinfo{eid}{A5}.

\bibtype{Article}%
\bibitem[Crowe and Garrison(1988)]{CrowGarr88}
\bibinfo{author}{Crowe RA} and  \bibinfo{author}{Garrison RF}
  (\bibinfo{year}{1988}).
\bibinfo{title}{The visible spectra of southern hemisphere \uppercase{M}ira
  variable stars}.
\bibinfo{journal}{{\em ApJS}} \bibinfo{volume}{66}: \bibinfo{pages}{69}.

\bibtype{Article}%
\bibitem[Crowther and Bohannan(1997)]{CrowBoha97}
\bibinfo{author}{Crowther PA} and  \bibinfo{author}{Bohannan B}
  (\bibinfo{year}{1997}).
\bibinfo{title}{The distinction between \uppercase{O}iafpe and
  \uppercase{WNL}ha stars. \uppercase{A} spectral analysis of \uppercase{HD
  151\,804, HD} 152\,408 and \uppercase{HDE} 313\,846.}
\bibinfo{journal}{{\em A\&A}} \bibinfo{volume}{317}: \bibinfo{pages}{532--547}.

\bibtype{Article}%
\bibitem[Crowther and Smith(1997)]{CrowSmit97}
\bibinfo{author}{Crowther PA} and  \bibinfo{author}{Smith LJ}
  (\bibinfo{year}{1997}).
\bibinfo{title}{Fundamental parameters of \uppercase{W}olf-\uppercase{R}ayet
  stars.\uppercase{ VI. L}arge \uppercase{M}agellanic \uppercase{C}loud
  \uppercase{WNL} stars.}
\bibinfo{journal}{{\em A\&A}} \bibinfo{volume}{320}: \bibinfo{pages}{500--524}.

\bibtype{Article}%
\bibitem[Crowther and Walborn(2011)]{CrowWalb11}
\bibinfo{author}{Crowther PA} and  \bibinfo{author}{Walborn NR}
  (\bibinfo{year}{2011}).
\bibinfo{title}{Spectral classification of \uppercase{O2-3.5
  I}f*/\uppercase{WN}5-7 stars}.
\bibinfo{journal}{{\em MNRAS}} \bibinfo{volume}{416}:
  \bibinfo{pages}{1311--1323}.

\bibtype{Article}%
\bibitem[Crowther et al.(1998)]{Crowetal98}
\bibinfo{author}{Crowther PA}, \bibinfo{author}{De~Marco O} and
  \bibinfo{author}{Barlow MJ} (\bibinfo{year}{1998}).
\bibinfo{title}{Quantitative classification of \uppercase{WC} and
  \uppercase{WO} stars}.
\bibinfo{journal}{{\em MNRAS}} \bibinfo{volume}{296} (\bibinfo{number}{2}):
  \bibinfo{pages}{367--378}.

\bibtype{Article}%
\bibitem[Crowther et al.(2010)]{Crowetal10}
\bibinfo{author}{Crowther PA}, \bibinfo{author}{Schnurr O},
  \bibinfo{author}{Hirschi R}, \bibinfo{author}{Yusof N},
  \bibinfo{author}{Parker RJ}, \bibinfo{author}{Goodwin SP} and
  \bibinfo{author}{Kassim HA} (\bibinfo{year}{2010}).
\bibinfo{title}{The \uppercase{R136} star cluster hosts several stars whose
  individual masses greatly exceed the accepted 150 \uppercase{M}$_\odot$
  stellar mass limit}.
\bibinfo{journal}{{\em MNRAS}} \bibinfo{volume}{408}:
  \bibinfo{pages}{731--751}.

\bibtype{Article}%
\bibitem[Curtiss(1932)]{Curt32}
\bibinfo{author}{Curtiss RH} (\bibinfo{year}{1932}).
\bibinfo{title}{Classification and description of stellar spectra}.
\bibinfo{journal}{{\em Handbuch der Astrophysik}} \bibinfo{volume}{5}:
  \bibinfo{pages}{1}.

\bibtype{Article}%
\bibitem[Cushing et al.(2011)]{Cushetal11}
\bibinfo{author}{Cushing MC}, \bibinfo{author}{Kirkpatrick JD},
  \bibinfo{author}{Gelino CR}, \bibinfo{author}{Griffith RL},
  \bibinfo{author}{Skrutskie MF}, \bibinfo{author}{Mainzer A},
  \bibinfo{author}{Marsh KA}, \bibinfo{author}{Beichman CA},
  \bibinfo{author}{Burgasser AJ}, \bibinfo{author}{Prato LA},
  \bibinfo{author}{Simcoe RA}, \bibinfo{author}{Marley MS},
  \bibinfo{author}{Saumon D}, \bibinfo{author}{Freedman RS},
  \bibinfo{author}{Eisenhardt PR} and  \bibinfo{author}{Wright EL}
  (\bibinfo{year}{2011}).
\bibinfo{title}{The discovery of \uppercase{Y} dwarfs using data from the
  \uppercase{W}ide-field \uppercase{I}nfrared \uppercase{S}urvey
  \uppercase{E}xplorer (\uppercase{WISE})}.
\bibinfo{journal}{{\em ApJ}} \bibinfo{volume}{743} (\bibinfo{number}{1}),
  \bibinfo{eid}{50}.

\bibtype{Article}%
\bibitem[Dai~et al.(2024)]{Daietal24}
\bibinfo{author}{Dai~et al. } (\bibinfo{year}{2024}).
\bibinfo{title}{An earth-sized planet on the verge of tidal disruption}.
\bibinfo{journal}{{\em AJ}} \bibinfo{volume}{168} (\bibinfo{number}{3}),
  \bibinfo{eid}{101}.

\bibtype{Article}%
\bibitem[De~Angeli~et al.(2023)]{DeAnetal23}
\bibinfo{author}{De~Angeli~et al. } (\bibinfo{year}{2023}).
\bibinfo{title}{\uppercase{G}aia \uppercase{D}ata \uppercase{R}elease 3.
  \uppercase{P}rocessing and validation of \uppercase{BP/RP} low-resolution
  spectral data}.
\bibinfo{journal}{{\em A\&A}} \bibinfo{volume}{674}, \bibinfo{eid}{A2}.

\bibtype{Article}%
\bibitem[de~Burgos et al.(2024)]{deBuetal24}
\bibinfo{author}{de~Burgos A}, \bibinfo{author}{Sim{\'o}n-D{\'\i}az S},
  \bibinfo{author}{Urbaneja MA} and  \bibinfo{author}{Puls J}
  (\bibinfo{year}{2024}).
\bibinfo{title}{The \uppercase{IACOB} project. \uppercase{X. L}arge-scale
  quantitative spectroscopic analysis of \uppercase{G}alactic luminous blue
  stars}.
\bibinfo{journal}{{\em A\&A}} \bibinfo{volume}{687}, \bibinfo{eid}{A228}.

\bibtype{Inproceedings}%
\bibitem[De~Greve(1996)]{DeGr96}
\bibinfo{author}{De~Greve JP} (\bibinfo{year}{1996}), \bibinfo{title}{Binary
  \uppercase{WR}: an alternative channel for \uppercase{C}onti's scenario},
  \bibinfo{booktitle}{WR stars in the Framework of Stellar Evolution},
  pp.~\bibinfo{pages}{55}.

\bibtype{Article}%
\bibitem[Delfosse et al.(1997)]{Delfetal97}
\bibinfo{author}{Delfosse X}, \bibinfo{author}{Tinney CG},
  \bibinfo{author}{Forveille T}, \bibinfo{author}{Epchtein N},
  \bibinfo{author}{Bertin E}, \bibinfo{author}{Borsenberger J},
  \bibinfo{author}{Copet E}, \bibinfo{author}{de~Batz B},
  \bibinfo{author}{Fouque P}, \bibinfo{author}{Kimeswenger S},
  \bibinfo{author}{Le~Bertre T}, \bibinfo{author}{Lacombe F},
  \bibinfo{author}{Rouan D} and  \bibinfo{author}{Tiphene D}
  (\bibinfo{year}{1997}).
\bibinfo{title}{Field brown dwarfs found by \uppercase{DENIS}}.
\bibinfo{journal}{{\em A\&A}} \bibinfo{volume}{327}: \bibinfo{pages}{L25--L28}.

\bibtype{Article}%
\bibitem[Delorme et al.(2008)]{Deloetal08}
\bibinfo{author}{Delorme P}, \bibinfo{author}{Delfosse X},
  \bibinfo{author}{Albert L}, \bibinfo{author}{Artigau E},
  \bibinfo{author}{Forveille T}, \bibinfo{author}{Reyl{\'e} C},
  \bibinfo{author}{Allard F}, \bibinfo{author}{Homeier D},
  \bibinfo{author}{Robin AC}, \bibinfo{author}{Willott CJ},
  \bibinfo{author}{Liu MC} and  \bibinfo{author}{Dupuy TJ}
  (\bibinfo{year}{2008}).
\bibinfo{title}{\uppercase{CFBDS J005910.90-011401.3: R}eaching the
  \uppercase{T-Y} brown dwarf transition?}
\bibinfo{journal}{{\em A\&A}} \bibinfo{volume}{482} (\bibinfo{number}{3}):
  \bibinfo{pages}{961--971}.

\bibtype{Article}%
\bibitem[Dimitrov et al.(2018)]{Dimietal18}
\bibinfo{author}{Dimitrov DP}, \bibinfo{author}{Kjurkchieva DP} and
  \bibinfo{author}{Ivanov EI} (\bibinfo{year}{2018}).
\bibinfo{title}{A study of the \uppercase{H}{$\alpha$} variability of
  \uppercase{B}e stars}.
\bibinfo{journal}{{\em AJ}} \bibinfo{volume}{156}, \bibinfo{eid}{61}.

\bibtype{Article}%
\bibitem[Dorda et al.(2016)]{Dordetal16b}
\bibinfo{author}{Dorda R}, \bibinfo{author}{Negueruela I},
  \bibinfo{author}{Gonz{\'a}lez-Fern{\'a}ndez C} and
  \bibinfo{author}{Tabernero HM} (\bibinfo{year}{2016}).
\bibinfo{title}{Spectral type, temperature, and evolutionary stage in cool
  supergiants}.
\bibinfo{journal}{{\em A\&A}} \bibinfo{volume}{592}, \bibinfo{eid}{A16}.

\bibtype{Article}%
\bibitem[Dorda et al.(2018)]{Dordetal18b}
\bibinfo{author}{Dorda R}, \bibinfo{author}{Negueruela I},
  \bibinfo{author}{Gonz{\'a}lez-Fern{\'a}ndez C} and  \bibinfo{author}{Marco A}
  (\bibinfo{year}{2018}).
\bibinfo{title}{An atlas of cool supergiants from the \uppercase{M}agellanic
  \uppercase{C}louds and typical interlopers. \uppercase{A} guide for the
  classification of luminous red stars}.
\bibinfo{journal}{{\em A\&A}} \bibinfo{volume}{618}, \bibinfo{eid}{A137}.

\bibtype{Article}%
\bibitem[Drilling et al.(2013)]{Driletal13}
\bibinfo{author}{Drilling JS}, \bibinfo{author}{Jeffery CS},
  \bibinfo{author}{Heber U}, \bibinfo{author}{Moehler S} and
  \bibinfo{author}{Napiwotzki R} (\bibinfo{year}{2013}).
\bibinfo{title}{An \uppercase{MK}-like system of spectral classification for
  hot subdwarfs}.
\bibinfo{journal}{{\em A\&A}} \bibinfo{volume}{551}, \bibinfo{eid}{A31}.

\bibtype{Article}%
\bibitem[Drissen et al.(1995)]{Drisetal95}
\bibinfo{author}{Drissen L}, \bibinfo{author}{Moffat AFJ},
  \bibinfo{author}{Walborn NR} and  \bibinfo{author}{Shara MM}
  (\bibinfo{year}{1995}).
\bibinfo{title}{The dense galactic starburst \uppercase{NGC 3603. I. HST/FOS}
  spectroscopy of individual stars in the core and the source of ionization and
  kinetic energy}.
\bibinfo{journal}{{\em AJ}} \bibinfo{volume}{110}: \bibinfo{pages}{2235--2241}.

\bibtype{Article}%
\bibitem[Duque-Arribas et al.(2023)]{Duquetal23}
\bibinfo{author}{Duque-Arribas C}, \bibinfo{author}{Montes D},
  \bibinfo{author}{Tabernero HM}, \bibinfo{author}{Caballero JA},
  \bibinfo{author}{Gorgas J} and  \bibinfo{author}{Marfil E}
  (\bibinfo{year}{2023}).
\bibinfo{title}{Photometric calibrations of \uppercase{M}-dwarf metallicity
  with \uppercase{M}arkov chain \uppercase{M}onte \uppercase{C}arlo and
  \uppercase{B}ayesian inference}.
\bibinfo{journal}{{\em ApJ}} \bibinfo{volume}{944} (\bibinfo{number}{1}),
  \bibinfo{eid}{106}.

\bibtype{Article}%
\bibitem[Evans and Howarth(2003)]{EvanHowa03}
\bibinfo{author}{Evans CJ} and  \bibinfo{author}{Howarth ID}
  (\bibinfo{year}{2003}).
\bibinfo{title}{Characteristics and classification of \uppercase{A}-type
  supergiants in the \uppercase{S}mall \uppercase{M}agellanic
  \uppercase{C}loud}.
\bibinfo{journal}{{\em MNRAS}} \bibinfo{volume}{345} (\bibinfo{number}{4}):
  \bibinfo{pages}{1223--1235}.

\bibtype{Article}%
\bibitem[Falc{\'o}n-Barroso et al.(2011)]{Falcetal11}
\bibinfo{author}{Falc{\'o}n-Barroso J},
  \bibinfo{author}{S{\'a}nchez-Bl{\'a}zquez P}, \bibinfo{author}{Vazdekis A},
  \bibinfo{author}{Ricciardelli E}, \bibinfo{author}{Cardiel N},
  \bibinfo{author}{Cenarro AJ}, \bibinfo{author}{Gorgas J} and
  \bibinfo{author}{Peletier RF} (\bibinfo{year}{2011}).
\bibinfo{title}{An updated \uppercase{MILES} stellar library and stellar
  population models}.
\bibinfo{journal}{{\em A\&A}} \bibinfo{volume}{532}, \bibinfo{eid}{A95}.

\bibtype{Article}%
\bibitem[Fang et al.(2020)]{Fangetal20}
\bibinfo{author}{Fang M}, \bibinfo{author}{Hillenbrand LA},
  \bibinfo{author}{Kim JS}, \bibinfo{author}{Findeisen K},
  \bibinfo{author}{Herczeg GJ}, \bibinfo{author}{Carpenter JM},
  \bibinfo{author}{Rebull LM} and  \bibinfo{author}{Wang H}
  (\bibinfo{year}{2020}).
\bibinfo{title}{The first extensive spectroscopic study of young stars in the
  \uppercase{N}orth \uppercase{A}merica and \uppercase{P}elican nebulae}.
\bibinfo{journal}{{\em ApJ}} \bibinfo{volume}{904} (\bibinfo{number}{2}),
  \bibinfo{eid}{146}.

\bibtype{Article}%
\bibitem[Feast et al.(1960)]{Feasetal60}
\bibinfo{author}{Feast MW}, \bibinfo{author}{Thackeray AD} and
  \bibinfo{author}{Wesselink AJ} (\bibinfo{year}{1960}).
\bibinfo{title}{The brightest stars in the \uppercase{M}agellanic
  \uppercase{C}louds}.
\bibinfo{journal}{{\em MNRAS}} \bibinfo{volume}{121}: \bibinfo{pages}{337}.

\bibtype{Article}%
\bibitem[Garc{\'\i}a-Hern{\'a}ndez et al.(2007)]{GarHetal07}
\bibinfo{author}{Garc{\'\i}a-Hern{\'a}ndez DA},
  \bibinfo{author}{Garc{\'\i}a-Lario P}, \bibinfo{author}{Plez B},
  \bibinfo{author}{Manchado A}, \bibinfo{author}{D'Antona F},
  \bibinfo{author}{Lub J} and  \bibinfo{author}{Habing H}
  (\bibinfo{year}{2007}).
\bibinfo{title}{Lithium and zirconium abundances in massive
  \uppercase{G}alactic \uppercase{O}-rich \uppercase{AGB} stars}.
\bibinfo{journal}{{\em A\&A}} \bibinfo{volume}{462} (\bibinfo{number}{2}):
  \bibinfo{pages}{711--730}.

\bibtype{Inproceedings}%
\bibitem[Garrison(1994)]{Garr94}
\bibinfo{author}{Garrison RF} (\bibinfo{year}{1994}), \bibinfo{title}{A
  hierarchy of standards for the \uppercase{MK} process},
  \bibinfo{editor}{Corbally CJ}, \bibinfo{editor}{Gray RO} and
  \bibinfo{editor}{Garrison RF}, (Eds.), \bibinfo{booktitle}{The MK Process at
  50 Years: A Powerful Tool for Astrophysical Insight},
  \bibinfo{series}{Astronomical Society of the Pacific Conference Series},
  \bibinfo{volume}{60}, pp.~\bibinfo{pages}{3}.

\bibtype{Article}%
\bibitem[Garrison and Gray(1994)]{GarrGray94}
\bibinfo{author}{Garrison RF} and  \bibinfo{author}{Gray RO}
  (\bibinfo{year}{1994}).
\bibinfo{title}{The late \uppercase{B}-type stars: \uppercase{R}efined
  \uppercase{MK} classification, confrontation with \uppercase{S}tr\"omgren
  photometry, and the effects of rotation}.
\bibinfo{journal}{{\em AJ}} \bibinfo{volume}{107}: \bibinfo{pages}{1556--1564}.

\bibtype{Article}%
\bibitem[Geballe~et al.(2002)]{Gebaetal02}
\bibinfo{author}{Geballe~et al. } (\bibinfo{year}{2002}).
\bibinfo{title}{Toward spectral classification of \uppercase{L} and
  \uppercase{T} dwarfs: \uppercase{I}nfrared and optical spectroscopy and
  analysis}.
\bibinfo{journal}{{\em ApJ}} \bibinfo{volume}{564} (\bibinfo{number}{1}):
  \bibinfo{pages}{466--481}.

\bibtype{Article}%
\bibitem[Gilmore~et al.(2012)]{Gilmetal12}
\bibinfo{author}{Gilmore~et al. } (\bibinfo{year}{2012}).
\bibinfo{title}{The \uppercase{G}aia-\uppercase{ESO} public spectroscopic
  survey}.
\bibinfo{journal}{{\em The Messenger}} \bibinfo{volume}{147}:
  \bibinfo{pages}{25--31}.

\bibtype{Article}%
\bibitem[Gizis(1997)]{Gizi97}
\bibinfo{author}{Gizis JE} (\bibinfo{year}{1997}).
\bibinfo{title}{M-subdwarfs: \uppercase{S}pectroscopic classification and the
  metallicity scale}.
\bibinfo{journal}{{\em AJ}} \bibinfo{volume}{113}: \bibinfo{pages}{806--822}.

\bibtype{Article}%
\bibitem[Gizis et al.(2002)]{Gizietal02}
\bibinfo{author}{Gizis JE}, \bibinfo{author}{Reid IN} and
  \bibinfo{author}{Hawley SL} (\bibinfo{year}{2002}).
\bibinfo{title}{The \uppercase{P}alomar/\uppercase{MSU} nearby-star
  spectroscopic survey. \uppercase{III. C}hromospheric activity, \uppercase{M}
  dwarf ages, and the local star formation history}.
\bibinfo{journal}{{\em AJ}} \bibinfo{volume}{123} (\bibinfo{number}{6}):
  \bibinfo{pages}{3356--3369}.

\bibtype{Article}%
\bibitem[Gorny and Stasi{\'n}ska(1995)]{GornStas95}
\bibinfo{author}{Gorny SK} and  \bibinfo{author}{Stasi{\'n}ska G}
  (\bibinfo{year}{1995}).
\bibinfo{title}{On the status of planetary nebulae with \uppercase{WR}-type
  nuclei.}
\bibinfo{journal}{{\em A\&A}} \bibinfo{volume}{303}: \bibinfo{pages}{893}.

\bibtype{Article}%
\bibitem[G{\"o}tberg et al.(2018)]{Gotbetal18}
\bibinfo{author}{G{\"o}tberg Y}, \bibinfo{author}{de~Mink SE},
  \bibinfo{author}{Groh JH}, \bibinfo{author}{Kupfer T},
  \bibinfo{author}{Crowther PA}, \bibinfo{author}{Zapartas E} and
  \bibinfo{author}{Renzo M} (\bibinfo{year}{2018}).
\bibinfo{title}{Spectral models for binary products: \uppercase{U}nifying
  subdwarfs and \uppercase{W}olf-\uppercase{R}ayet stars as a sequence of
  stripped-envelope stars}.
\bibinfo{journal}{{\em A\&A}} \bibinfo{volume}{615}, \bibinfo{eid}{A78}.

\bibtype{Article}%
\bibitem[G{\"o}tberg et al.(2023)]{Gotbetal23}
\bibinfo{author}{G{\"o}tberg Y}, \bibinfo{author}{Drout MR},
  \bibinfo{author}{Ji AP}, \bibinfo{author}{Groh JH}, \bibinfo{author}{Ludwig
  BA}, \bibinfo{author}{Crowther PA}, \bibinfo{author}{Smith N},
  \bibinfo{author}{de~Koter A} and  \bibinfo{author}{de~Mink SE}
  (\bibinfo{year}{2023}).
\bibinfo{title}{Stellar properties of observed stars stripped in binaries in
  the \uppercase{M}agellanic \uppercase{C}louds}.
\bibinfo{journal}{{\em ApJ}} \bibinfo{volume}{959} (\bibinfo{number}{2}),
  \bibinfo{eid}{125}.

\bibtype{Book}%
\bibitem[Gray and Corbally(2009)]{GrayCorb09}
\bibinfo{author}{Gray RO} and  \bibinfo{author}{Corbally J. C}
  (\bibinfo{year}{2009}).
\bibinfo{title}{Stellar Spectral Classification, ISBN: 978-0-691-12511-4}.

\bibtype{Article}%
\bibitem[Gray and Garrison(1987)]{GrayGarr87}
\bibinfo{author}{Gray RO} and  \bibinfo{author}{Garrison RF}
  (\bibinfo{year}{1987}).
\bibinfo{title}{The early \uppercase{A}-type stars: \uppercase{R}efined
  \uppercase{MK} classification, confrontation with \uppercase{S}tr\"omgren
  photometry, and the effects of rotation}.
\bibinfo{journal}{{\em ApJS}} \bibinfo{volume}{65}: \bibinfo{pages}{581}.

\bibtype{Article}%
\bibitem[Gray and Garrison(1989{\natexlab{a}})]{GrayGarr89a}
\bibinfo{author}{Gray RO} and  \bibinfo{author}{Garrison RF}
  (\bibinfo{year}{1989}{\natexlab{a}}).
\bibinfo{title}{The early \uppercase{F}-type stars: \uppercase{R}efined
  classification, confrontation with \uppercase{S}tr\"omgren photometry, and
  the effects of rotation}.
\bibinfo{journal}{{\em ApJS}} \bibinfo{volume}{69}: \bibinfo{pages}{301}.

\bibtype{Article}%
\bibitem[Gray and Garrison(1989{\natexlab{b}})]{GrayGarr89b}
\bibinfo{author}{Gray RO} and  \bibinfo{author}{Garrison RF}
  (\bibinfo{year}{1989}{\natexlab{b}}).
\bibinfo{title}{The late \uppercase{A}-type stars: \uppercase{R}efined
  \uppercase{MK} classification, confrontation with \uppercase{S}tr\"omgren
  photometry, and the effects of rotation}.
\bibinfo{journal}{{\em ApJS}} \bibinfo{volume}{70}: \bibinfo{pages}{623}.

\bibtype{Article}%
\bibitem[Gray et al.(2003)]{Grayetal03}
\bibinfo{author}{Gray RO}, \bibinfo{author}{Corbally CJ},
  \bibinfo{author}{Garrison RF}, \bibinfo{author}{McFadden MT} and
  \bibinfo{author}{Robinson PE} (\bibinfo{year}{2003}).
\bibinfo{title}{Contributions to the \uppercase{N}earby \uppercase{S}tars
  (\uppercase{NS}tars) project: \uppercase{S}pectroscopy of stars earlier than
  \uppercase{M}0 within 40 parsecs: \uppercase{T}he northern sample.
  \uppercase{I}.}
\bibinfo{journal}{{\em AJ}} \bibinfo{volume}{126}: \bibinfo{pages}{2048--2059}.

\bibtype{Article}%
\bibitem[Hawley et al.(1996)]{Hawletal96}
\bibinfo{author}{Hawley SL}, \bibinfo{author}{Gizis JE} and
  \bibinfo{author}{Reid IN} (\bibinfo{year}{1996}).
\bibinfo{title}{The \uppercase{P}alomar/\uppercase{MSU} nearby-star
  spectroscopic survey. \uppercase{II. T}he southern \uppercase{M} dwarfs and
  investigation of magnetic activity}.
\bibinfo{journal}{{\em AJ}} \bibinfo{volume}{112}: \bibinfo{pages}{2799}.

\bibtype{Article}%
\bibitem[Hawley~et al.(2002)]{Hawletal02}
\bibinfo{author}{Hawley~et al. } (\bibinfo{year}{2002}).
\bibinfo{title}{Characterization of \uppercase{M, L,} and \uppercase{T} dwarfs
  in the \uppercase{S}loan \uppercase{D}igital \uppercase{S}ky
  \uppercase{S}urvey}.
\bibinfo{journal}{{\em AJ}} \bibinfo{volume}{123} (\bibinfo{number}{6}):
  \bibinfo{pages}{3409--3427}.

\bibtype{Article}%
\bibitem[Hayashi and Nakano(1963)]{HayaNaka63}
\bibinfo{author}{Hayashi C} and  \bibinfo{author}{Nakano T}
  (\bibinfo{year}{1963}).
\bibinfo{title}{Evolution of stars of small masses in the
  \uppercase{P}re-\uppercase{M}ain-\uppercase{S}equence stages}.
\bibinfo{journal}{{\em Progress of Theoretical Physics}} \bibinfo{volume}{30}
  (\bibinfo{number}{4}): \bibinfo{pages}{460--474}.

\bibtype{Book}%
\bibitem[Hearnshaw(1990)]{Hear90}
\bibinfo{author}{Hearnshaw JB} (\bibinfo{year}{1990}).
\bibinfo{title}{The Analysis of Starlight, ISBN: 0-521-39916-5}.

\bibtype{Article}%
\bibitem[Heber(2016)]{Hebe16}
\bibinfo{author}{Heber U} (\bibinfo{year}{2016}).
\bibinfo{title}{Hot subluminous stars}.
\bibinfo{journal}{{\em PASP}} \bibinfo{volume}{128} (\bibinfo{number}{8}):
  \bibinfo{pages}{082001}.

\bibtype{Article}%
\bibitem[Herwig(2005)]{Herw05}
\bibinfo{author}{Herwig F} (\bibinfo{year}{2005}).
\bibinfo{title}{Evolution of \uppercase{A}symptotic \uppercase{G}iant
  \uppercase{B}ranch stars}.
\bibinfo{journal}{{\em ARA\&A}} \bibinfo{volume}{43} (\bibinfo{number}{1}):
  \bibinfo{pages}{435--479}.

\bibtype{Article}%
\bibitem[Holgado et al.(2018)]{Holgetal18}
\bibinfo{author}{Holgado G}, \bibinfo{author}{Sim{\'o}n-D{\'\i}az S},
  \bibinfo{author}{Barb{\'a} RH}, \bibinfo{author}{Puls J},
  \bibinfo{author}{Herrero A}, \bibinfo{author}{Castro N},
  \bibinfo{author}{Garcia M}, \bibinfo{author}{Ma{\'\i}z~Apell{\'a}niz J},
  \bibinfo{author}{Negueruela I} and  \bibinfo{author}{Sab{\'\i}n-Sanjuli{\'a}n
  C} (\bibinfo{year}{2018}).
\bibinfo{title}{The \uppercase{IACOB} project. \uppercase{V. S}pectroscopic
  parameters of the \uppercase{O}-type stars in the modern grid of standards
  for spectral classification}.
\bibinfo{journal}{{\em A\&A}} \bibinfo{volume}{613}, \bibinfo{eid}{A65}.

\bibtype{Article}%
\bibitem[Holgado et al.(2022)]{Holgetal22}
\bibinfo{author}{Holgado G}, \bibinfo{author}{Sim{\'o}n-D{\'\i}az S},
  \bibinfo{author}{Herrero A} and  \bibinfo{author}{Barb{\'a} RH}
  (\bibinfo{year}{2022}).
\bibinfo{title}{The \uppercase{IACOB} project. \uppercase{VII}. \uppercase{T}he
  rotational properties of \uppercase{G}alactic massive \uppercase{O}-type
  stars revisited}.
\bibinfo{journal}{{\em A\&A}} \bibinfo{volume}{665}, \bibinfo{eid}{A150}.

\bibtype{Article}%
\bibitem[Hubrig et al.(2024)]{Hubretal24}
\bibinfo{author}{Hubrig S}, \bibinfo{author}{Sch{\"o}ller M},
  \bibinfo{author}{J{\"a}rvinen SP}, \bibinfo{author}{Cikota A},
  \bibinfo{author}{Abdul-Masih M}, \bibinfo{author}{Escorza A} and
  \bibinfo{author}{Jayaraman R} (\bibinfo{year}{2024}).
\bibinfo{title}{Detection of extragalactic magnetic massive stars}.
\bibinfo{journal}{{\em A\&A}} \bibinfo{volume}{686}, \bibinfo{eid}{L4}.

\bibtype{Article}%
\bibitem[H{\"u}mmerich et al.(2020)]{Hummetal20}
\bibinfo{author}{H{\"u}mmerich S}, \bibinfo{author}{Paunzen E} and
  \bibinfo{author}{Bernhard K} (\bibinfo{year}{2020}).
\bibinfo{title}{A plethora of new, magnetic chemically peculiar stars from
  \uppercase{LAMOST DR4}}.
\bibinfo{journal}{{\em A\&A}} \bibinfo{volume}{640}, \bibinfo{eid}{A40}.

\bibtype{Article}%
\bibitem[Humphreys and Davidson(1979)]{HumpDavi79}
\bibinfo{author}{Humphreys RM} and  \bibinfo{author}{Davidson K}
  (\bibinfo{year}{1979}).
\bibinfo{title}{Studies of luminous stars in nearby galaxies. \uppercase{III.
  C}omments on the evolution of the most massive stars in the \uppercase{M}ilky
  \uppercase{W}ay and the \uppercase{L}arge \uppercase{M}agellanic
  \uppercase{C}loud.}
\bibinfo{journal}{{\em ApJ}} \bibinfo{volume}{232}: \bibinfo{pages}{409--420}.

\bibtype{Article}%
\bibitem[Jao et al.(2008)]{Jaoetal08}
\bibinfo{author}{Jao WC}, \bibinfo{author}{Henry TJ}, \bibinfo{author}{Beaulieu
  TD} and  \bibinfo{author}{Subasavage JP} (\bibinfo{year}{2008}).
\bibinfo{title}{Cool subdwarf investigations. \uppercase{I. N}ew thoughts on
  the spectral types of \uppercase{K} and \uppercase{M} subdwarfs}.
\bibinfo{journal}{{\em AJ}} \bibinfo{volume}{136} (\bibinfo{number}{2}):
  \bibinfo{pages}{840--880}.

\bibtype{Article}%
\bibitem[Jeffers et al.(2018)]{Jeffetal18}
\bibinfo{author}{Jeffers SV}, \bibinfo{author}{Sch{\"o}fer P},
  \bibinfo{author}{Lamert A}, \bibinfo{author}{Reiners A},
  \bibinfo{author}{Montes D}, \bibinfo{author}{Caballero JA},
  \bibinfo{author}{Cort{\'e}s-Contreras M}, \bibinfo{author}{Marvin CJ},
  \bibinfo{author}{Passegger VM}, \bibinfo{author}{Zechmeister M},
  \bibinfo{author}{Quirrenbach A}, \bibinfo{author}{Alonso-Floriano FJ},
  \bibinfo{author}{Amado PJ}, \bibinfo{author}{Bauer FF},
  \bibinfo{author}{Casal E}, \bibinfo{author}{Diez~Alonso E},
  \bibinfo{author}{Herrero E}, \bibinfo{author}{Morales JC},
  \bibinfo{author}{Mundt R}, \bibinfo{author}{Ribas I} and
  \bibinfo{author}{Sarmiento LF} (\bibinfo{year}{2018}).
\bibinfo{title}{\uppercase{CARMENES} input catalogue of \uppercase{M} dwarfs.
  \uppercase{III. R}otation and activity from high-resolution spectroscopic
  observations}.
\bibinfo{journal}{{\em A\&A}} \bibinfo{volume}{614}, \bibinfo{eid}{A76}.

\bibtype{Article}%
\bibitem[Johnson and Morgan(1953)]{JohnMorg53}
\bibinfo{author}{Johnson HL} and  \bibinfo{author}{Morgan WW}
  (\bibinfo{year}{1953}).
\bibinfo{title}{Fundamental stellar photometry for standards of spectral type
  on the revised system of the \uppercase{Y}erkes spectral atlas}.
\bibinfo{journal}{{\em ApJ}} \bibinfo{volume}{117}: \bibinfo{pages}{313--352}.

\bibtype{Article}%
\bibitem[Joy(1947)]{Joy47}
\bibinfo{author}{Joy AH} (\bibinfo{year}{1947}).
\bibinfo{title}{Radial velocities and spectral types of 181 dwarf stars.}
\bibinfo{journal}{{\em ApJ}} \bibinfo{volume}{105}: \bibinfo{pages}{96}.

\bibtype{Article}%
\bibitem[Joy and Abt(1974)]{JoyAbt74}
\bibinfo{author}{Joy AH} and  \bibinfo{author}{Abt HA} (\bibinfo{year}{1974}).
\bibinfo{title}{Spectral types of \uppercase{M} dwarf stars}.
\bibinfo{journal}{{\em ApJS}} \bibinfo{volume}{28}: \bibinfo{pages}{1}.

\bibtype{Article}%
\bibitem[Keenan(1942)]{Keen42}
\bibinfo{author}{Keenan PC} (\bibinfo{year}{1942}).
\bibinfo{title}{Luminosities of the \uppercase{M}-type variables of small
  range.}
\bibinfo{journal}{{\em ApJ}} \bibinfo{volume}{95}: \bibinfo{pages}{461}.

\bibtype{Inproceedings}%
\bibitem[Keenan(1984)]{Keen84}
\bibinfo{author}{Keenan PC} (\bibinfo{year}{1984}), \bibinfo{title}{What is
  wrong with the \uppercase{MK} system?}, \bibinfo{editor}{Garrison RF}, (Ed.),
  \bibinfo{booktitle}{The MK Process and Stellar Classification},
  pp.~\bibinfo{pages}{29}.

\bibtype{Article}%
\bibitem[Keenan(1993)]{Keen93}
\bibinfo{author}{Keenan PC} (\bibinfo{year}{1993}).
\bibinfo{title}{Revised \uppercase{MK} spectral classification of the red
  carbon stars}.
\bibinfo{journal}{{\em PASP}} \bibinfo{volume}{105}: \bibinfo{pages}{905}.

\bibtype{Article}%
\bibitem[Keenan and McNeil(1989)]{KeenMcNe89}
\bibinfo{author}{Keenan PC} and  \bibinfo{author}{McNeil RC}
  (\bibinfo{year}{1989}).
\bibinfo{title}{The \uppercase{P}erkins catalog of revised \uppercase{MK} types
  for the cooler stars}.
\bibinfo{journal}{{\em ApJS}} \bibinfo{volume}{71}: \bibinfo{pages}{245}.

\bibtype{Article}%
\bibitem[Keenan and Schroeder(1952)]{KeenSchr52}
\bibinfo{author}{Keenan PC} and  \bibinfo{author}{Schroeder LW}
  (\bibinfo{year}{1952}).
\bibinfo{title}{An infrared system of bands of \uppercase{VO} in
  \uppercase{M}-type stars.}
\bibinfo{journal}{{\em ApJ}} \bibinfo{volume}{115}: \bibinfo{pages}{82}.

\bibtype{Article}%
\bibitem[Keenan et al.(1974)]{Keenetal74}
\bibinfo{author}{Keenan PC}, \bibinfo{author}{Garrison RF} and
  \bibinfo{author}{Deutsch AJ} (\bibinfo{year}{1974}).
\bibinfo{title}{Revised catalog of spectra of \uppercase{M}ira variables of
  types \uppercase{ME} and \uppercase{S}e}.
\bibinfo{journal}{{\em ApJS}} \bibinfo{volume}{28}: \bibinfo{pages}{271}.

\bibtype{Article}%
\bibitem[Kirkpatrick(2005)]{Kirk05}
\bibinfo{author}{Kirkpatrick JD} (\bibinfo{year}{2005}).
\bibinfo{title}{New spectral types \uppercase{L} and \uppercase{T}}.
\bibinfo{journal}{{\em ARA\&A}} \bibinfo{volume}{43} (\bibinfo{number}{1}):
  \bibinfo{pages}{195--245}.

\bibtype{Article}%
\bibitem[Kirkpatrick et al.(1991)]{Kirketal91}
\bibinfo{author}{Kirkpatrick JD}, \bibinfo{author}{Henry TJ} and
  \bibinfo{author}{McCarthy Donald~W. J} (\bibinfo{year}{1991}).
\bibinfo{title}{A standard stellar spectral sequence in the red/near-infrared:
  Classes \uppercase{K5} to \uppercase{M9}}.
\bibinfo{journal}{{\em ApJS}} \bibinfo{volume}{77}: \bibinfo{pages}{417}.

\bibtype{Article}%
\bibitem[Kirkpatrick et al.(1995)]{Kirketal95}
\bibinfo{author}{Kirkpatrick JD}, \bibinfo{author}{Henry TJ} and
  \bibinfo{author}{Simons DA} (\bibinfo{year}{1995}).
\bibinfo{title}{The solar neighborhood. \uppercase{II. T}he first list of
  dwarfs with spectral types of \uppercase{M}7 and cooler}.
\bibinfo{journal}{{\em AJ}} \bibinfo{volume}{109}: \bibinfo{pages}{797}.

\bibtype{Article}%
\bibitem[Kirkpatrick et al.(1999)]{Kirketal99}
\bibinfo{author}{Kirkpatrick JD}, \bibinfo{author}{Reid IN},
  \bibinfo{author}{Liebert J}, \bibinfo{author}{Cutri RM},
  \bibinfo{author}{Nelson B}, \bibinfo{author}{Beichman CA},
  \bibinfo{author}{Dahn CC}, \bibinfo{author}{Monet DG}, \bibinfo{author}{Gizis
  JE} and  \bibinfo{author}{Skrutskie MF} (\bibinfo{year}{1999}).
\bibinfo{title}{Dwarfs cooler than \uppercase{``M``: T}he definition of
  spectral type \uppercase{``L''} using discoveries from the 2
  \uppercase{M}icron \uppercase{A}ll-\uppercase{S}ky \uppercase{S}urvey
  (\uppercase{2MASS})}.
\bibinfo{journal}{{\em ApJ}} \bibinfo{volume}{519} (\bibinfo{number}{2}):
  \bibinfo{pages}{802--833}.

\bibtype{Article}%
\bibitem[Kirkpatrick et al.(2012)]{Kirketal12}
\bibinfo{author}{Kirkpatrick JD}, \bibinfo{author}{Gelino CR},
  \bibinfo{author}{Cushing MC}, \bibinfo{author}{Mace GN},
  \bibinfo{author}{Griffith RL}, \bibinfo{author}{Skrutskie MF},
  \bibinfo{author}{Marsh KA}, \bibinfo{author}{Wright EL},
  \bibinfo{author}{Eisenhardt PR}, \bibinfo{author}{McLean IS},
  \bibinfo{author}{Mainzer AK}, \bibinfo{author}{Burgasser AJ},
  \bibinfo{author}{Tinney CG}, \bibinfo{author}{Parker S} and
  \bibinfo{author}{Salter G} (\bibinfo{year}{2012}).
\bibinfo{title}{Further defining spectral type \uppercase{``Y''} and exploring
  the low-mass end of the field brown dwarf mass function}.
\bibinfo{journal}{{\em ApJ}} \bibinfo{volume}{753} (\bibinfo{number}{2}),
  \bibinfo{eid}{156}.

\bibtype{Article}%
\bibitem[Kirkpatrick~et al.(2021)]{Kirketal21}
\bibinfo{author}{Kirkpatrick~et al. } (\bibinfo{year}{2021}).
\bibinfo{title}{The field substellar mass function based on the full-sky 20 pc
  census of 525 \uppercase{L, T, and Y} dwarfs}.
\bibinfo{journal}{{\em ApJS}} \bibinfo{volume}{253} (\bibinfo{number}{1}),
  \bibinfo{eid}{7}.

\bibtype{Article}%
\bibitem[Kirkpatrick~et al.(2024)]{Kirketal24}
\bibinfo{author}{Kirkpatrick~et al. } (\bibinfo{year}{2024}).
\bibinfo{title}{The initial mass function based on the full-sky 20 pc census of
  {\ensuremath{\sim}}3600 stars and brown dwarfs}.
\bibinfo{journal}{{\em ApJS}} \bibinfo{volume}{271} (\bibinfo{number}{2}),
  \bibinfo{eid}{55}.

\bibtype{Article}%
\bibitem[Kraft(1967)]{Kraf67}
\bibinfo{author}{Kraft RP} (\bibinfo{year}{1967}).
\bibinfo{title}{Studies of stellar rotation. \uppercase{V. T}he dependence of
  rotation on age among solar-type stars}.
\bibinfo{journal}{{\em ApJ}} \bibinfo{volume}{150}: \bibinfo{pages}{551}.

\bibtype{Article}%
\bibitem[Kuiper(1942)]{Kuip42}
\bibinfo{author}{Kuiper GP} (\bibinfo{year}{1942}).
\bibinfo{title}{The nearest stars.}
\bibinfo{journal}{{\em ApJ}} \bibinfo{volume}{95}: \bibinfo{pages}{201}.

\bibtype{Article}%
\bibitem[Kumar(1963)]{Kuma63}
\bibinfo{author}{Kumar SS} (\bibinfo{year}{1963}).
\bibinfo{title}{The structure of stars of very low mass.}
\bibinfo{journal}{{\em ApJ}} \bibinfo{volume}{137}: \bibinfo{pages}{1121}.

\bibtype{Article}%
\bibitem[Lamers et al.(1998)]{Lameetal98}
\bibinfo{author}{Lamers HJGLM}, \bibinfo{author}{Zickgraf FJ},
  \bibinfo{author}{de~Winter D}, \bibinfo{author}{Houziaux L} and
  \bibinfo{author}{Zorec J} (\bibinfo{year}{1998}).
\bibinfo{title}{An improved classification of \uppercase{B}[e]-type stars}.
\bibinfo{journal}{{\em A\&A}} \bibinfo{volume}{340}: \bibinfo{pages}{117--128}.

\bibtype{Article}%
\bibitem[Leavitt and Pickering(1912)]{LeavPick12}
\bibinfo{author}{Leavitt HS} and  \bibinfo{author}{Pickering EC}
  (\bibinfo{year}{1912}).
\bibinfo{title}{Periods of 25 variable stars in the \uppercase{S}mall
  \uppercase{M}agellanic \uppercase{C}loud.}
\bibinfo{journal}{{\em Harvard College Observatory Circular}}
  \bibinfo{volume}{173}: \bibinfo{pages}{1--3}.

\bibtype{Article}%
\bibitem[Lebzelter et al.(2019)]{Lebzetal19}
\bibinfo{author}{Lebzelter T}, \bibinfo{author}{Trabucchi M},
  \bibinfo{author}{Mowlavi N}, \bibinfo{author}{Wood PR},
  \bibinfo{author}{Marigo P}, \bibinfo{author}{Pastorelli G} and
  \bibinfo{author}{Lecoeur-Ta{\"\i}bi I} (\bibinfo{year}{2019}).
\bibinfo{title}{Period-luminosity diagram of long period variables in the
  \uppercase{M}agellanic \uppercase{C}louds. \uppercase{N}ew aspects revealed
  from \uppercase{G}aia \uppercase{D}ata \uppercase{R}elease 2}.
\bibinfo{journal}{{\em A\&A}} \bibinfo{volume}{631}, \bibinfo{eid}{A24}.

\bibtype{Article}%
\bibitem[Lee(1984)]{Lee84}
\bibinfo{author}{Lee SG} (\bibinfo{year}{1984}).
\bibinfo{title}{Spectral classification of high-proper-motion stars.}
\bibinfo{journal}{{\em AJ}} \bibinfo{volume}{89}: \bibinfo{pages}{702--719}.

\bibtype{Article}%
\bibitem[Lennon(1997)]{Lenn97}
\bibinfo{author}{Lennon DJ} (\bibinfo{year}{1997}).
\bibinfo{title}{Revised spectral types for 64 \uppercase{B}-supergiants in the
  \uppercase{S}mall \uppercase{M}agellanic \uppercase{C}loud: metallicity
  effects.}
\bibinfo{journal}{{\em A\&A}} \bibinfo{volume}{317}: \bibinfo{pages}{871--882}.

\bibtype{Inproceedings}%
\bibitem[Lennon(1999)]{Lenn99}
\bibinfo{author}{Lennon DJ} (\bibinfo{year}{1999}), \bibinfo{title}{Spectral
  morphology and classification of massive stars in the \uppercase{S}mall
  \uppercase{M}agellanic \uppercase{C}loud: effects of metallicity.},
  \bibinfo{editor}{Morrell NI}, \bibinfo{editor}{Niemela VS} and
  \bibinfo{editor}{Barb{\'a} RH}, (Eds.), \bibinfo{booktitle}{Rev. Mex. Astron.
  Astrof{\'\i}s. (conference series)}, \bibinfo{volume}{8},
  \bibinfo{pages}{21--28}.

\bibtype{Article}%
\bibitem[L{\'e}pine and Gaidos(2011)]{LepiGaid11}
\bibinfo{author}{L{\'e}pine S} and  \bibinfo{author}{Gaidos E}
  (\bibinfo{year}{2011}).
\bibinfo{title}{An all-sky catalog of bright \uppercase{M} dwarfs}.
\bibinfo{journal}{{\em AJ}} \bibinfo{volume}{142} (\bibinfo{number}{4}),
  \bibinfo{eid}{138}.

\bibtype{Article}%
\bibitem[L{\'e}pine et al.(2003)]{Lepietal03}
\bibinfo{author}{L{\'e}pine S}, \bibinfo{author}{Rich RM} and
  \bibinfo{author}{Shara MM} (\bibinfo{year}{2003}).
\bibinfo{title}{Spectroscopy of new high proper motion stars in the northern
  sky. \uppercase{I. N}ew nearby stars, new high-velocity stars, and an
  enhanced classification scheme for \uppercase{M} dwarfs}.
\bibinfo{journal}{{\em AJ}} \bibinfo{volume}{125} (\bibinfo{number}{3}):
  \bibinfo{pages}{1598--1622}.

\bibtype{Article}%
\bibitem[L{\'e}pine et al.(2007)]{Lepietal07}
\bibinfo{author}{L{\'e}pine S}, \bibinfo{author}{Rich RM} and
  \bibinfo{author}{Shara MM} (\bibinfo{year}{2007}).
\bibinfo{title}{Revised metallicity classes for low-mass stars:
  \uppercase{D}warfs (d\uppercase{M}), subdwarfs (sd\uppercase{M}), extreme
  subdwarfs (esd\uppercase{M}), and ultrasubdwarfs (usd\uppercase{M})}.
\bibinfo{journal}{{\em ApJ}} \bibinfo{volume}{669} (\bibinfo{number}{2}):
  \bibinfo{pages}{1235--1247}.

\bibtype{Article}%
\bibitem[L{\'e}pine et al.(2013)]{Lepietal13}
\bibinfo{author}{L{\'e}pine S}, \bibinfo{author}{Hilton EJ},
  \bibinfo{author}{Mann AW}, \bibinfo{author}{Wilde M},
  \bibinfo{author}{Rojas-Ayala B}, \bibinfo{author}{Cruz KL} and
  \bibinfo{author}{Gaidos E} (\bibinfo{year}{2013}).
\bibinfo{title}{A spectroscopic catalog of the brightest (\uppercase{J} $<$ 9)
  \uppercase{M} dwarfs in the northern sky}.
\bibinfo{journal}{{\em AJ}} \bibinfo{volume}{145} (\bibinfo{number}{4}),
  \bibinfo{eid}{102}.

\bibtype{Article}%
\bibitem[Lesh(1968)]{Lesh68}
\bibinfo{author}{Lesh JR} (\bibinfo{year}{1968}).
\bibinfo{title}{The kinematics of the \uppercase{G}ould \uppercase{B}elt: an
  expanding group?}
\bibinfo{journal}{{\em ApJS}} \bibinfo{volume}{17}: \bibinfo{pages}{371--444}.

\bibtype{Article}%
\bibitem[Luhman et al.(2024)]{Luhmetal24}
\bibinfo{author}{Luhman KL}, \bibinfo{author}{Tremblin P},
  \bibinfo{author}{Alves~de Oliveira C}, \bibinfo{author}{Birkmann SM},
  \bibinfo{author}{Baraffe I}, \bibinfo{author}{Chabrier G},
  \bibinfo{author}{Manjavacas E}, \bibinfo{author}{Parker RJ} and
  \bibinfo{author}{Valenti J} (\bibinfo{year}{2024}).
\bibinfo{title}{\uppercase{JWST/NIRS}pec observations of the coldest known
  brown dwarf}.
\bibinfo{journal}{{\em AJ}} \bibinfo{volume}{167} (\bibinfo{number}{1}),
  \bibinfo{eid}{5}.

\bibtype{Article}%
\bibitem[Luyten(1922)]{Luyt22}
\bibinfo{author}{Luyten WJ} (\bibinfo{year}{1922}).
\bibinfo{title}{A list of new and suspected dwarfs of class \uppercase{M}}.
\bibinfo{journal}{{\em PASP}} \bibinfo{volume}{34} (\bibinfo{number}{202}):
  \bibinfo{pages}{342}.

\bibtype{Inproceedings}%
\bibitem[Maeder(1996)]{Maed96b}
\bibinfo{author}{Maeder A} (\bibinfo{year}{1996}), \bibinfo{title}{The
  \uppercase{C}onti scenario for forming \uppercase{WR} stars: past, present
  and future}, \bibinfo{booktitle}{WR stars in the Framework of Stellar
  Evolution}, pp.~\bibinfo{pages}{39}.

\bibtype{Article}%
\bibitem[Ma{\'\i}z~Apell{\'a}niz(2024)]{Maiz24}
\bibinfo{author}{Ma{\'\i}z~Apell{\'a}niz J} (\bibinfo{year}{2024}).
\bibinfo{title}{Extinction, the elephant in the room that hinders optical
  \uppercase{G}alactic observations} : \bibinfo{pages}{arXiv:2401.01116}.

\bibtype{Article}%
\bibitem[Ma{\'\i}z~Apell{\'a}niz and Weiler(2018)]{MaizWeil18}
\bibinfo{author}{Ma{\'\i}z~Apell{\'a}niz J} and  \bibinfo{author}{Weiler M}
  (\bibinfo{year}{2018}).
\bibinfo{title}{Reanalysis of the \uppercase{G}aia \uppercase{D}ata
  \uppercase{R}elease 2 photometric sensitivity curves using
  \uppercase{HST/STIS} spectrophotometry}.
\bibinfo{journal}{{\em A\&A}} \bibinfo{volume}{619}, \bibinfo{eid}{A180}.

\bibtype{Inproceedings}%
\bibitem[Ma{\'\i}z~Apell{\'a}niz et al.(2011)]{Maizetal11}
\bibinfo{author}{Ma{\'\i}z~Apell{\'a}niz J}, \bibinfo{author}{Sota A},
  \bibinfo{author}{Walborn NR}, \bibinfo{author}{Alfaro EJ},
  \bibinfo{author}{Barb{\'a} RH}, \bibinfo{author}{Morrell NI},
  \bibinfo{author}{Gamen RC} and  \bibinfo{author}{Arias JI}
  (\bibinfo{year}{2011}), \bibinfo{title}{The \uppercase{G}alactic
  \uppercase{O}-\uppercase{S}tar \uppercase{S}pectroscopic survey
  (\uppercase{GOSSS})}, \bibinfo{booktitle}{Highlights of Spanish Astrophysics
  VI},  \bibinfo{pages}{467--472}, \eprint{arXiv:1010.5680}.

\bibtype{Inproceedings}%
\bibitem[Ma{\'\i}z~Apell{\'a}niz et al.(2012)]{Maizetal12}
\bibinfo{author}{Ma{\'\i}z~Apell{\'a}niz J}, \bibinfo{author}{Pellerin A},
  \bibinfo{author}{Barb{\'a} RH}, \bibinfo{author}{Sim{\'o}n-D{\'\i}az S},
  \bibinfo{author}{Alfaro EJ}, \bibinfo{author}{Morrell NI},
  \bibinfo{author}{Sota A}, \bibinfo{author}{Penad{\'e}s~Ordaz M} and
  \bibinfo{author}{Gallego~Calvente AT} (\bibinfo{year}{2012}),
  \bibinfo{title}{The \uppercase{G}alactic \uppercase{O}-\uppercase{S}tar
  \uppercase{S}pectroscopic (\uppercase{GOSSS}) and \uppercase{N}orthern
  \uppercase{M}assive \uppercase{D}im \uppercase{S}tars
  (\uppercase{N}o\uppercase{M}a\uppercase{DS}) surveys, the
  \uppercase{G}alactic \uppercase{O}-\uppercase{S}tar \uppercase{C}atalog
  (\uppercase{GOSC}), and \uppercase{M}arxist \uppercase{G}host
  \uppercase{B}uster (\uppercase{MGB})}, \bibinfo{editor}{Drissen L},
  \bibinfo{editor}{Robert C}, \bibinfo{editor}{St-Louis N} and
  \bibinfo{editor}{Moffat AFJ}, (Eds.), \bibinfo{booktitle}{Astronomical
  Society of the Pacific Conference Series}, \bibinfo{volume}{465}, pp.
  \bibinfo{pages}{484}.

\bibtype{Inproceedings}%
\bibitem[Ma{\'\i}z~Apell{\'a}niz et al.(2014)]{Maizetal14b}
\bibinfo{author}{Ma{\'\i}z~Apell{\'a}niz J}, \bibinfo{author}{Sota A},
  \bibinfo{author}{Barb{\'a} RH}, \bibinfo{author}{Morrell NI},
  \bibinfo{author}{Pellerin A}, \bibinfo{author}{Alfaro EJ} and
  \bibinfo{author}{Sim{\'o}n-D{\'\i}az S} (\bibinfo{year}{2014}),
  \bibinfo{title}{First results from a study of \uppercase{DIB}s with thousands
  of high-quality massive-star spectra}, \bibinfo{booktitle}{IAUS},
  \bibinfo{volume}{297},  \bibinfo{pages}{117--120}.

\bibtype{Article}%
\bibitem[Ma{\'\i}z~Apell{\'a}niz et al.(2016)]{Maizetal16}
\bibinfo{author}{Ma{\'\i}z~Apell{\'a}niz J}, \bibinfo{author}{Sota A},
  \bibinfo{author}{Arias JI}, \bibinfo{author}{Barb{\'a} RH},
  \bibinfo{author}{Walborn NR}, \bibinfo{author}{Sim{\'o}n-D{\'\i}az S},
  \bibinfo{author}{Negueruela I}, \bibinfo{author}{Marco A},
  \bibinfo{author}{Le{\~a}o JRS}, \bibinfo{author}{Herrero A},
  \bibinfo{author}{Gamen RC} and  \bibinfo{author}{Alfaro EJ}
  (\bibinfo{year}{2016}), \bibinfo{month}{May}.
\bibinfo{title}{The \uppercase{G}alactic \uppercase{O-S}tar
  \uppercase{S}pectroscopic \uppercase{S}urvey (\uppercase{GOSSS}).
  \uppercase{III}. 142 additional \uppercase{O}-type systems.}
\bibinfo{journal}{{\em ApJS}} \bibinfo{volume}{224}, \bibinfo{eid}{4}.

\bibtype{Inproceedings}%
\bibitem[Ma{\'\i}z~Apell{\'a}niz et al.(2019{\natexlab{a}})]{Maizetal19a}
\bibinfo{author}{Ma{\'\i}z~Apell{\'a}niz J},
  \bibinfo{author}{Trigueros~P{\'a}ez E},
  \bibinfo{author}{Jim{\'e}nez~Mart{\'\i}nez I}, \bibinfo{author}{Barb{\'a}
  RH}, \bibinfo{author}{Sim{\'o}n-D{\'\i}az S}, \bibinfo{author}{Pellerin A},
  \bibinfo{author}{Negueruela I} and  \bibinfo{author}{Souza~Le{\~a}o JR}
  (\bibinfo{year}{2019}{\natexlab{a}}),
  \bibinfo{title}{\uppercase{L}i\uppercase{L}i\uppercase{M}a\uppercase{R}lin, a
  \textbf{\uppercase{l}i}brary of \textbf{\uppercase{l}i}braries of
  \textbf{\uppercase{m}a}ssive-\uppercase{S}tar
  \uppercase{H}igh-\textbf{\uppercase{r}}eso\textbf{l}ut\textbf{i}o\textbf{n}
  \uppercase{S}pectra with applications to \uppercase{OWN}, \uppercase{MONOS},
  and \uppercase{C}oll\uppercase{DIB}s}, \bibinfo{booktitle}{Highlights of
  Spanish Astrophysics X}, pp. \bibinfo{pages}{420}, \eprint{arXiv:1810.10943}.

\bibtype{Article}%
\bibitem[Ma{\'\i}z~Apell{\'a}niz et al.(2019{\natexlab{b}})]{Maizetal19b}
\bibinfo{author}{Ma{\'\i}z~Apell{\'a}niz J},
  \bibinfo{author}{Trigueros~P{\'a}ez E}, \bibinfo{author}{Negueruela I},
  \bibinfo{author}{Barb{\'a} RH}, \bibinfo{author}{Sim{\'o}n-D{\'\i}az S},
  \bibinfo{author}{Lorenzo J}, \bibinfo{author}{Sota A}, \bibinfo{author}{Gamen
  RC}, \bibinfo{author}{Fari{\~n}a C}, \bibinfo{author}{Salas J},
  \bibinfo{author}{Caballero JA}, \bibinfo{author}{Morrell NI},
  \bibinfo{author}{Pellerin A}, \bibinfo{author}{Alfaro EJ},
  \bibinfo{author}{Herrero A}, \bibinfo{author}{Arias JI} and
  \bibinfo{author}{Marco A} (\bibinfo{year}{2019}{\natexlab{b}}).
\bibinfo{title}{\uppercase{MONOS}: Multiplicity of northern \uppercase{O}-type
  spectroscopic systems. \uppercase{I. P}roject description and spectral
  classifications and visual multiplicity of previously known objects}.
\bibinfo{journal}{{\em A\&A}} \bibinfo{volume}{626}, \bibinfo{eid}{A20}.

\bibtype{Article}%
\bibitem[Ma{\'\i}z~Apell{\'a}niz et al.(2023)]{Maizetal23}
\bibinfo{author}{Ma{\'\i}z~Apell{\'a}niz J}, \bibinfo{author}{Holgado G},
  \bibinfo{author}{Pantaleoni~Gonz{\'a}lez M} and  \bibinfo{author}{Caballero
  JA} (\bibinfo{year}{2023}).
\bibinfo{title}{Stellar variability in \uppercase{G}aia \uppercase{DR3. I.
  T}hree-band photometric dispersions for 145 million sources}.
\bibinfo{journal}{{\em A\&A}} \bibinfo{volume}{677}, \bibinfo{eid}{A137}.

\bibtype{Article}%
\bibitem[Marco and Negueruela(2013)]{MarcNegu13}
\bibinfo{author}{Marco A} and  \bibinfo{author}{Negueruela I}
  (\bibinfo{year}{2013}).
\bibinfo{title}{\uppercase{NGC} 7419 as a template for red supergiant
  clusters}.
\bibinfo{journal}{{\em A\&A}} \bibinfo{volume}{552}, \bibinfo{eid}{A92}.

\bibtype{Article}%
\bibitem[Marfil~et al.(2021)]{Marfetal21}
\bibinfo{author}{Marfil~et al. } (\bibinfo{year}{2021}).
\bibinfo{title}{The \uppercase{CARMENES} search for exoplanets around
  \uppercase{M} dwarfs. stellar atmospheric parameters of target stars with
  \uppercase{S}te\uppercase{P}ar\uppercase{S}yn}.
\bibinfo{journal}{{\em A\&A}} \bibinfo{volume}{656}, \bibinfo{eid}{A162}.

\bibtype{Article}%
\bibitem[Martin and Kun(1996)]{MartKun96}
\bibinfo{author}{Martin EL} and  \bibinfo{author}{Kun M}
  (\bibinfo{year}{1996}).
\bibinfo{title}{Spectroscopy of possible \uppercase{H}{\ensuremath{\alpha}}
  emission stars in regions of high \uppercase{G}alactic latitude molecular
  clouds.}
\bibinfo{journal}{{\em A\&AS}} \bibinfo{volume}{116}:
  \bibinfo{pages}{467--471}.

\bibtype{Article}%
\bibitem[Mart{\'\i}n et al.(1996)]{Martetal96}
\bibinfo{author}{Mart{\'\i}n EL}, \bibinfo{author}{Rebolo R} and
  \bibinfo{author}{Zapatero-Osorio MR} (\bibinfo{year}{1996}).
\bibinfo{title}{Spectroscopy of new substellar candidates in the
  \uppercase{P}leiades: \uppercase{T}oward a spectral sequence for young brown
  dwarfs}.
\bibinfo{journal}{{\em ApJ}} \bibinfo{volume}{469}: \bibinfo{pages}{706}.

\bibtype{Article}%
\bibitem[Mart{\'\i}n et al.(1997)]{Martetal97}
\bibinfo{author}{Mart{\'\i}n EL}, \bibinfo{author}{Basri G},
  \bibinfo{author}{Delfosse X} and  \bibinfo{author}{Forveille T}
  (\bibinfo{year}{1997}).
\bibinfo{title}{Keck \uppercase{HIRES} spectra of the brown dwarf
  \uppercase{DENIS-P J1228.2-1547}}.
\bibinfo{journal}{{\em A\&A}} \bibinfo{volume}{327}: \bibinfo{pages}{L29--L32}.

\bibtype{Article}%
\bibitem[Mart{\'\i}n et al.(1999)]{Martetal99}
\bibinfo{author}{Mart{\'\i}n EL}, \bibinfo{author}{Delfosse X},
  \bibinfo{author}{Basri G}, \bibinfo{author}{Goldman B},
  \bibinfo{author}{Forveille T} and  \bibinfo{author}{Zapatero~Osorio MR}
  (\bibinfo{year}{1999}).
\bibinfo{title}{Spectroscopic classification of late-\uppercase{M} and
  \uppercase{L} field dwarfs}.
\bibinfo{journal}{{\em AJ}} \bibinfo{volume}{118} (\bibinfo{number}{5}):
  \bibinfo{pages}{2466--2482}.

\bibtype{Article}%
\bibitem[Mart{\'\i}n et al.(2024)]{Martetal24}
\bibinfo{author}{Mart{\'\i}n EL}, \bibinfo{author}{Zhang JY},
  \bibinfo{author}{Lanchas H}, \bibinfo{author}{Lodieu N},
  \bibinfo{author}{Shahbaz T} and  \bibinfo{author}{Pavlenko YV}
  (\bibinfo{year}{2024}).
\bibinfo{title}{Optical properties of \uppercase{Y} dwarfs observed with the
  \uppercase{G}ran \uppercase{T}elescopio \uppercase{C}anarias}.
\bibinfo{journal}{{\em A\&A}} \bibinfo{volume}{686}, \bibinfo{eid}{A73}.

\bibtype{Article}%
\bibitem[Martins and Palacios(2017)]{MartPala17}
\bibinfo{author}{Martins F} and  \bibinfo{author}{Palacios A}
  (\bibinfo{year}{2017}).
\bibinfo{title}{Spectroscopic evolution of massive stars on the main sequence}.
\bibinfo{journal}{{\em A\&A}} \bibinfo{volume}{598}, \bibinfo{eid}{A56}.

\bibtype{Article}%
\bibitem[Maury and Pickering(1897)]{MaurPick97}
\bibinfo{author}{Maury AC} and  \bibinfo{author}{Pickering EC}
  (\bibinfo{year}{1897}).
\bibinfo{title}{Spectra of bright stars photographed with the 11-inch
  \uppercase{D}raper telescope as part of the \uppercase{H}enry
  \uppercase{D}raper memorial.}
\bibinfo{journal}{{\em Annals of Harvard College Observatory}}
  \bibinfo{volume}{28}: \bibinfo{pages}{1--128}.

\bibtype{Article}%
\bibitem[Mayor and Queloz(1995)]{MayoQuel95}
\bibinfo{author}{Mayor M} and  \bibinfo{author}{Queloz D}
  (\bibinfo{year}{1995}).
\bibinfo{title}{A \uppercase{J}upiter-mass companion to a solar-type star}.
\bibinfo{journal}{{\em Nature}} \bibinfo{volume}{378} (\bibinfo{number}{6555}):
  \bibinfo{pages}{355--359}.

\bibtype{Article}%
\bibitem[Mermilliod and Mayor(1989)]{MermMayo89}
\bibinfo{author}{Mermilliod JC} and  \bibinfo{author}{Mayor M}
  (\bibinfo{year}{1989}).
\bibinfo{title}{Red giants in open clusters. \uppercase{I. B}inarity and
  stellar evolution in five \uppercase{H}yades-generation clusters:
  \uppercase{NGC} 2447, 2539, 2632, 6633 and 6940.}
\bibinfo{journal}{{\em A\&A}} \bibinfo{volume}{219}: \bibinfo{pages}{125--141}.

\bibtype{Article}%
\bibitem[Morgan(1937)]{Morg37}
\bibinfo{author}{Morgan WW} (\bibinfo{year}{1937}).
\bibinfo{title}{On the spectral classification of the stars of types
  \uppercase{A} to \uppercase{K}}.
\bibinfo{journal}{{\em ApJ}} \bibinfo{volume}{85}: \bibinfo{pages}{380}.

\bibtype{Article}%
\bibitem[Morgan(1938{\natexlab{a}})]{Morg38a}
\bibinfo{author}{Morgan WW} (\bibinfo{year}{1938}{\natexlab{a}}).
\bibinfo{title}{On the determination of color indices of stars from a
  classification of their spectra}.
\bibinfo{journal}{{\em ApJ}} \bibinfo{volume}{87}: \bibinfo{pages}{460}.

\bibtype{Article}%
\bibitem[Morgan(1938{\natexlab{b}})]{Morg38b}
\bibinfo{author}{Morgan WW} (\bibinfo{year}{1938}{\natexlab{b}}).
\bibinfo{title}{On the spectral types and luminosities of the \uppercase{M}
  dwarfs}.
\bibinfo{journal}{{\em ApJ}} \bibinfo{volume}{87}: \bibinfo{pages}{589}.

\bibtype{Article}%
\bibitem[Morgan(1951)]{Morg51}
\bibinfo{author}{Morgan WW} (\bibinfo{year}{1951}).
\bibinfo{title}{Application of the principle of natural groups to the
  classification of stellar spectra.}
\bibinfo{journal}{{\em Publications of Michigan Observatory}}
  \bibinfo{volume}{10}: \bibinfo{pages}{33}.

\bibtype{Article}%
\bibitem[Morgan and Hiltner(1965)]{MorgHilt65}
\bibinfo{author}{Morgan WW} and  \bibinfo{author}{Hiltner WA}
  (\bibinfo{year}{1965}).
\bibinfo{title}{Studies in spectral classification. \uppercase{I. T}he
  \uppercase{HR D}iagram of the \uppercase{H}yades.}
\bibinfo{journal}{{\em ApJ}} \bibinfo{volume}{141}: \bibinfo{pages}{177}.

\bibtype{Article}%
\bibitem[Morgan and Keenan(1973)]{MorgKeen73}
\bibinfo{author}{Morgan WW} and  \bibinfo{author}{Keenan PC}
  (\bibinfo{year}{1973}).
\bibinfo{title}{Spectral classification}.
\bibinfo{journal}{{\em ARA\&A}} \bibinfo{volume}{11}: \bibinfo{pages}{29}.

\bibtype{Book}%
\bibitem[Morgan et al.(1943)]{Morgetal43}
\bibinfo{author}{Morgan WW}, \bibinfo{author}{Keenan PC} and
  \bibinfo{author}{Kellman E} (\bibinfo{year}{1943}).
\bibinfo{title}{An atlas of stellar spectra, with an outline of spectral
  classification}.

\bibtype{Book}%
\bibitem[Morgan et al.(1978)]{Morgetal78}
\bibinfo{author}{Morgan WW}, \bibinfo{author}{Abt HA} and
  \bibinfo{author}{Tapscott JW} (\bibinfo{year}{1978}).
\bibinfo{title}{Revised \uppercase{MK} Spectral Atlas for stars earlier than
  the \uppercase{S}un}.

\bibtype{Article}%
\bibitem[Munari(2019)]{Muna19}
\bibinfo{author}{Munari U} (\bibinfo{year}{2019}).
\bibinfo{title}{The symbiotic stars} : \bibinfo{pages}{arXiv:1909.01389}.

\bibtype{Inproceedings}%
\bibitem[Mu{\~n}oz et al.(2020)]{Munoetal20}
\bibinfo{author}{Mu{\~n}oz MS}, \bibinfo{author}{Wade GA},
  \bibinfo{author}{Faes D} and  \bibinfo{author}{Carciofi A}
  (\bibinfo{year}{2020}), \bibinfo{title}{The photometric and polarimetric
  variability of magnetic \uppercase{O}-type stars}, \bibinfo{editor}{Wade G},
  \bibinfo{editor}{Alecian E}, \bibinfo{editor}{Bohlender D} and
  \bibinfo{editor}{Sigut A}, (Eds.), \bibinfo{booktitle}{Stellar Magnetism: A
  Workshop in Honour of the Career and Contributions of John D. Landstreet},
  \bibinfo{volume}{11},  \bibinfo{pages}{148--155}, \eprint{arXiv:2004.13594}.

\bibtype{Article}%
\bibitem[Nakajima et al.(1995)]{Nakaetal95}
\bibinfo{author}{Nakajima T}, \bibinfo{author}{Oppenheimer BR},
  \bibinfo{author}{Kulkarni SR}, \bibinfo{author}{Golimowski DA},
  \bibinfo{author}{Matthews K} and  \bibinfo{author}{Durrance ST}
  (\bibinfo{year}{1995}).
\bibinfo{title}{Discovery of a cool brown dwarf}.
\bibinfo{journal}{{\em Nature}} \bibinfo{volume}{378} (\bibinfo{number}{6556}):
  \bibinfo{pages}{463--465}.

\bibtype{Article}%
\bibitem[Naz{\'e} et al.(2016)]{Nazeetal16a}
\bibinfo{author}{Naz{\'e} Y}, \bibinfo{author}{Barb{\'a} R},
  \bibinfo{author}{Bagnulo S}, \bibinfo{author}{Morrell N},
  \bibinfo{author}{Gamen R}, \bibinfo{author}{Petit V} and
  \bibinfo{author}{Neiner C} (\bibinfo{year}{2016}).
\bibinfo{title}{The puzzling properties of the magnetic \uppercase{O} star
  \uppercase{T}r16-22}.
\bibinfo{journal}{{\em A\&A}} \bibinfo{volume}{596}, \bibinfo{eid}{A44}.

\bibtype{Article}%
\bibitem[Naz{\'e} et al.(2017)]{Nazeetal17b}
\bibinfo{author}{Naz{\'e} Y}, \bibinfo{author}{Neiner C},
  \bibinfo{author}{Grunhut J}, \bibinfo{author}{Bagnulo S},
  \bibinfo{author}{Alecian E}, \bibinfo{author}{Rauw G}, \bibinfo{author}{Wade
  GA} and  \bibinfo{author}{{BinaMIcS Collaboration}} (\bibinfo{year}{2017}).
\bibinfo{title}{How unique is \uppercase{P}laskett's star? \uppercase{A} search
  for organized magnetic fields in short period, interacting or
  post-interaction massive binary systems}.
\bibinfo{journal}{{\em MNRAS}} \bibinfo{volume}{467}:
  \bibinfo{pages}{501--511}.

\bibtype{Article}%
\bibitem[Negueruela et al.(2004)]{Neguetal04}
\bibinfo{author}{Negueruela I}, \bibinfo{author}{Steele IA} and
  \bibinfo{author}{Bernabeu G} (\bibinfo{year}{2004}).
\bibinfo{title}{On the class of \uppercase{O}e stars}.
\bibinfo{journal}{{\em AN}} \bibinfo{volume}{325}: \bibinfo{pages}{749--760}.

\bibtype{Article}%
\bibitem[Negueruela et al.(2012)]{Neguetal12}
\bibinfo{author}{Negueruela I}, \bibinfo{author}{Marco A},
  \bibinfo{author}{Gonz{\'a}lez-Fern{\'a}ndez C},
  \bibinfo{author}{Jim{\'e}nez-Esteban F}, \bibinfo{author}{Clark JS},
  \bibinfo{author}{Garcia M} and  \bibinfo{author}{Solano E}
  (\bibinfo{year}{2012}).
\bibinfo{title}{Red supergiants around the obscured open cluster
  \uppercase{S}tephenson 2}.
\bibinfo{journal}{{\em A\&A}} \bibinfo{volume}{547}, \bibinfo{eid}{A15}.

\bibtype{Article}%
\bibitem[Negueruela et al.(2024)]{Neguetal24}
\bibinfo{author}{Negueruela I}, \bibinfo{author}{Sim{\'o}n-D{\'\i}az S},
  \bibinfo{author}{de~Burgos A}, \bibinfo{author}{Casasbuenas A} and
  \bibinfo{author}{Beck PG} (\bibinfo{year}{2024}).
\bibinfo{title}{The \uppercase{IACOB} project: \uppercase{XII. N}ew grid of
  northern standards for the spectral classification of \uppercase{B}-type
  stars}.
\bibinfo{journal}{{\em A\&A}} \bibinfo{volume}{690}, \bibinfo{eid}{A176}.

\bibtype{Article}%
\bibitem[Passegger et al.(2022)]{Passetal22}
\bibinfo{author}{Passegger VM}, \bibinfo{author}{Bello-Garc{\'\i}a A},
  \bibinfo{author}{Ordieres-Mer{\'e} J}, \bibinfo{author}{Antoniadis-Karnavas
  A}, \bibinfo{author}{Marfil E}, \bibinfo{author}{Duque-Arribas C},
  \bibinfo{author}{Amado PJ}, \bibinfo{author}{Delgado-Mena E},
  \bibinfo{author}{Montes D}, \bibinfo{author}{Rojas-Ayala B},
  \bibinfo{author}{Schweitzer A}, \bibinfo{author}{Tabernero HM},
  \bibinfo{author}{B{\'e}jar VJS}, \bibinfo{author}{Caballero JA},
  \bibinfo{author}{Hatzes AP}, \bibinfo{author}{Henning T},
  \bibinfo{author}{Pedraz S}, \bibinfo{author}{Quirrenbach A},
  \bibinfo{author}{Reiners A} and  \bibinfo{author}{Ribas I}
  (\bibinfo{year}{2022}).
\bibinfo{title}{Metallicities in \uppercase{M} dwarfs:
  \uppercase{I}nvestigating different determination techniques}.
\bibinfo{journal}{{\em A\&A}} \bibinfo{volume}{658}, \bibinfo{eid}{A194}.

\bibtype{Article}%
\bibitem[Paunzen et al.(2021)]{Paunetal21}
\bibinfo{author}{Paunzen E}, \bibinfo{author}{H{\"u}mmerich S} and
  \bibinfo{author}{Bernhard K} (\bibinfo{year}{2021}).
\bibinfo{title}{New mercury-manganese stars and candidates from
  \uppercase{LAMOST DR4}}.
\bibinfo{journal}{{\em A\&A}} \bibinfo{volume}{645}, \bibinfo{eid}{A34}.

\bibtype{Phdthesis}%
\bibitem[Payne(1925)]{Payn25}
\bibinfo{author}{Payne CH} (\bibinfo{year}{1925}).
\bibinfo{title}{Stellar Atmospheres; a Contribution to the Observational Study
  of High Temperature in the Reversing Layers of Stars.}
\bibinfo{comment}{Ph.D. thesis}, \bibinfo{school}{Radcliffe College}.

\bibtype{Article}%
\bibitem[Pickering(1890)]{Pick90}
\bibinfo{author}{Pickering EC} (\bibinfo{year}{1890}).
\bibinfo{title}{The \uppercase{D}raper \uppercase{C}atalogue of stellar spectra
  photographed with the 8-inch \uppercase{B}ache telescope as a part of the
  \uppercase{H}enry \uppercase{D}raper memorial}.
\bibinfo{journal}{{\em Annals of Harvard College Observatory}}
  \bibinfo{volume}{27}: \bibinfo{pages}{1--388}.

\bibtype{Article}%
\bibitem[Pickering(1897)]{Pick97}
\bibinfo{author}{Pickering EC} (\bibinfo{year}{1897}).
\bibinfo{title}{The spectrum of {\ensuremath{\zeta}} \uppercase{P}uppis}.
\bibinfo{journal}{{\em Harvard College Observatory Circular}}
  \bibinfo{volume}{16}: \bibinfo{pages}{1--2}.

\bibtype{Article}%
\bibitem[Pickering and Fleming(1897)]{PickFlem97}
\bibinfo{author}{Pickering EC} and  \bibinfo{author}{Fleming M}
  (\bibinfo{year}{1897}).
\bibinfo{title}{Miscellaneous investigations of the \uppercase{H}enry
  \uppercase{D}raper memorial}.
\bibinfo{journal}{{\em Annals of Harvard College Observatory}}
  \bibinfo{volume}{26}: \bibinfo{pages}{193--P.XI.2}.

\bibtype{Article}%
\bibitem[Porter and Rivinius(2003)]{PortRivi03}
\bibinfo{author}{Porter JM} and  \bibinfo{author}{Rivinius T}
  (\bibinfo{year}{2003}).
\bibinfo{title}{Classical \uppercase{B}e stars}.
\bibinfo{journal}{{\em PASP}} \bibinfo{volume}{115}:
  \bibinfo{pages}{1153--1170}.

\bibtype{Article}%
\bibitem[Prusti~et al.(2016)]{Prusetal16}
\bibinfo{author}{Prusti~et al. } (\bibinfo{year}{2016}).
\bibinfo{title}{The \uppercase{G}aia mission}.
\bibinfo{journal}{{\em A\&A}} \bibinfo{volume}{595}, \bibinfo{eid}{A1}.

\bibtype{Article}%
\bibitem[Randich~et al.(2022)]{Randetal22}
\bibinfo{author}{Randich~et al. } (\bibinfo{year}{2022}).
\bibinfo{title}{The \uppercase{G}aia-\uppercase{ESO} public spectroscopic
  survey: \uppercase{I}mplementation, data products, open cluster survey,
  science, and legacy}.
\bibinfo{journal}{{\em A\&A}} \bibinfo{volume}{666}, \bibinfo{eid}{A121}.

\bibtype{Article}%
\bibitem[Rebolo et al.(1995)]{Reboetal95}
\bibinfo{author}{Rebolo R}, \bibinfo{author}{Zapatero~Osorio MR} and
  \bibinfo{author}{Mart{\'\i}n EL} (\bibinfo{year}{1995}).
\bibinfo{title}{Discovery of a brown dwarf in the \uppercase{P}leiades star
  cluster}.
\bibinfo{journal}{{\em Nature}} \bibinfo{volume}{377} (\bibinfo{number}{6545}):
  \bibinfo{pages}{129--131}.

\bibtype{Article}%
\bibitem[Recio-Blanco~et al.(2023)]{Recietal23}
\bibinfo{author}{Recio-Blanco~et al. } (\bibinfo{year}{2023}).
\bibinfo{title}{\uppercase{G}aia \uppercase{D}ata \uppercase{R}elease 3.
  \uppercase{A}nalysis of \uppercase{RVS} spectra using the general stellar
  parametriser from spectroscopy}.
\bibinfo{journal}{{\em A\&A}} \bibinfo{volume}{674}, \bibinfo{eid}{A29}.

\bibtype{Article}%
\bibitem[Reid et al.(1995)]{Reidetal95}
\bibinfo{author}{Reid IN}, \bibinfo{author}{Hawley SL} and
  \bibinfo{author}{Gizis JE} (\bibinfo{year}{1995}).
\bibinfo{title}{The \uppercase{P}alomar/\uppercase{MSU} nearby-star
  spectroscopic survey. \uppercase{I. T}he northern \uppercase{M} dwarfs -
  bandstrengths and kinematics}.
\bibinfo{journal}{{\em AJ}} \bibinfo{volume}{110}: \bibinfo{pages}{1838}.

\bibtype{Article}%
\bibitem[Reid et al.(2002)]{Reidetal02}
\bibinfo{author}{Reid IN}, \bibinfo{author}{Gizis JE} and
  \bibinfo{author}{Hawley SL} (\bibinfo{year}{2002}).
\bibinfo{title}{The \uppercase{P}alomar/\uppercase{MSU} nearby-star
  spectroscopic survey. \uppercase{IV. T}he luminosity function in the solar
  neighborhood and \uppercase{M} dwarf kinematics}.
\bibinfo{journal}{{\em AJ}} \bibinfo{volume}{124} (\bibinfo{number}{5}):
  \bibinfo{pages}{2721--2738}.

\bibtype{Article}%
\bibitem[Reiners~et al.(2018)]{Reinetal18}
\bibinfo{author}{Reiners~et al. } (\bibinfo{year}{2018}).
\bibinfo{title}{The \uppercase{CARMENES} search for exoplanets around
  \uppercase{M} dwarfs. \uppercase{HD} 147\,379 b: \uppercase{A} nearby
  \uppercase{N}eptune in the temperate zone of an early-\uppercase{M} dwarf}.
\bibinfo{journal}{{\em A\&A}} \bibinfo{volume}{609}, \bibinfo{eid}{L5}.

\bibtype{Article}%
\bibitem[Reyl{\'e} et al.(2021)]{Reyletal21}
\bibinfo{author}{Reyl{\'e} C}, \bibinfo{author}{Jardine K},
  \bibinfo{author}{Fouqu{\'e} P}, \bibinfo{author}{Caballero JA},
  \bibinfo{author}{Smart RL} and  \bibinfo{author}{Sozzetti A}
  (\bibinfo{year}{2021}).
\bibinfo{title}{The 10 parsec sample in the \uppercase{G}aia era}.
\bibinfo{journal}{{\em A\&A}} \bibinfo{volume}{650}, \bibinfo{eid}{A201}.

\bibtype{Article}%
\bibitem[Riaz et al.(2006)]{Riazetal06}
\bibinfo{author}{Riaz B}, \bibinfo{author}{Gizis JE} and
  \bibinfo{author}{Harvin J} (\bibinfo{year}{2006}).
\bibinfo{title}{Identification of new \uppercase{M} dwarfs in the solar
  neighborhood}.
\bibinfo{journal}{{\em AJ}} \bibinfo{volume}{132} (\bibinfo{number}{2}):
  \bibinfo{pages}{866--872}.

\bibtype{Article}%
\bibitem[Ribas~et al.(2023)]{Ribaetal23}
\bibinfo{author}{Ribas~et al. } (\bibinfo{year}{2023}).
\bibinfo{title}{The \uppercase{CARMENES} search for exoplanets around
  \uppercase{M} dwarfs. \uppercase{G}uaranteed time observations data release 1
  (2016-2020)}.
\bibinfo{journal}{{\em A\&A}} \bibinfo{volume}{670}, \bibinfo{eid}{A139}.

\bibtype{Article}%
\bibitem[Roman(1950)]{Roma50}
\bibinfo{author}{Roman NG} (\bibinfo{year}{1950}).
\bibinfo{title}{A correlation between the spectroscopic and dynamical
  characteristics of the late \uppercase{F} - and early \uppercase{G} - type
  stars.}
\bibinfo{journal}{{\em ApJ}} \bibinfo{volume}{112}: \bibinfo{pages}{554}.

\bibtype{Article}%
\bibitem[Rountree and Sonneborn(1991)]{RounSonn91}
\bibinfo{author}{Rountree J} and  \bibinfo{author}{Sonneborn G}
  (\bibinfo{year}{1991}).
\bibinfo{title}{Criteria for the spectral classification of \uppercase{B} stars
  in the ultraviolet}.
\bibinfo{journal}{{\em ApJ}} \bibinfo{volume}{369}: \bibinfo{pages}{515}.

\bibtype{Article}%
\bibitem[Royer et al.(2007)]{Royeetal07}
\bibinfo{author}{Royer F}, \bibinfo{author}{Zorec J} and
  \bibinfo{author}{G{\'o}mez AE} (\bibinfo{year}{2007}).
\bibinfo{title}{Rotational velocities of \uppercase{A}-type stars.
  \uppercase{III. V}elocity distributions}.
\bibinfo{journal}{{\em A\&A}} \bibinfo{volume}{463} (\bibinfo{number}{2}):
  \bibinfo{pages}{671--682}.

\bibtype{Article}%
\bibitem[Ruiz et al.(1997)]{Ruizetal97}
\bibinfo{author}{Ruiz MT}, \bibinfo{author}{Leggett SK} and
  \bibinfo{author}{Allard F} (\bibinfo{year}{1997}).
\bibinfo{title}{Kelu-1: \uppercase{A} free-floating brown dwarf in the solar
  neighborhood}.
\bibinfo{journal}{{\em ApJL}} \bibinfo{volume}{491} (\bibinfo{number}{2}):
  \bibinfo{pages}{L107--L110}.

\bibtype{Article}%
\bibitem[Saha(1921)]{Saha21}
\bibinfo{author}{Saha MN} (\bibinfo{year}{1921}).
\bibinfo{title}{On a physical theory of stellar spectra}.
\bibinfo{journal}{{\em Proceedings of the Royal Society of London Series A}}
  \bibinfo{volume}{99} (\bibinfo{number}{697}): \bibinfo{pages}{135--153}.

\bibtype{Article}%
\bibitem[Secchi(1866)]{Secc66}
\bibinfo{author}{Secchi A} (\bibinfo{year}{1866}).
\bibinfo{title}{Spectrum of {\ensuremath{\alpha}} \uppercase{O}rionis}.
\bibinfo{journal}{{\em MNRAS}} \bibinfo{volume}{26}: \bibinfo{pages}{214}.

\bibtype{Article}%
\bibitem[Shkolnik et al.(2009)]{Shkoetal09}
\bibinfo{author}{Shkolnik E}, \bibinfo{author}{Liu MC} and
  \bibinfo{author}{Reid IN} (\bibinfo{year}{2009}).
\bibinfo{title}{Identifying the young low-mass stars within 25 pc.
  \uppercase{I. S}pectroscopic observations}.
\bibinfo{journal}{{\em ApJ}} \bibinfo{volume}{699} (\bibinfo{number}{1}):
  \bibinfo{pages}{649--666}.

\bibtype{Article}%
\bibitem[Simon and Sturm(1994)]{SimoStur94}
\bibinfo{author}{Simon KP} and  \bibinfo{author}{Sturm E}
  (\bibinfo{year}{1994}).
\bibinfo{title}{Disentangling of composite spectra}.
\bibinfo{journal}{{\em A\&A}} \bibinfo{volume}{281}: \bibinfo{pages}{286--291}.

\bibtype{Article}%
\bibitem[Sim{\'o}n-D{\'\i}az et al.(2015)]{SimDetal15a}
\bibinfo{author}{Sim{\'o}n-D{\'\i}az S}, \bibinfo{author}{Caballero JA},
  \bibinfo{author}{Lorenzo J}, \bibinfo{author}{Ma{\'\i}z~Apell{\'a}niz J},
  \bibinfo{author}{Schneider FRN}, \bibinfo{author}{Negueruela I},
  \bibinfo{author}{Barb{\'a} RH}, \bibinfo{author}{Dorda R},
  \bibinfo{author}{Marco A}, \bibinfo{author}{Montes D},
  \bibinfo{author}{Pellerin A}, \bibinfo{author}{Sanchez-Bermudez J},
  \bibinfo{author}{S{\'o}dor {\'A}} and  \bibinfo{author}{Sota A}
  (\bibinfo{year}{2015}).
\bibinfo{title}{Orbital and physical properties of the {$\sigma$}
  \uppercase{O}ri \uppercase{A}a, \uppercase{A}b, \uppercase{B} triple system}.
\bibinfo{journal}{{\em ApJ}} \bibinfo{volume}{799}, \bibinfo{eid}{169}.

\bibtype{Article}%
\bibitem[Smith(2011)]{Smit11a}
\bibinfo{author}{Smith N} (\bibinfo{year}{2011}).
\bibinfo{title}{Explosions triggered by violent binary-star collisions:
  application to $\eta$ \uppercase{C}arinae and other eruptive transients}.
\bibinfo{journal}{{\em MNRAS}} \bibinfo{volume}{415}:
  \bibinfo{pages}{2020--2024}.

\bibtype{Article}%
\bibitem[Smith and Frew(2011)]{SmitFrew11}
\bibinfo{author}{Smith N} and  \bibinfo{author}{Frew DJ}
  (\bibinfo{year}{2011}).
\bibinfo{title}{A revised historical light curve of $\eta$ \uppercase{C}arinae
  and the timing of close periastron encounters}.
\bibinfo{journal}{{\em MNRAS}} \bibinfo{volume}{415}:
  \bibinfo{pages}{2009--2019}.

\bibtype{Article}%
\bibitem[Smith et al.(1995)]{Smitetal95}
\bibinfo{author}{Smith VV}, \bibinfo{author}{Plez B}, \bibinfo{author}{Lambert
  DL} and  \bibinfo{author}{Lubowich DA} (\bibinfo{year}{1995}).
\bibinfo{title}{A survey of lithium in the red giants of the
  \uppercase{M}agellanic \uppercase{C}louds}.
\bibinfo{journal}{{\em ApJ}} \bibinfo{volume}{441}: \bibinfo{pages}{735}.

\bibtype{Article}%
\bibitem[Smith et al.(1996)]{Smitetal96}
\bibinfo{author}{Smith LF}, \bibinfo{author}{Shara MM} and
  \bibinfo{author}{Moffat AFJ} (\bibinfo{year}{1996}).
\bibinfo{title}{A three-dimensional classification for \uppercase{WN} stars}.
\bibinfo{journal}{{\em MNRAS}} \bibinfo{volume}{281} (\bibinfo{number}{1}):
  \bibinfo{pages}{163--191}.

\bibtype{Article}%
\bibitem[Smith~Neubig and Bruhweiler(1997)]{SmiNBruh97}
\bibinfo{author}{Smith~Neubig MM} and  \bibinfo{author}{Bruhweiler FC}
  (\bibinfo{year}{1997}).
\bibinfo{title}{\uppercase{UV} spectral classification of \uppercase{O} and
  \uppercase{B} stars in the \uppercase{S}mall \uppercase{M}agellanic
  \uppercase{C}loud}.
\bibinfo{journal}{{\em AJ}} \bibinfo{volume}{114}: \bibinfo{pages}{1951}.

\bibtype{Article}%
\bibitem[Smith~Neubig and Bruhweiler(1999)]{SmiNBruh99}
\bibinfo{author}{Smith~Neubig MM} and  \bibinfo{author}{Bruhweiler FC}
  (\bibinfo{year}{1999}).
\bibinfo{title}{Ultraviolet spectral classification of \uppercase{O} and
  \uppercase{B} stars in the \uppercase{L}arge \uppercase{M}agellanic
  \uppercase{C}loud}.
\bibinfo{journal}{{\em AJ}} \bibinfo{volume}{117}: \bibinfo{pages}{2856}.

\bibtype{Article}%
\bibitem[Solf(1978)]{Solf78}
\bibinfo{author}{Solf J} (\bibinfo{year}{1978}).
\bibinfo{title}{Spectral type and luminosity classification of late-type
  \uppercase{M} stars from near-infrared image tube coud{\'e} spectrograms.}
\bibinfo{journal}{{\em A\&AS}} \bibinfo{volume}{34}: \bibinfo{pages}{409--416}.

\bibtype{Article}%
\bibitem[Sota et al.(2011)]{Sotaetal11a}
\bibinfo{author}{Sota A}, \bibinfo{author}{Ma{\'\i}z~Apell{\'a}niz J},
  \bibinfo{author}{Walborn NR}, \bibinfo{author}{Alfaro EJ},
  \bibinfo{author}{Barb{\'a} RH}, \bibinfo{author}{Morrell NI},
  \bibinfo{author}{Gamen RC} and  \bibinfo{author}{Arias JI}
  (\bibinfo{year}{2011}).
\bibinfo{title}{The \uppercase{G}alactic \uppercase{O}-\uppercase{S}tar
  \uppercase{S}pectroscopic \uppercase{S}urvey. \uppercase{I}.
  \uppercase{C}lassification system and bright northern stars in the
  blue-violet at \uppercase{R}$\sim$2500}.
\bibinfo{journal}{{\em ApJS}} \bibinfo{volume}{193}: \bibinfo{pages}{24}.

\bibtype{Article}%
\bibitem[Sota et al.(2014)]{Sotaetal14}
\bibinfo{author}{Sota A}, \bibinfo{author}{Ma{\'\i}z~Apell{\'a}niz J},
  \bibinfo{author}{Morrell NI}, \bibinfo{author}{Barb{\'a} RH},
  \bibinfo{author}{Walborn NR}, \bibinfo{author}{Gamen RC},
  \bibinfo{author}{Arias JI} and  \bibinfo{author}{Alfaro EJ}
  (\bibinfo{year}{2014}).
\bibinfo{title}{The \uppercase{G}alactic \uppercase{O}-\uppercase{S}tar
  \uppercase{S}pectroscopic \uppercase{S}urvey (\uppercase{GOSSS}).
  \uppercase{II}. \uppercase{B}right southern stars}.
\bibinfo{journal}{{\em ApJS}} \bibinfo{volume}{211}, \bibinfo{eid}{10}.

\bibtype{Article}%
\bibitem[Struve(1929)]{Stru29}
\bibinfo{author}{Struve O} (\bibinfo{year}{1929}).
\bibinfo{title}{The \uppercase{S}tark effect as a means of determining
  comparative absolute magnitudes}.
\bibinfo{journal}{{\em ApJ}} \bibinfo{volume}{70}: \bibinfo{pages}{237--242}.

\bibtype{Article}%
\bibitem[Tabernero et al.(2021)]{Tabeetal21}
\bibinfo{author}{Tabernero HM}, \bibinfo{author}{Dorda R},
  \bibinfo{author}{Negueruela I} and  \bibinfo{author}{Marfil E}
  (\bibinfo{year}{2021}).
\bibinfo{title}{The nature of \uppercase{VX S}agitarii. is it a
  \uppercase{T{\.Z}O}, a \uppercase{RSG}, or a high-mass \uppercase{AGB} star?}
\bibinfo{journal}{{\em A\&A}} \bibinfo{volume}{646}, \bibinfo{eid}{A98}.

\bibtype{Inproceedings}%
\bibitem[Tarter(1976)]{Tart76}
\bibinfo{author}{Tarter JC} (\bibinfo{year}{1976}), \bibinfo{title}{Brown
  dwarfs, lilliputian stars, giant planets and missing mass problems.},
  \bibinfo{booktitle}{Bulletin of the American Astronomical Society},
  \bibinfo{volume}{8}, pp. \bibinfo{pages}{517}.

\bibtype{Article}%
\bibitem[Valdes et al.(2004)]{Valdetal04}
\bibinfo{author}{Valdes F}, \bibinfo{author}{Gupta R}, \bibinfo{author}{Rose
  JA}, \bibinfo{author}{Singh HP} and  \bibinfo{author}{Bell DJ}
  (\bibinfo{year}{2004}).
\bibinfo{title}{The \uppercase{I}ndo-\uppercase{US} library of
  \uppercase{C}oud{\'e} feed stellar spectra}.
\bibinfo{journal}{{\em ApJS}} \bibinfo{volume}{152}: \bibinfo{pages}{251--259}.

\bibtype{Article}%
\bibitem[Vink~et al.(2023)]{Vinketal23}
\bibinfo{author}{Vink~et al. } (\bibinfo{year}{2023}).
\bibinfo{title}{X-\uppercase{S}hooting \uppercase{ULLYSES}: \uppercase{M}assive
  stars at low metallicity. \uppercase{I}. \uppercase{P}roject description}.
\bibinfo{journal}{{\em A\&A}} \bibinfo{volume}{675}, \bibinfo{eid}{A154}.

\bibtype{Article}%
\bibitem[Wade et al.(2020)]{Wadeetal20}
\bibinfo{author}{Wade GA}, \bibinfo{author}{Bagnulo S},
  \bibinfo{author}{Keszthelyi Z}, \bibinfo{author}{Folsom CP},
  \bibinfo{author}{Alecian E}, \bibinfo{author}{Castro N},
  \bibinfo{author}{David-Uraz A}, \bibinfo{author}{Fossati L},
  \bibinfo{author}{Petit V}, \bibinfo{author}{Shultz ME} and
  \bibinfo{author}{Sikora J} (\bibinfo{year}{2020}).
\bibinfo{title}{No evidence of a sudden change of spectral appearance or
  magnetic field strength of the \uppercase{O9.7}~v star \uppercase{HD}
  54\,879}.
\bibinfo{journal}{{\em MNRAS}} \bibinfo{volume}{492} (\bibinfo{number}{1}):
  \bibinfo{pages}{L1--L5}.

\bibtype{Article}%
\bibitem[Walborn(1971{\natexlab{a}})]{Walb71c}
\bibinfo{author}{Walborn NR} (\bibinfo{year}{1971}{\natexlab{a}}).
\bibinfo{title}{On the existence of \uppercase{OB} stars with anomalous
  nitrogen and carbon spectra}.
\bibinfo{journal}{{\em ApJL}} \bibinfo{volume}{164}: \bibinfo{pages}{L67--L69}.

\bibtype{Article}%
\bibitem[Walborn(1971{\natexlab{b}})]{Walb71b}
\bibinfo{author}{Walborn NR} (\bibinfo{year}{1971}{\natexlab{b}}).
\bibinfo{title}{Some extremely early \uppercase{O} stars near $\eta$
  \uppercase{C}arinae}.
\bibinfo{journal}{{\em ApJL}} \bibinfo{volume}{167}: \bibinfo{pages}{L31--L33}.

\bibtype{Article}%
\bibitem[Walborn(1971{\natexlab{c}})]{Walb71a}
\bibinfo{author}{Walborn NR} (\bibinfo{year}{1971}{\natexlab{c}}).
\bibinfo{title}{Some spectroscopic characteristics of the \uppercase{OB} stars:
  an investigation of the space distribution of certain \uppercase{OB} stars
  and the reference frame of the classification}.
\bibinfo{journal}{{\em ApJS}} \bibinfo{volume}{23}: \bibinfo{pages}{257--282}.

\bibtype{Article}%
\bibitem[Walborn(1972)]{Walb72}
\bibinfo{author}{Walborn NR} (\bibinfo{year}{1972}).
\bibinfo{title}{Spectral classification of \uppercase{OB} stars in both
  hemispheres and the absolute magnitude calibration}.
\bibinfo{journal}{{\em AJ}} \bibinfo{volume}{77}: \bibinfo{pages}{312--318}.

\bibtype{Article}%
\bibitem[Walborn(1974)]{Walb74}
\bibinfo{author}{Walborn NR} (\bibinfo{year}{1974}).
\bibinfo{title}{Some morphological properties of \uppercase{WN} spectra}.
\bibinfo{journal}{{\em ApJ}} \bibinfo{volume}{189}: \bibinfo{pages}{269--271}.

\bibtype{Article}%
\bibitem[Walborn(1976)]{Walb76}
\bibinfo{author}{Walborn NR} (\bibinfo{year}{1976}).
\bibinfo{title}{The \uppercase{OBN} and \uppercase{OBC} stars}.
\bibinfo{journal}{{\em ApJ}} \bibinfo{volume}{205}: \bibinfo{pages}{419--425}.

\bibtype{Article}%
\bibitem[Walborn(1982{\natexlab{a}})]{Walb82b}
\bibinfo{author}{Walborn NR} (\bibinfo{year}{1982}{\natexlab{a}}).
\bibinfo{title}{The \uppercase{O}3 stars}.
\bibinfo{journal}{{\em ApJL}} \bibinfo{volume}{254}: \bibinfo{pages}{L15--L17}.

\bibtype{Article}%
\bibitem[Walborn(1982{\natexlab{b}})]{Walb82c}
\bibinfo{author}{Walborn NR} (\bibinfo{year}{1982}{\natexlab{b}}).
\bibinfo{title}{\uppercase{O}fpe/\uppercase{WN}9 circumstellar shells in the
  \uppercase{L}arge \uppercase{M}agellanic \uppercase{C}loud}.
\bibinfo{journal}{{\em ApJ}} \bibinfo{volume}{256}: \bibinfo{pages}{452--459}.

\bibtype{Inproceedings}%
\bibitem[Walborn(2011)]{Walb11b}
\bibinfo{author}{Walborn NR} (\bibinfo{year}{2011}),
  \bibinfo{title}{Morphological characteristics of \uppercase{OB} spectra and
  environments}, \bibinfo{booktitle}{Highlights of Spanish Astrophysics VI},
  \bibinfo{pages}{25--35}.

\bibtype{Article}%
\bibitem[Walborn and Fitzpatrick(1990)]{WalbFitz90}
\bibinfo{author}{Walborn NR} and  \bibinfo{author}{Fitzpatrick EL}
  (\bibinfo{year}{1990}).
\bibinfo{title}{Contemporary optical spectral classification of the
  \uppercase{OB} stars - \uppercase{A} digital atlas}.
\bibinfo{journal}{{\em PASP}} \bibinfo{volume}{102}: \bibinfo{pages}{379--411}.

\bibtype{Article}%
\bibitem[Walborn and Fitzpatrick(2000)]{WalbFitz00}
\bibinfo{author}{Walborn NR} and  \bibinfo{author}{Fitzpatrick EL}
  (\bibinfo{year}{2000}).
\bibinfo{title}{The \uppercase{OB} zoo: \uppercase{A} digital atlas of peculiar
  spectra}.
\bibinfo{journal}{{\em PASP}} \bibinfo{volume}{112}: \bibinfo{pages}{50--64}.

\bibtype{Book}%
\bibitem[Walborn et al.(1985)]{Walbetal85}
\bibinfo{author}{Walborn NR}, \bibinfo{author}{Nichols-Bohlin J} and
  \bibinfo{author}{Panek RJ} (\bibinfo{year}{1985}).
\bibinfo{title}{\uppercase{IUE} Atlas of \uppercase{O}-Type Spectra from 1200
  to 1900 Angstroms}, \bibinfo{publisher}{NASA RP-1155}.

\bibtype{Book}%
\bibitem[Walborn et al.(1995)]{Walbetal95a}
\bibinfo{author}{Walborn NR}, \bibinfo{author}{Parker JW} and
  \bibinfo{author}{Nichols-Bohlin J} (\bibinfo{year}{1995}).
\bibinfo{title}{\uppercase{IUE} Atlas of \uppercase{B}-Type Spectra from 1200
  to 1900 Angstroms}, \bibinfo{publisher}{NASA RP-1363}.

\bibtype{Article}%
\bibitem[Walborn et al.(2002)]{Walbetal02b}
\bibinfo{author}{Walborn NR}, \bibinfo{author}{Howarth ID},
  \bibinfo{author}{Lennon DJ}, \bibinfo{author}{Massey P}, \bibinfo{author}{Oey
  MS}, \bibinfo{author}{Moffat AFJ}, \bibinfo{author}{Skalkowski G},
  \bibinfo{author}{Morrell NI}, \bibinfo{author}{Drissen L} and
  \bibinfo{author}{Parker JW} (\bibinfo{year}{2002}).
\bibinfo{title}{A new spectral classification system for the earliest
  \uppercase{O} stars: \uppercase{D}efinition of type \uppercase{O}2}.
\bibinfo{journal}{{\em AJ}} \bibinfo{volume}{123}: \bibinfo{pages}{2754--2771}.

\bibtype{Article}%
\bibitem[Walborn et al.(2010)]{Walbetal10a}
\bibinfo{author}{Walborn NR}, \bibinfo{author}{Sota A},
  \bibinfo{author}{Ma{\'\i}z~Apell{\'a}niz J}, \bibinfo{author}{Alfaro EJ},
  \bibinfo{author}{Morrell NI}, \bibinfo{author}{Barb{\'a} RH},
  \bibinfo{author}{Arias JI} and  \bibinfo{author}{Gamen RC}
  (\bibinfo{year}{2010}).
\bibinfo{title}{Early results from the \uppercase{G}alactic \uppercase{O-S}tar
  \uppercase{S}pectroscopic \uppercase{S}urvey: \uppercase{C III} emission
  lines in \uppercase{O}f spectra}.
\bibinfo{journal}{{\em ApJL}} \bibinfo{volume}{711}:
  \bibinfo{pages}{L143--L147}.

\bibtype{Article}%
\bibitem[Walborn et al.(2011)]{Walbetal11}
\bibinfo{author}{Walborn NR}, \bibinfo{author}{Ma{\'\i}z~Apell{\'a}niz J},
  \bibinfo{author}{Sota A}, \bibinfo{author}{Alfaro EJ},
  \bibinfo{author}{Morrell NI}, \bibinfo{author}{Barb{\'a} RH},
  \bibinfo{author}{Arias JI} and  \bibinfo{author}{Gamen RC}
  (\bibinfo{year}{2011}).
\bibinfo{title}{Further results from the \uppercase{G}alactic
  \uppercase{O-S}tar \uppercase{S}pectroscopic \uppercase{S}urvey:
  \uppercase{R}apidly rotating late \uppercase{ON} giants}.
\bibinfo{journal}{{\em AJ}} \bibinfo{volume}{142}, \bibinfo{eid}{150}.

\bibtype{Article}%
\bibitem[Walborn et al.(2014)]{Walbetal14}
\bibinfo{author}{Walborn NR}, \bibinfo{author}{Sana H},
  \bibinfo{author}{Sim{\'o}n-D{\'\i}az S},
  \bibinfo{author}{Ma{\'\i}z~Apell{\'a}niz J}, \bibinfo{author}{Taylor WD},
  \bibinfo{author}{Evans CJ}, \bibinfo{author}{Markova N},
  \bibinfo{author}{Lennon DJ} and  \bibinfo{author}{de~Koter A}
  (\bibinfo{year}{2014}).
\bibinfo{title}{The \uppercase{VLT-FLAMES} \uppercase{T}arantula
  \uppercase{S}urvey. \uppercase{XIV}. the \uppercase{O}-type stellar content
  of 30 \uppercase{D}oradus}.
\bibinfo{journal}{{\em A\&A}} \bibinfo{volume}{564}, \bibinfo{eid}{A40}.

\bibtype{Article}%
\bibitem[Walborn et al.(2015{\natexlab{a}})]{Walbetal15b}
\bibinfo{author}{Walborn NR}, \bibinfo{author}{Morrell NI},
  \bibinfo{author}{Naz{\'e} Y}, \bibinfo{author}{Wade GA},
  \bibinfo{author}{Bagnulo S}, \bibinfo{author}{Barb{\'a} RH},
  \bibinfo{author}{Ma{\'\i}z~Apell{\'a}niz J}, \bibinfo{author}{Howarth ID},
  \bibinfo{author}{Evans CJ} and  \bibinfo{author}{Sota A}
  (\bibinfo{year}{2015}{\natexlab{a}}).
\bibinfo{title}{Spectral variations of \uppercase{O}f?p oblique magnetic
  rotator candidates in the \uppercase{M}agellanic \uppercase{C}louds}.
\bibinfo{journal}{{\em AJ}} \bibinfo{volume}{150}, \bibinfo{eid}{99}.

\bibtype{Article}%
\bibitem[Walborn et al.(2015{\natexlab{b}})]{Walbetal15a}
\bibinfo{author}{Walborn NR}, \bibinfo{author}{Sana H}, \bibinfo{author}{Evans
  CJ}, \bibinfo{author}{Taylor WD}, \bibinfo{author}{Sabbi E},
  \bibinfo{author}{Barb{\'a} RH}, \bibinfo{author}{Morrell NI},
  \bibinfo{author}{Ma{\'\i}z~Apell{\'a}niz J}, \bibinfo{author}{Sota A},
  \bibinfo{author}{Dufton PL}, \bibinfo{author}{McEvoy CM},
  \bibinfo{author}{Clark JS}, \bibinfo{author}{Markova N} and
  \bibinfo{author}{Ulaczyk K} (\bibinfo{year}{2015}{\natexlab{b}}).
\bibinfo{title}{Broad \uppercase{B}almer wings in \uppercase{BA}
  hyper/supergiants distorted by \uppercase{D}iffuse \uppercase{I}nterstellar
  \uppercase{B}ands: \uppercase{F}ive examples in the 30 \uppercase{D}oradus
  region from the \uppercase{VLT-FLAMES} \uppercase{T}arantula
  \uppercase{S}urvey}.
\bibinfo{journal}{{\em ApJ}} \bibinfo{volume}{809}, \bibinfo{eid}{109}.

\bibtype{Article}%
\bibitem[Walborn et al.(2017)]{Walbetal17}
\bibinfo{author}{Walborn NR}, \bibinfo{author}{Gamen RC},
  \bibinfo{author}{Morrell NI}, \bibinfo{author}{Barb{\'a} RH},
  \bibinfo{author}{Fern{\'a}ndez~Laj{\'u}s E} and  \bibinfo{author}{Angeloni R}
  (\bibinfo{year}{2017}).
\bibinfo{title}{Active \uppercase{L}uminous \uppercase{B}lue
  \uppercase{V}ariables in the \uppercase{L}arge \uppercase{M}agellanic
  \uppercase{C}loud}.
\bibinfo{journal}{{\em AJ}} \bibinfo{volume}{154}, \bibinfo{eid}{15}.

\bibtype{Article}%
\bibitem[Weiler et al.(2023)]{Weiletal23}
\bibinfo{author}{Weiler M}, \bibinfo{author}{Carrasco JM},
  \bibinfo{author}{Fabricius C} and  \bibinfo{author}{Jordi C}
  (\bibinfo{year}{2023}).
\bibinfo{title}{Analysing spectral lines in \uppercase{G}aia low-resolution
  spectra}.
\bibinfo{journal}{{\em A\&A}} \bibinfo{volume}{671}, \bibinfo{eid}{A52}.

\bibtype{Article}%
\bibitem[White and Basri(2003)]{WhitBasr03}
\bibinfo{author}{White RJ} and  \bibinfo{author}{Basri G}
  (\bibinfo{year}{2003}).
\bibinfo{title}{Very low mass stars and brown dwarfs in
  \uppercase{T}aurus-\uppercase{A}uriga}.
\bibinfo{journal}{{\em ApJ}} \bibinfo{volume}{582} (\bibinfo{number}{2}):
  \bibinfo{pages}{1109--1122}.

\bibtype{Article}%
\bibitem[Wildt(1936)]{Wild36}
\bibinfo{author}{Wildt R} (\bibinfo{year}{1936}).
\bibinfo{title}{Low-dispersion spectra of red stars}.
\bibinfo{journal}{{\em ApJ}} \bibinfo{volume}{84}: \bibinfo{pages}{303}.

\bibtype{Article}%
\bibitem[Wilking et al.(2005)]{Wilketal05}
\bibinfo{author}{Wilking BA}, \bibinfo{author}{Meyer MR},
  \bibinfo{author}{Robinson JG} and  \bibinfo{author}{Greene TP}
  (\bibinfo{year}{2005}).
\bibinfo{title}{Optical spectroscopy of the surface population of the
  {\ensuremath{\rho}} \uppercase{O}phiuchi molecular cloud: \uppercase{T}he
  first wave of star formation}.
\bibinfo{journal}{{\em AJ}} \bibinfo{volume}{130} (\bibinfo{number}{4}):
  \bibinfo{pages}{1733--1751}.

\bibtype{Article}%
\bibitem[Wolf and Rayet(1867)]{WolfRaye67}
\bibinfo{author}{Wolf C} and  \bibinfo{author}{Rayet G} (\bibinfo{year}{1867}).
\bibinfo{title}{Spectroscopie stellaire}.
\bibinfo{journal}{{\em Academie des Sciences Paris Comptes Rendus}}
  \bibinfo{volume}{65}: \bibinfo{pages}{292--296}.

\bibtype{Article}%
\bibitem[Wright~et al.(2010)]{Wrigetal10b}
\bibinfo{author}{Wright~et al. } (\bibinfo{year}{2010}).
\bibinfo{title}{The \uppercase{W}ide-field \uppercase{I}nfrared
  \uppercase{S}urvey \uppercase{E}xplorer \uppercase{(WISE): M}ission
  description and initial on-orbit performance}.
\bibinfo{journal}{{\em AJ}} \bibinfo{volume}{140}, \bibinfo{eid}{1868}.

\end{thebibliography*}

\end{document}